\documentclass{aa}  

\usepackage{graphicx}
\usepackage{txfonts}
\usepackage{booktabs}
\usepackage{color}
\definecolor{mhi}{rgb}{0.6,0,0.6}

\definecolor{msp}{rgb}{0.0,0.6,0.6}

\begin{document}

   \title{The dwarf galaxy satellite system of Centaurus A\thanks{Based on observations collected at the European Organisation for Astronomical Research in the Southern Hemisphere 
under ESO program 0101.A-0193(A). The photometric catalogs are available in electronic form at the
CDS via anonymous ftp to cdsarc.u-strasbg.fr (130.79.128.5) or via
http://cdsweb.u-strasbg.fr/cgi-bin/qcat?J/A+A/.}}
   \author{
          Oliver M\"uller\inst{1}
          \and
          Marina Rejkuba\inst{2}
          \and 
          Marcel S. Pawlowski\inst{3}
          \and 
          Rodrigo Ibata\inst{1}
          \and 
          Federico Lelli\inst{2}
          \and
          Michael Hilker\inst{2}
          \and  
          Helmut Jerjen\inst{4}
          }

 \institute{Observatoire Astronomique de Strasbourg  (ObAS),
Universite de Strasbourg - CNRS, UMR 7550 Strasbourg, France\\
 \email{oliver.muller@astro.unistra.fr}
\and
 European Southern Observatory, Karl-Schwarzschild Strasse 2, 85748, Garching, Germany
\and
Leibniz-Institut fur Astrophysik Potsdam (AIP), An der Sternwarte 16, D-14482 Potsdam, Germany
\and
 Research School of Astronomy and Astrophysics, Australian National University, Canberra,
ACT 2611, Australia
}

   \date{Received tba; accepted tba}

 
  \abstract{Dwarf galaxy satellite systems are essential probes to test models of structure formation, making it necessary to establish a census of dwarf galaxies outside of our own 
  Local Group.
  {We present deep {FORS2} $VI$ band images from the ESO Very Large Telescope (VLT) for 15 dwarf galaxy candidates in the Centaurus group of galaxies. We} confirm nine dwarfs to be members of Cen\,A by measuring their distances
  {using a Bayesian approach to determine the tip of the red giant branch luminosity.}
  {We have also fit theoretical isochrones to measure their mean metallicities.} The properties of the {new} dwarfs are similar to those in the Local Group {in terms of their sizes, luminosities, and mean metallicities.} {Within our photometric precision, there is no evidence of a metallicity spread, but we do observe possible extended star formation in several galaxies, as evidenced by a population of asymptotic giant branch stars brighter than the red giant branch tip.} The new dwarfs do not show  any signs of tidal disruption.
Together with the recently reported dwarf galaxies by the complementary PISCeS survey, we study the luminosity function and 3D structure of the group. {By comparing the observed luminosity function 
to the high-resolution cosmological simulation IllustrisTNG, we find agreement within a 90$\%$ confidence interval. However, Cen\,A seems to be missing its brightest satellites and has an overabundance of the faintest dwarfs in comparison to its simulated analogs. In terms of the overall 3D distribution of the observed satellites, we find that the whole structure is flattened along the line-of-sight{, with a root-mean-square (rms) height of 130\,kpc and an rms semi-major axis length of 330\,kpc}.
Future distance measurements of the remaining dwarf galaxy candidates 
are needed to complete the census of dwarf galaxies in the Centaurus group.} 
}

   \maketitle
%

\section{Introduction}
The hunt for 
dwarf galaxies 
in the Local Volume ({LV,} $D<11$\,Mpc, \citealt{1979AN....300..181K,2004AJ....127.2031K,2013AJ....145..101K}) has become widely popular, with several independent teams searching for nearby dwarf galaxies \citep{2009AJ....137.3009C,2014ApJ...787L..37M,2015A&A...583A..79M,2017A&A...597A...7M,2017A&A...602A.119M,2018A&A...615A.105M,2016ApJ...828L...5C,2016ApJ...823...19C,2016A&A...588A..89J,2017ApJ...848...19P,2017Ap.....60..295B,2018ApJ...863..152S}. {The increased pace of new discoveries reflects the improvements in survey capabilities, but also the growing interest in dwarf galaxies as}
direct probes of cosmological models on small scales \citep[e.g.,][]{2017ARA&A..55..343B}. Key motivators are the long-standing missing satellite problem \citep{1999ApJ...524L..19M}, the too-big-to-fail problem \citep{2010A&A...523A..32K,2011MNRAS.415L..40B}, the cusp-core problem \citep{2010AdAst2010E...5D}, the trinity of cosmological problems (which could be solved with baryonic physics), and lately the plane-of-satellites problem {(e.g., \citealt{2005A&A...431..517K,2006AJ....131.1405K,2013MNRAS.435.2116P,2013Natur.493...62I,2014ApJ...784L...6I,2018A&A...614A..59B,2018MNRAS.477.4768B,2018MNRAS.475.2754H}, see \citealt{2018MPLA...3330004P} for a recent review)}. The latter phenomenon is a more fundamental problem mostly unaffected by the baryonic processes within the galaxies \citep{2015ApJ...815...19P}. 

All these studies are mainly based on the dwarf galaxy system in the Local Group. Therefore, it remains a crucial task to test whether these problems persist in other galaxy groups. One of the closest galaxy groups in reach for such studies is the Centaurus group. 
{It consists of two main concentrations around the dominant early-type galaxy Cen\,A at $D\simeq3.8$ Mpc \citep{2004A&A...413..903R} and the secondary late-type galaxy M\,83 at $D\simeq4.8$ Mpc \citep{2008ApJ...683..630H}.}
Another, less massive galaxy (NGC\,4945),  with roughly 1/3 of the total luminosity of Cen\,A,  {is located}
at the outskirt of Cen\,A's virial radius {of $r_{vir}=409$\,kpc (\citealt{2015AJ....149...54T}, using a Hubble constant of 71\,km s$^{-1}$ Mpc$^{-1}$, \citealt{2011ApJS..192...18K})}. 
The Centaurus group resides at the border of the Local Void and is well isolated (see e.g., Fig.\,7 in \citealt{2018AJ....156..105A}). 
{Its} isolation makes it a well suited target for studies of the dwarf galaxy population since the contamination of background galaxies is low. However, due to the group's low Galactic latitude, Galactic cirrus becomes a major problem as its small-scale morphology can visually mimic low surface brightness dwarf galaxies. 
Therefore, it is expected that some dwarf galaxy candidates are in fact patches of Galactic cirrus. Only deeper follow-up observations can test the true membership  {within} the galaxy group. 

The Centaurus group has recently been targeted by three different surveys with the aim of discovering faint dwarf galaxies. 
The Magellan and Megacam-based PISCeS survey \citep{2014ApJ...795L..35C,2016ApJ...823...19C} screened an area of 50 square degrees around Cen\,A, 
{exploiting} the advantage
{of the 6.5m Magellan Clay telescope and long integration times. This enabled them to resolve individual bright red giant stars in the Cen\,A halo and its dwarf galaxy satellites.} 
Our own survey \citep{2015A&A...583A..79M,2017A&A...597A...7M}, conducted with the Dark Energy Camera (DECam), 
covered {a much larger area of} 550 square degrees {in $gr$ bands}, including most of the Centaurus association.
The SCABS survey \citep{2016arXiv160807285T,2017MNRAS.469.3444T,2018ApJ...867L..15T} covers an area of 21 square degrees around Cen\,A, employing five band imaging ($ugriz$) also with the DECam. 
In total, these surveys have more than doubled the currently known dwarf galaxy population around Cen\,A, {assuming that the} dwarf galaxy candidates are confirmed with follow-up measurements. Their surface brightnesses and estimated absolute magnitudes are in the range of the classical dwarfs, allowing {us} to perform similar cosmological tests as in the Local Group.

In the Centaurus group there is some evidence for planar structures \citep{2015ApJ...802L..25T,2015MNRAS.452.1052L,2016A&A...595A.119M,MuellerTRGB2018}, which is similar to the Local Group. 
{\citet{2015ApJ...802L..25T} suggested the existence of two, almost parallel planes of satellites,} but the subsequent discovery of new dwarf galaxies in the group \citep{2014ApJ...795L..35C,2016ApJ...823...19C,2017A&A...597A...7M} 
weakened the case of a strict separation between the two planes \citep{2016A&A...595A.119M}.
Furthermore, \citet{2018Sci...359..534M} studied the kinematics of the dwarf galaxies around Cen\,A and found a phase-space correlation of the satellites, with 14 out of 16 satellites seemingly co-rotating around their host. A comparison to $\Lambda$CDM simulations finds a similar degree of conflict as earlier studies 
{for the satellite systems in the Local Group}
\citep[e.g.,][]{2014ApJ...784L...6I}.

For many dwarf galaxies around Cen\,A, no velocity and distance measurements are available to date. 
Their memberships are mainly based on morphological arguments from integrated-light observations \citep{2015A&A...583A..79M,2017A&A...597A...7M,2018ApJ...867L..15T}. Deep follow-up {observations} are needed to 
derive distances with a  five percent accuracy, which
enables us to study the 3D structure of this galaxy group \citep[e.g.,\,][]{2012MNRAS.425..709M,2013AJ....146..126C,2017MNRAS.465.5026C,2017ApJ...837..136D,MuellerTRGB2018,2018ApJ...868...96C,2019ApJ...872...80C}. In this paper, we present follow-up observations for 15 dwarf galaxy candidates in the region of Cen\,A using the ESO Very Large Telescope {(VLT)}. 

This paper is structured as follows. In Section \ref{observations}, we present the observations, the data reduction, and photometry; in Section \ref{CMD}, we use the photometric catalog to produce color-magnitude diagrams and measure the distances of the dwarf galaxies; in Section \ref{sec:properties}, we characterize the newly confirmed dwarf galaxies; in Section \ref{LFcena}, we discuss the luminosity function of Cen\,A; in Section \ref{spatial}, we 
study the spatial distribution of the satellite system; and finally in Section \ref{summary}, we give a summary and conclusion.

\section{Observations and data reduction}
\label{observations}

\begin{table*}[ht]
\small
\caption{Observation summary for 15 dwarf galaxy candidates. 
}
\centering                          
\begin{tabular}{l l l l r l l l}        
\hline\hline                 
$\alpha_{2000}$ & $\delta_{2000}$ & Observing & Instrument & Exposure & Filter & Airmass & Image quality  \\    
(hh:mm:ss) & (dd:mm:ss) & Date &  & time (s) & & & (arcsec)\\    
(1) & (2) & (3)  & (4)  & (5) & (6) & (7) & (8)\\ 
\hline      \\[-2mm]                  
\bf{dw1315-45}\\
13:15:56 & $-$45:45:02 & 16/17 May 2018 & FORS2 & $3\times800 $ & V & 1.28 & $0.5-0.6$\\
 &  & 04/05 May 2018 & FORS2 & $12\times230 $ & I & 1.08 & $0.3-0.4$\\

\bf{KKs54}\\
13:21:32 & $-$31:53:11 & 14/15 Apr 2018 & FORS2 & $3\times800 $ & V & 1.01 & $0.6-0.7$\\
 &  & 13/14 Apr 2018 & FORS2 & $12\times230 $ & I & 1.05 & $0.3-0.4$ \& 0.5\\

 \bf{dw1318-44}\\
 13:18:58 & $-$44:53:41 & 15/16 May 2018 & FORS2 & $2\times800 $ & V &  & $>0.6$\\
& & 16/17 May 2018 & FORS2 & $3\times800 $ & V & 1.07 & 0.4\\
 &  & 04/05 May 2018 & FORS2 & $12\times230 $ & I & 1.07 & $0.3-0.4$\\
 
 \bf{dw1322-39}\\
13:22:32 & $-$39:54:20 & 15/16 May 2018 & FORS2 & $2\times800 $ & V &  & $>0.6$\\
& &16/17 May 2018 & FORS2 & $3\times800 $ & V & 1.04 & $0.4-0.5$\\
 &  & 04/05 May 2018 & FORS2 & $12\times230 $ & I & 1.11 & $<0.3$\\

 \bf{KK\,198}\\
13:22:56 & $-$33:34:22 & 18/19 Apr 2018 & FORS2 & $3\times800 $ & V & 1.21 & $>0.6$\\
& &19/20 Apr 2018 & FORS2 & $1\times136^*$ & V &  &  {not measured}\\
 &  & 04/05 May 2018 & FORS2 & $12\times230 $ & I & 1.04 & $0.5-0.6$\\
 
  \bf{dw1323-40c}\\
13:23:37 & $-$40:43:17 & 14/15 Apr 2018 & FORS2 & $3\times800 $ & V & 1.13 & $0.6-0.7$\\
 &  & 14/15 Apr 2018 & FORS2 & $12\times230 $ & I & 1.07 & $0.5-0.6$\\

\bf{dw1323-40b}\\
13:23:55 & $-$40:50:09 & 12/13 Apr 2018 & FORS2 & $3\times800 $ & V & 1.24 & $0.5-0.6$ \\
 &  & 12/13 Apr 2018 & FORS2 & $12\times230 $ & I & 1.13 & $0.3-0.4$\\
  
   \bf{dw1323-40}\\
13:24:53 & $-$40:45:41 & 09/10 Apr 2018 & FORS2 & $3\times800 $ & V & 1.31 & $0.5-0.6$\\
 &  & 08/09 Apr 2018 & FORS2 & $12\times230 $ & I & 1.05 & $0.4-0.5$\\

  \bf{dw1329-45}\\
13:29:10 & $-$45:10:31 & 12/13 Apr 2018 & FORS2 & $3\times800 $ & V & 1.12 & $0.5-0.6$\\
 &  & 09/10 Apr 2018 & FORS2 & $12\times230 $ & I & 1.11 & $0.3-0.4$\\
 
 \bf{dw1331-37}\\
13:31:32 & $-$37:03:29 & 09/10 Apr 2018 & FORS2 & $3\times800 $ & V & 1.03 & 0.4\\
 &  & 09/10 Apr 2018 & FORS2 & $12\times230 $ & I & 1.16 & 0.4\\

 \bf{dw1336-44}\\
13:36:44 & $-$44:26:50 & 12/13 Apr 2018 & FORS2 & $3\times800 $ & V & 1.08 & $0.3-0.4$\\
 &  & 09/10 Apr 2018 & FORS2 & $12\times230 $ & I & 1.08 & $0.3-0.4$\\
 
 \bf{dw1337-44}\\
13:37:34 & $-$44:13:07 & 09/10 Apr 2018 & FORS2 & $3\times800 $ & V & 1.16 & $0.5-0.6$\\
 &  & 08/09 Apr 2018 & FORS2 & $12\times230 $ & I & 1.07 & $0.3-0.4$\\

 \bf{dw1341-43}\\
13:41:37 & $-$43:51:17 & 08/09 Apr 2018 & FORS2 & $3\times800 $ & V & 1.11 & $0.5-0.6$\\
 &  & 08/09 Apr 2018 & FORS2 & $12\times230 $ & I & 1.19 & $0.4-0.5$\\
 
 \bf{dw1342-43}\\
13:42:44 & $-$43:15:19 & 16/17 May 2018 & FORS2 & $3\times800 $ & V & 1.12 & $0.4-0.5$\\
 &  & 04/05 May 2018 & FORS2 & $6\times230 $ & I & 1.13 & $0.3$ \\
 &  & 15/16 May 2018 & FORS2 & $12\times230 $ & I & 1.11 & $0.6-0.7$\\
 
 \bf{KKs58}\\
13:46:00 & $-$36:19:44 & 15/16 May 2018 & FORS2 & $3\times800 $ & V & 1.02 & $0.5$\\
 &  & 15/16 May 2018 & FORS2 & $12\times230 $ & I & 1.13 & $0.5, 0.6$\\
 
\hline
\end{tabular}
\tablefoot{(1) and (2): Coordinates of the dwarf galaxy candidates (epoch J2000); (3) Date of observation; 
(4): {Exposure time};
(5) Filter used for the observation; (6) Mean airmass during the observation; and (7): Average image quality.
{(*): KK198 V-band exposure on 19 and 20 Apr was aborted - here we report it for completeness, but the data was not used.}}
\label{fit} 
\end{table*}

Our targets were {12 out of the 57 dwarf galaxy candidates found by our team in the Cen\,A subgroup \citep{2017A&A...597A...7M} and three previously known dwarf galaxy candidates taken from the LV catalog \citep{2004AJ....127.2031K,2013AJ....145..101K}. 
They were selected based on our study of the planes around Cen\,A (see Table 2 of \citealt{2016A&A...595A.119M}) since the members could most likely simultaneously contribute to confirm, or dispute, the existence of the two planes.}

In total, 26 hours of VLT time in service mode were specifically allocated to the program 0101.A-0193(A). 
We requested excellent observing conditions, that is dark time and seeing better than $0\farcs 6$, which
is  necessary  to resolve the asymptotic giant branch (AGB) and the upper red giant branch (RGB) stars at a distance of up to $\sim 5.5$~Mpc \citep{MuellerTRGB2018}. 
Deep $V$ and $I$ CCD images were taken between April and May 2018 with the FOcal Reducer and low dispersion Spectrograph (FORS2) mounted on the UT1 of the VLT.
With a field of view of $6\farcm 8\times6\farcm 8,$ the FORS2 camera is equipped with a mosaic of two $2k \times 4k$ MIT CCDs.  When using the standard resolution (SR) and $2\times 2$ binning, this instrument offers a scale of $0\farcs 25$ per pixel. Since the dwarf galaxy candidates are small 
{on the sky,}
the targets were centered on Chip\,1.

In Table\,\ref{fit}, we provide a log
of the observations. We requested three $V$-band frames of 800\,sec, as well as 12 exposures in $I$-band of 230\,sec for each target. {We applied small offsets ($<$50 pixels) between individual exposures with the aim to improve sky subtraction and remove bad pixels. No rotation was applied.}
During some observations, the conditions suddenly worsened.\ This {led} to the cancellation and {partial} repetition of {some exposures}. In Table\,\ref{fit}, we also show the image quality measured {on the images}. {The sky transparency was clear or photometric during all observations, except for I-band data for KKs\,54.}
Mostly, the seeing remained constant over time. We indicate the range of measured image quality {if} there was a change during the image acquisition.

\subsection{Data reduction and calibration}
The data reduction was carried out using the python package Astropy \citep{astropy:2013,astropy:2018}, and the Image Reduction and Analysis Facility (IRAF) software \citep{1993ASPC...52..173T}. 
Within Astropy, a median combined stack of ten bias frames was used to create a master bias. This master bias was subtracted from the individual flat fields. The master flat fields were created by combining ten twilight flats for each filter using a median stack to reject any residual faint stars, bad pixels, and cosmic rays. Then the master flats were normalized by their median values.  
These final master calibration images, master bias and master flat, were used to process the individual science frames.

\begin{figure*}[ht]
\centering

\fbox{\includegraphics[width=3.3cm]{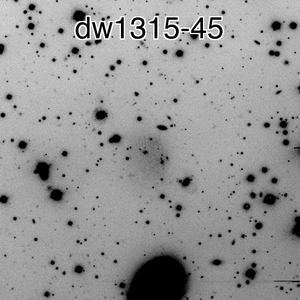}}
\fbox{\includegraphics[width=3.3cm]{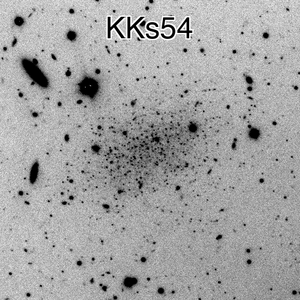}} 
\fbox{\includegraphics[width=3.3cm]{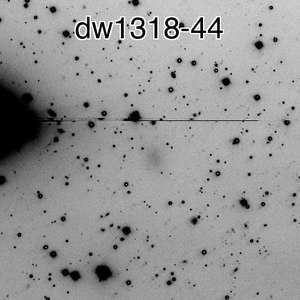}}
\fbox{\includegraphics[width=3.3cm]{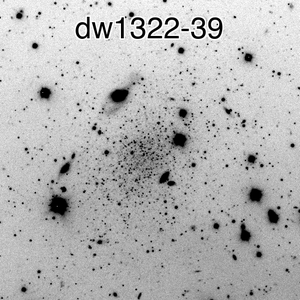}}
\fbox{\includegraphics[width=3.3cm]{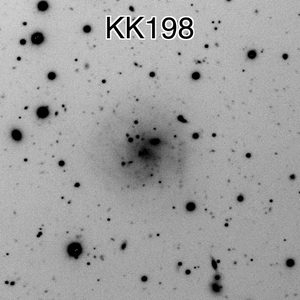}}

\fbox{\includegraphics[width=3.3cm]{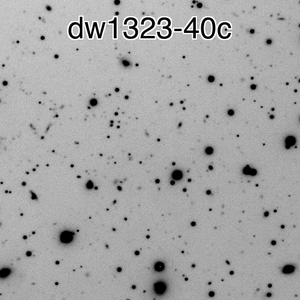}}
\fbox{\includegraphics[width=3.3cm]{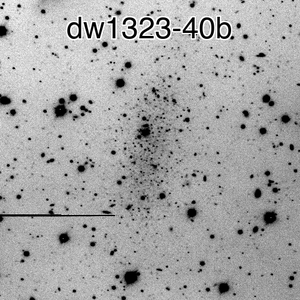}}
\fbox{\includegraphics[width=3.3cm]{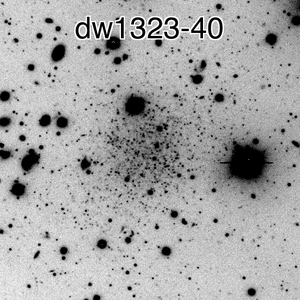}}
\fbox{\includegraphics[width=3.3cm]{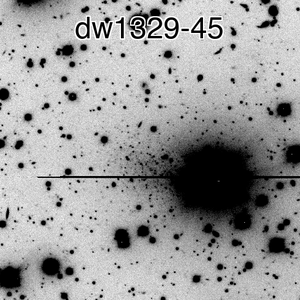}}
\fbox{\includegraphics[width=3.3cm]{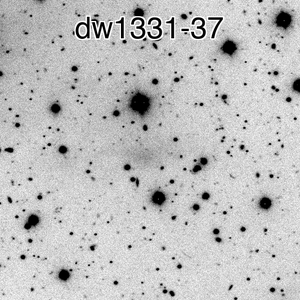}}

\fbox{\includegraphics[width=3.3cm]{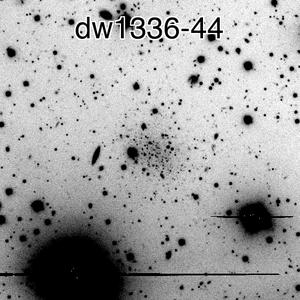}}
\fbox{\includegraphics[width=3.3cm]{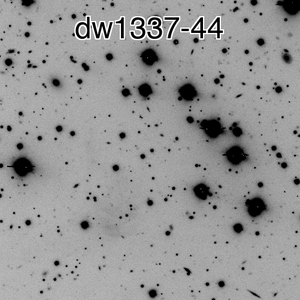}}
\fbox{\includegraphics[width=3.3cm]{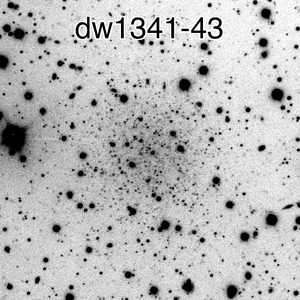}}
\fbox{\includegraphics[width=3.3cm]{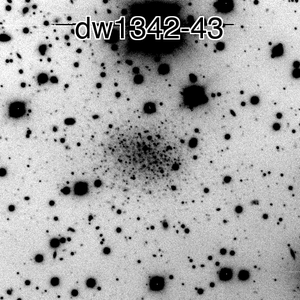}}
\fbox{\includegraphics[width=3.3cm]{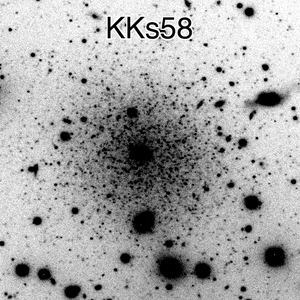}}
\caption{
FORS2\@ VLT images combined with MONTAGE. All images are oriented with north up and east to the left and {cover a $2\arcmin \times 2\arcmin$ area}.
}
\label{targetsVLT}
\end{figure*}

{Since the initial determination of image transformations with DAOMATCH failed, we decided to resample the individual, scientific, and fully calibrated images.}
To do so, we first generated  a combined image using IRAF with the command \textit{imcombine} utilizing all individual scientific images (using both photometric bands). 
This combined image contains the full coverage of the sky for the images for one galaxy. All values within this image were then set to zero using the \textit{imreplace} command to create an empty canvas. Again, using IRAF's \textit{imcombine} algorithm, we merged this canvas with each individual scientific image separately, resulting in a true correspondence between the pixel coordinate and the world coordinate, and additionally between each image. This gives us our final scientific images that are ready for photometry. We note that the resampling did not change the resolution of the individual images, nor did it change the flux. Cutouts of the final stacked images centered on the dwarf galaxy candidates are shown in Fig\,\ref{targetsVLT}.

We took the zero points, $ZP_{V}$ and $ZP_{I,}$ and the extinction coefficients, $k_{V}$ and  $k_{I,}$ from the ESO Quality Control webpages. In \citet{MuellerTRGB2018}, we checked and confirmed the consistency between these values 
and those derived from the standard fields by ourselves. Therefore, we use the former in the following {analysis.}
The photometry was calibrated using the formulae:
\begin{eqnarray*}
V &=& V_{instr} + ZP_V - k_V\cdot X_V - A_V + 2.5\cdot \log_{10}(t),\\
I &=& I_{instr} + ZP_I - k_I\cdot X_I - A_I + 2.5\cdot \log_{10}(t),
\end{eqnarray*}
\noindent where the exposure time $t$ is given for a single exposure (800 and 230 seconds for $V$ and $I$, respectively),  $X_V$ and $X_I$ are the mean airmasses during the observations, and  $A_V$ and $A_I$ are the Galactic extinction values based on the reddening map by \citet{1998ApJ...500..525S} and the correction coefficients from \citet{2011ApJ...737..103S}.

For one galaxy -- KKs\,54 -- a problem with the zero point became apparent during our analysis because all stars were systematically shifted toward a blueish color. {Indeed, by checking the Atmospheric Site Monitor\footnote{{http://www.eso.org/asm/ui/publicLog}} data,
we saw quite strong variation in sky transparency, indicative of clouds passing during the KKs\,54 I-band imaging observation}. Therefore, we had to 
{determine} the zero point for this galaxy {by calibrating our data with respect to public catalogs}. Most photometric surveys like Gaia \citep{2016A&A...595A...1G} or Skymapper \citep{2018PASA...35...10W} have no overlap with our {targets} in $I$-band: 
the faintest available stars in these photometric surveys are already saturated in our FORS2 images. Therefore, we used deep $ri$ images of the region around KKs\,54 from the DECam archive  (proposal ID: 2014A-0306) and calibrated them using APASS \citep{2014CoSka..43..518H}. 
In these DECam images, {there are} faint stars which {also} {have good quality magnitudes} in our FORS2 images. We could then derive the FORS2 offsets by directly matching the instrumental magnitudes {from}
FORS2 to the calibrated {magnitudes from} DECam, 
{adopting the following equation \citep{SloanConv}:}
\begin{eqnarray*} I = r - 1.2444\cdot(r - i)  - 0.3820
\end{eqnarray*}
{to transform to Peter Stetson's photometric calibration used by FORS2.}

\subsection{Photometry}
We carried out photometric measurements for all stars on the previously reduced science images using the stand-alone version of DAOPHOT2 \citep{1987PASP...99..191S} package, which is optimized for crowded fields. We followed the same procedure as in \citet{MuellerTRGB2018}, which includes the detection of point sources, aperture photometry with a small aperture on each such detection, and then point-spread function (PSF) modeling. 
The first approximation of the PSF model is done using 50 isolated bona fide stars across the field and, which is done on every individual science frame (i.e., 12 times in $I$ band and 3 times in $V$ band for every galaxy). 
To improve the PSF model, we checked the residuals of each star and rejected the stars with a strong residual, deriving a new PSF model, and repeatedly checked the residuals until there were no apparent {systematic} residuals in {subtracted images of stars used for the PSF model.}
To produce the deepest possible image (to find the faintest stars) we used MONTAGE2, 
{which stacks the $V$ and $I$ frames together.}
On this deep image, the DAOPHOT2 routines FIND, PHOT, and ALLSTAR were run
{producing} the deepest possible point source catalog. This catalog was then used as input for the simultaneous PSF fitting on every science frame  using ALLFRAME. {The advantages of using ALLFRAME to perform simultaneous consistent photometry on all images of a given field are described in detail by \citet{1994PASP..106..250S}. In essence, ALLFRAME photometry starts with the initial star list that is derived from the very deep image, made with MONTAGE2, and then it determines the brightness and centroids for all stars in each frame iteratively. In each iteration, the brightness and centroid corrections are computed from the residuals, after PSF subtraction, and are applied to improve the fit. Periodically, the underlying diffuse sky brightness is computed around each star's location after all stars have been subtracted from the frame. This procedure results in a more robust sky correction and stellar centroid determination, which is ultimately important for precise PSF photometry.}
The resulting catalogs for each individual image were then average combined, per filter, with the DAOMATCH and DAOMASTER routines. 
We kept only those 
objects that were measured on at least two thirds of all input images for the given filter, that is twice detected in the $V$ band and eight times in the $I$ band. The $V$ and $I$ catalogs were than merged with a 1~pix tolerance to create the final master catalog of all detected sources in our science images. 
 
\begin{figure*}[ht]
\centering
\includegraphics[width=18cm]{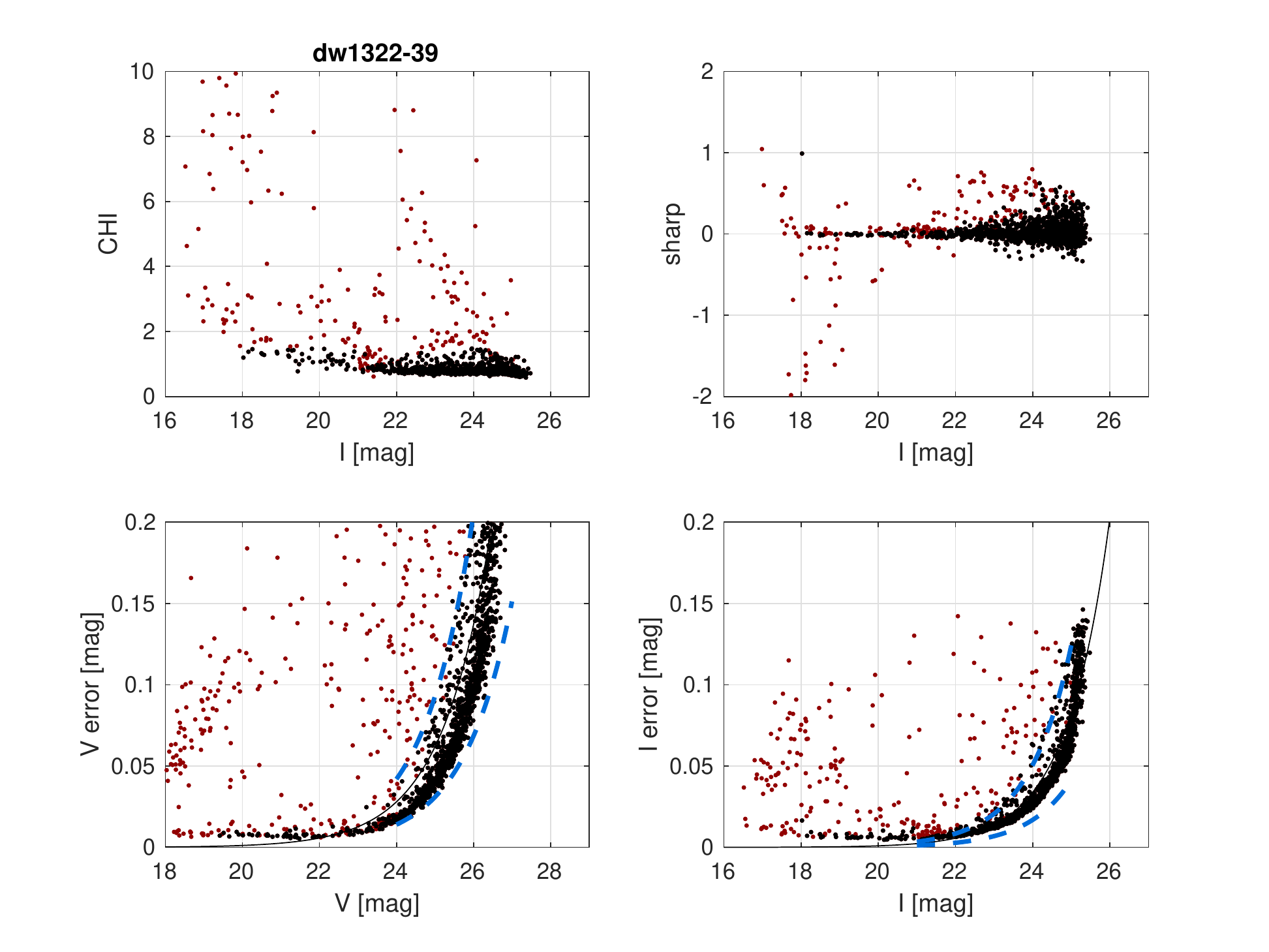}
\caption{
Quality cuts on photometric catalog of dw1322-39. The black dots correspond to stars fulfilling the constraints, while the red dots show rejected objects{; blue dashed lines in the bottom plots of magnitude vs. error indicate 50\% deviation from the best fitting value {(black line)} at the given magnitude}.}
\label{goodstars}
\end{figure*}

Subsequently, we purged this photometric catalog from non-stellar and blended sources. To do so, we applied quality cuts on the  $\chi$ and $sharp$ parameters, and the photometric errors.
 {We imposed that} $\chi$ had to be smaller than 1.5, $sharp$ had to be between $-2$ and $2$, and the magnitude errors were only allowed to deviate 50\% from the best fitting value at a given magnitude. 
In total, a detection has to fulfill these four constraints ($\chi$, $sharp$, $V_{err}$, and $I_{err}$) to remain in the catalog. See Fig.\,\ref{goodstars} for an example of the different constraints in the case of dw1322-39. 
 The red symbols are detections which were removed from the catalog 
 {since they did not satisfy}
 our quality constrains.

\subsection{Completeness and error analysis}
\label{artStarErr}
To assess the detection completeness and characterize the measurement errors for our PSF photometry, we performed 
artificial star experiment{s}. In each image we injected 900 artificial stars, all with the same magnitude, and repeated this in 0.5\, magnitude steps between 20--28\,mag {(i.e., 16 realizations)}. These stars were uniformly spread on a hexagonal grid over each field. The separation between stars corresponds to twice the PSF fitting radius (i.e., 40 pixels), such that the artificial stars do not increase the natural crowding, while {we} still 
{added a sufficiently large number of stars}
to get meaningful statistics out of the experiment. {We injected the same $V$ and $I$ instrumental magnitudes to the images, which translates into $(V-I)=0.6$\,mag when the zero points are applied. We did not make specific simulations that explore a wider range of colors.}
Since we resampled the scientific images, we could inject each star at the same pixel position throughout the different images of each galaxy. In this way, we created 16 mock-realizations for each galaxy. On these mock-realizations, we applied our PSF photometry pipeline as described before;\ the only difference was that we did not rederive the PSF model. The pipeline 
{produced}
new catalogs which include the photometry of the artificial stars. By comparing the number of detected artificial stars and their photometry, we 
derived the completeness and the photometric error. We present the completeness curves and photometric error as a function of the input magnitude for dw1322-39 in Fig.\,\ref{artTests} and all galaxies in the Appendix (Fig.\,\ref{app:artTests}). {At the faint end, below the 50\% completeness level, the detected artificial stars systematically deviate from the input magnitude.\ This is due to them overlapping with positive noise peaks, making them detectable in the first place, but also too bright.}
The 75\%, 50\%, and 25\% completeness levels are 
{listed} in Table \ref{depth}. We have used a linear interpolation between the measured points to estimate these limits.

\begin{table}
\caption{Completeness limits according to our artificial star test.}
\centering                          
\setlength{\tabcolsep}{4pt} %
\begin{tabular}{l l l l l l l}        
\hline\hline                 
Name & $V_{75}$ & $I_{75}$ & $V_{50}$ & $I_{50}$ & $V_{25}$ & $I_{25}$ \\    
 & mag & mag & mag & mag & mag & mag \\ 
\hline      \\[-2mm]                  
dw1315-45 & 26.00 & 24.55 & 26.31 & 24.73 & 26.58 & 24.90 \\
KKs54 & 26.01 & 24.32 & 26.20 & 24.58 & 26.39 & 24.84 \\
dw1318-44 & 26.13 & 24.16 & 26.46 & 24.31 & 26.69 & 24.46 \\
dw1322-39 & 26.28 & 24.96 & 26.51 & 25.21 & 26.70 & 25.40 \\
KK198 & 26.00 & 24.19 & 26.25 & 24.35 & 26.54 & 24.51 \\
dw1323-40c & 25.75 & 24.09 & 25.95 & 24.27 & 26.16 & 24.44 \\
dw1323-40b & 25.89 & 24.66 & 26.19 & 24.89 & 26.46 & 25.18 \\
dw1323-40 & 25.80 & 24.61 & 26.05 & 24.77 & 26.34 & 24.92 \\
dw1329-45 & 25.83 & 24.40 & 26.03 & 24.73 & 26.22 & 24.98 \\
dw1331-37 & 26.13 & 24.25 & 26.43 & 24.46 & 26.66 & 24.72 \\
dw1336-44 & 26.23 & 24.58 & 26.56 & 24.77 & 26.89 & 24.96 \\
dw1337-44 & 25.54 & 24.64 & 25.87 & 24.88 & 26.09 & 25.18 \\
dw1341-43 & 25.97 & 24.22 & 26.29 & 24.48 & 26.57 & 24.76 \\
dw1342-43 & 25.99 & 24.22 & 26.37 & 24.54 & 26.65 & 24.82 \\
KKs58 & 26.12 & 24.20 & 26.45 & 24.36 & 26.73 & 24.54 \\
\hline
\end{tabular}
\label{depth} 
\end{table}

\begin{figure}[ht]
\includegraphics[width=9cm]{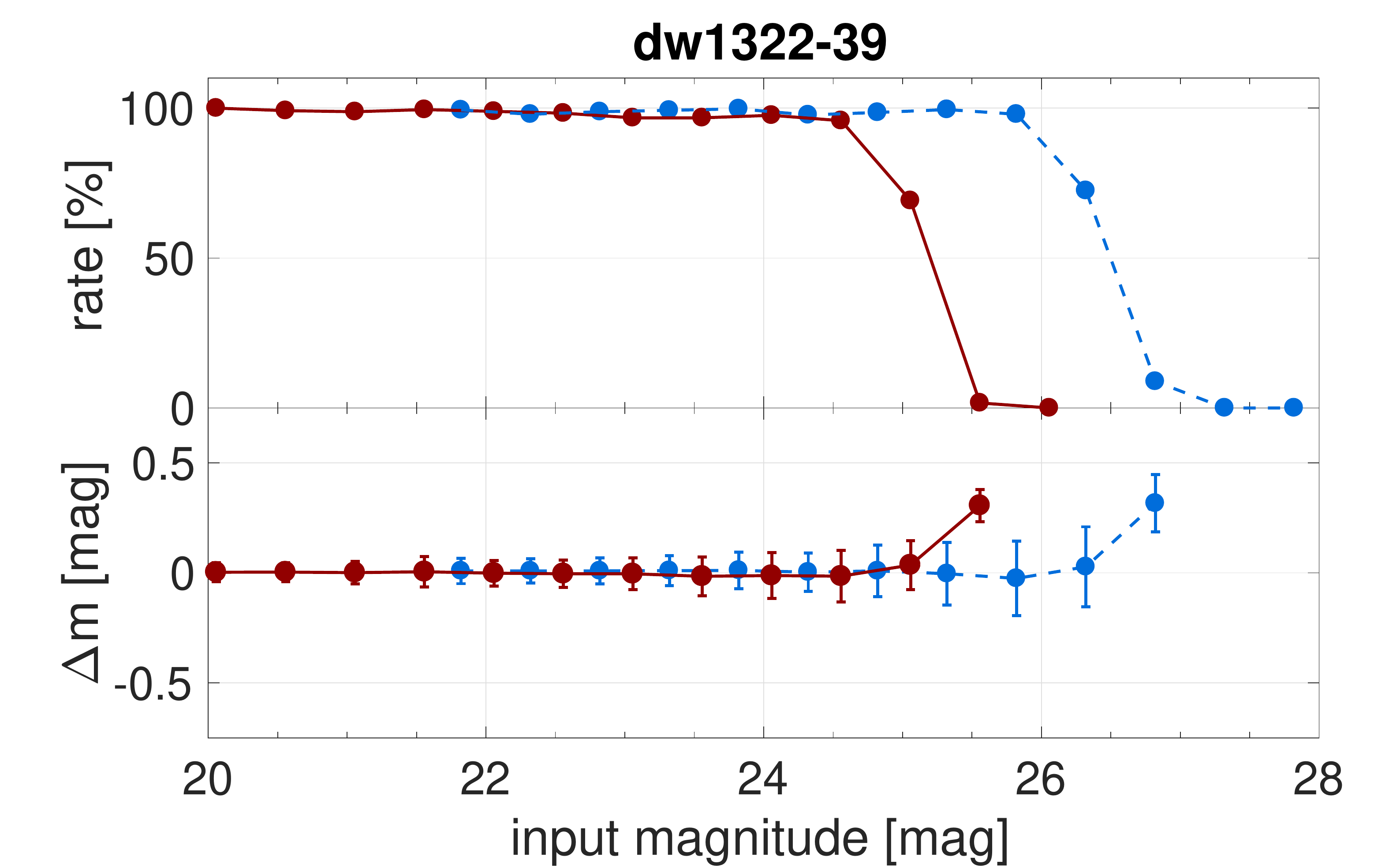}
\caption{{Results of our artificial star experiments for $I$ (red) and $V$ (dashed blue) bands for dwarf galaxy dw1322-39. The remaining galaxies are shown in Fig\,\ref{app:artTests} in the Appendix.} Upper {panel}: the recovery 
fraction of the artificial stars induced into the science frames as a function of the input magnitude.
Bottom {panel}: the difference between the input 
and measured magnitude as a function of input magnitude.  
}
\label{artTests}
\end{figure}

\section{Color magnitude diagrams}
\label{CMD}
In the previous section we presented the pipeline creating our photometric catalogs. In the following, we produce color magnitude diagrams (CMDs) to study the resolved red giant branch of the dwarf galaxies and measure {their} mean metallicity. Finally, 
we measure the distances for nine dwarf galaxies using a Bayesian approach. 

\subsection{The dwarf galaxy CMDs}
\begin{figure*}[ht]
\includegraphics[width=6cm]{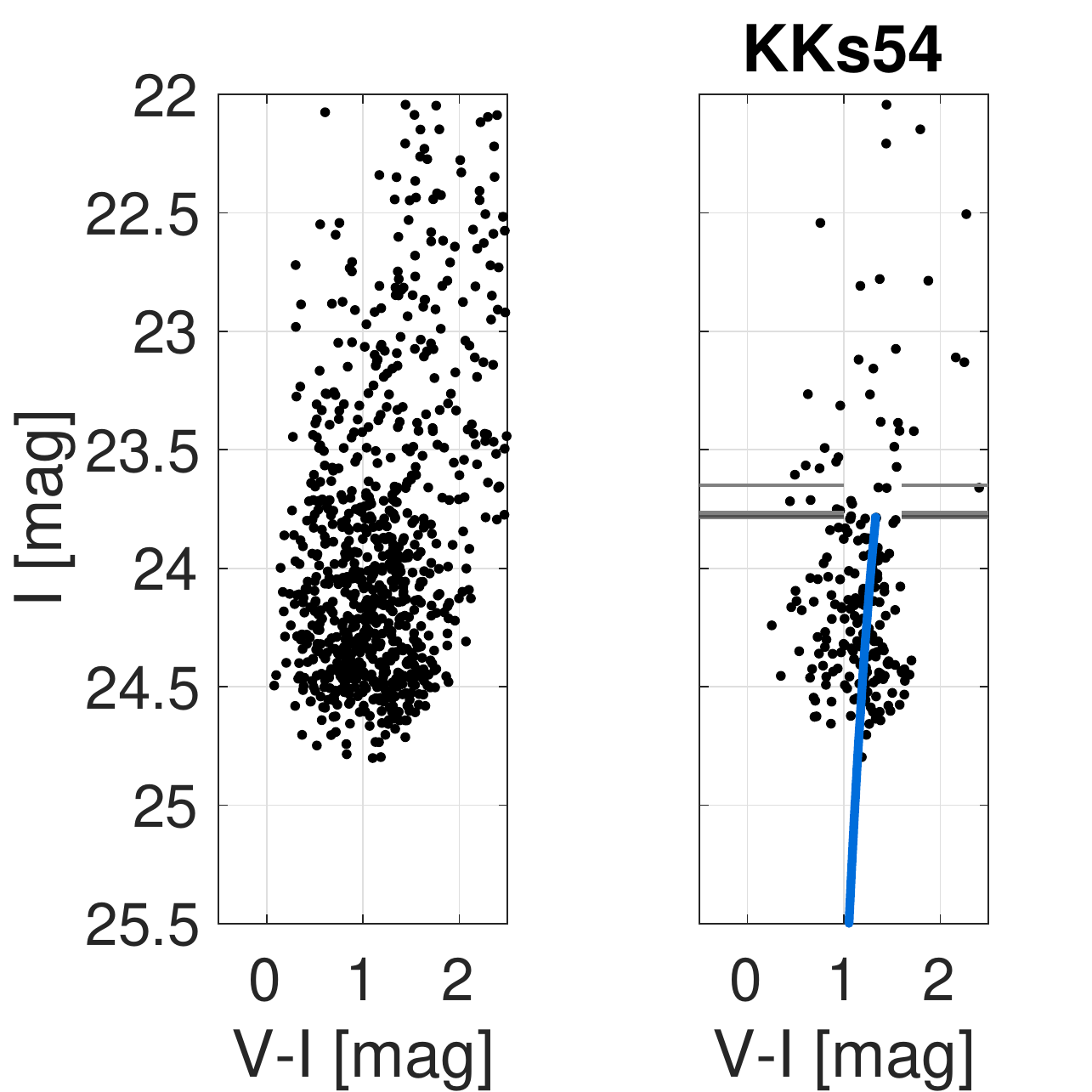}
\includegraphics[width=6cm]{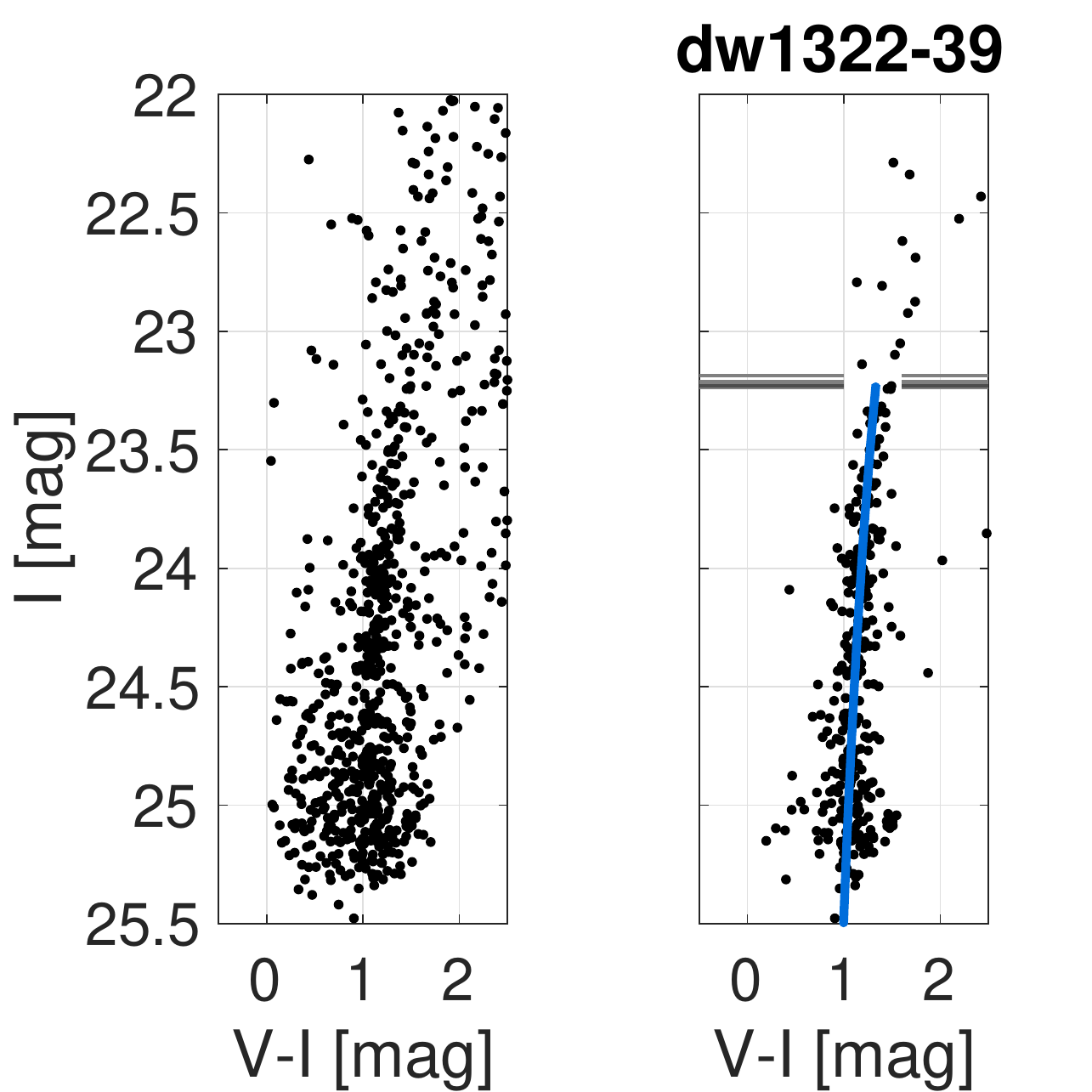}
\includegraphics[width=6cm]{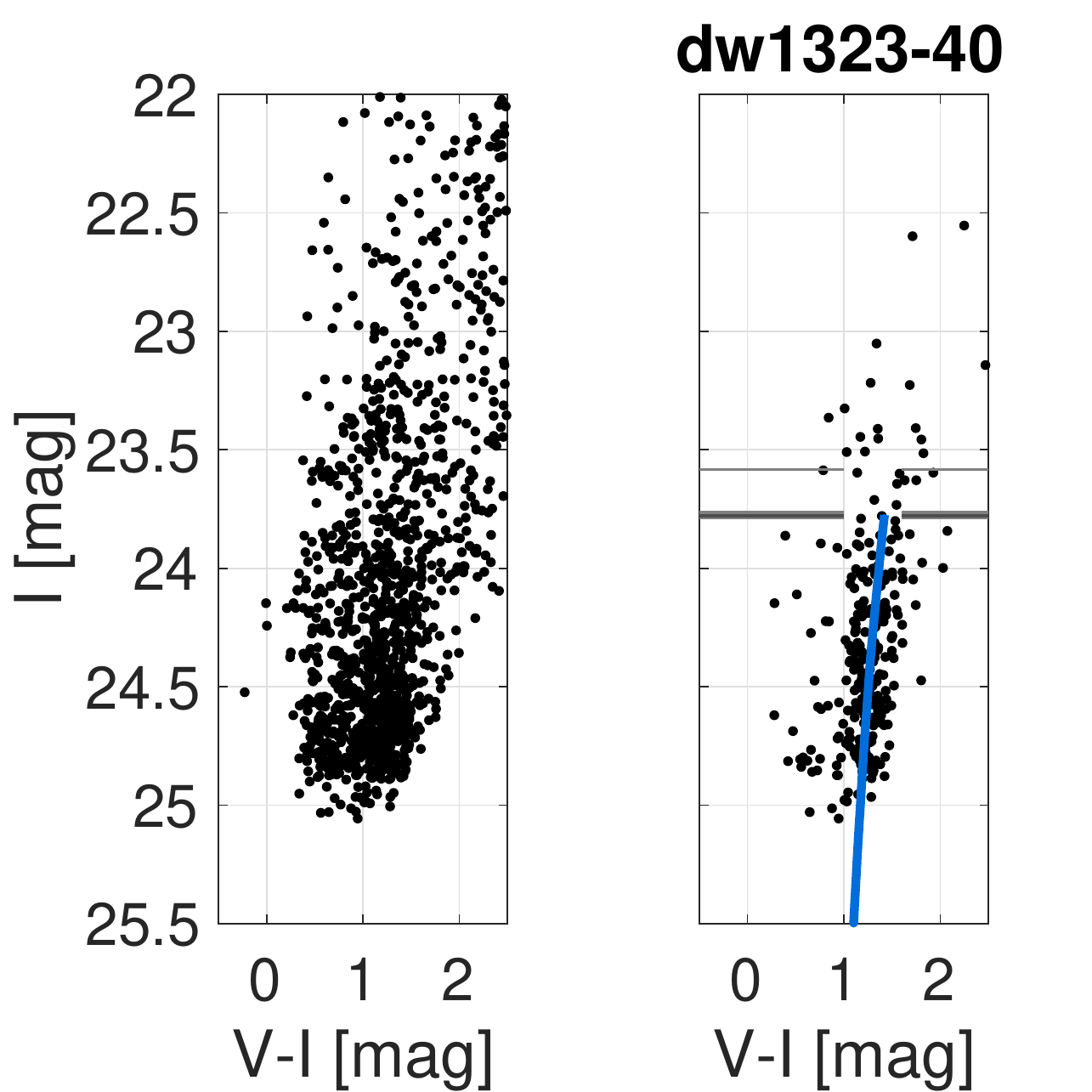}

\includegraphics[width=6cm]{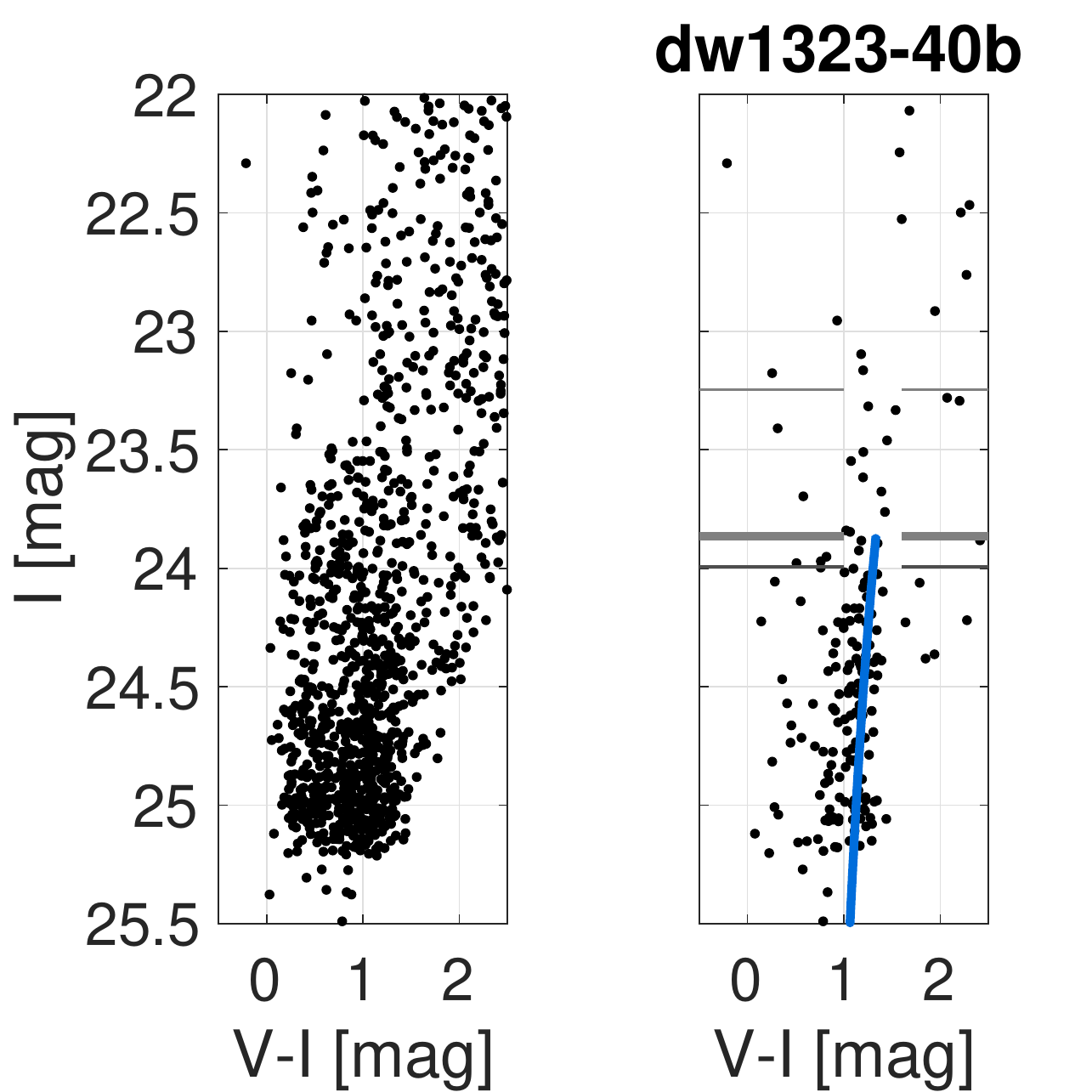}
\includegraphics[width=6cm]{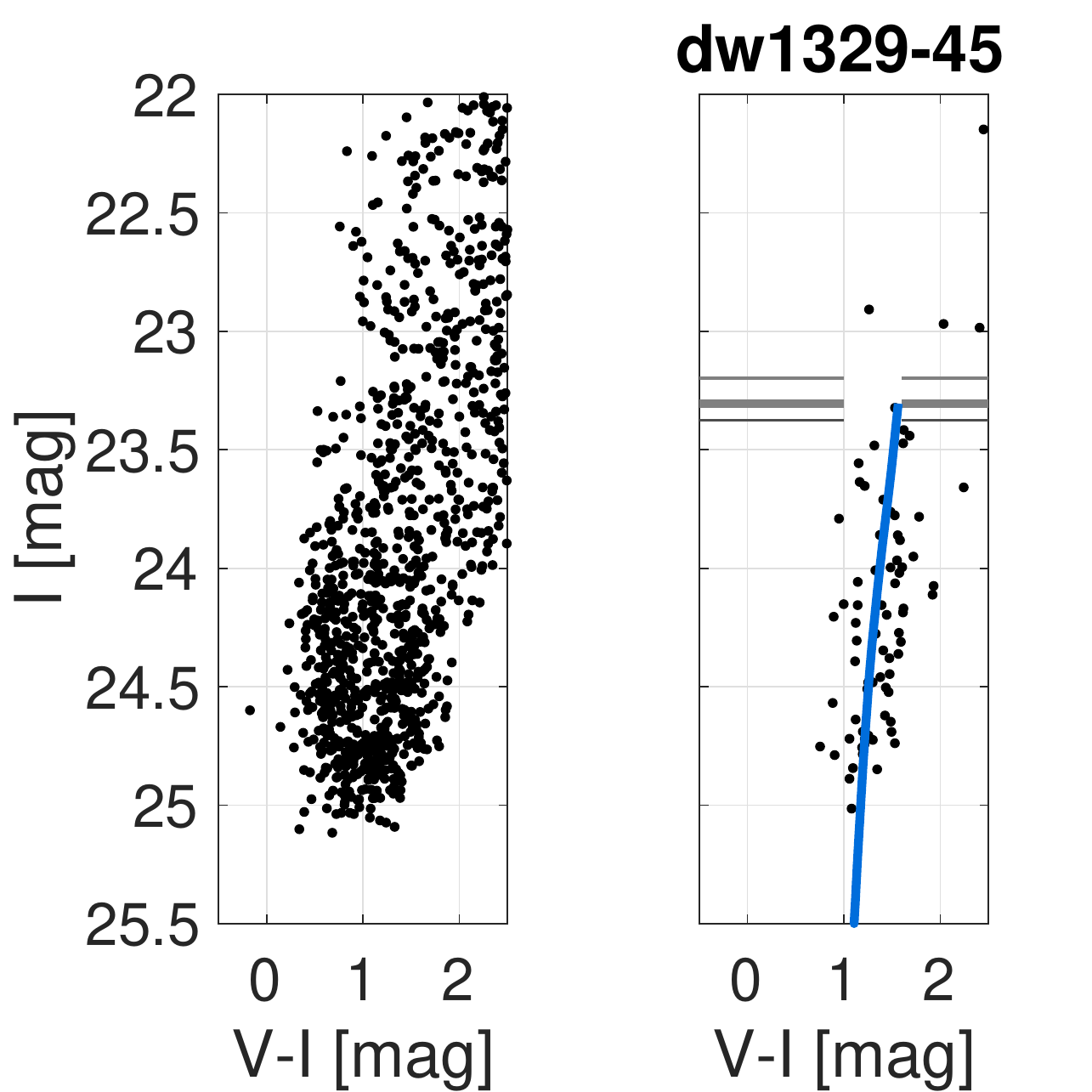}
\includegraphics[width=6cm]{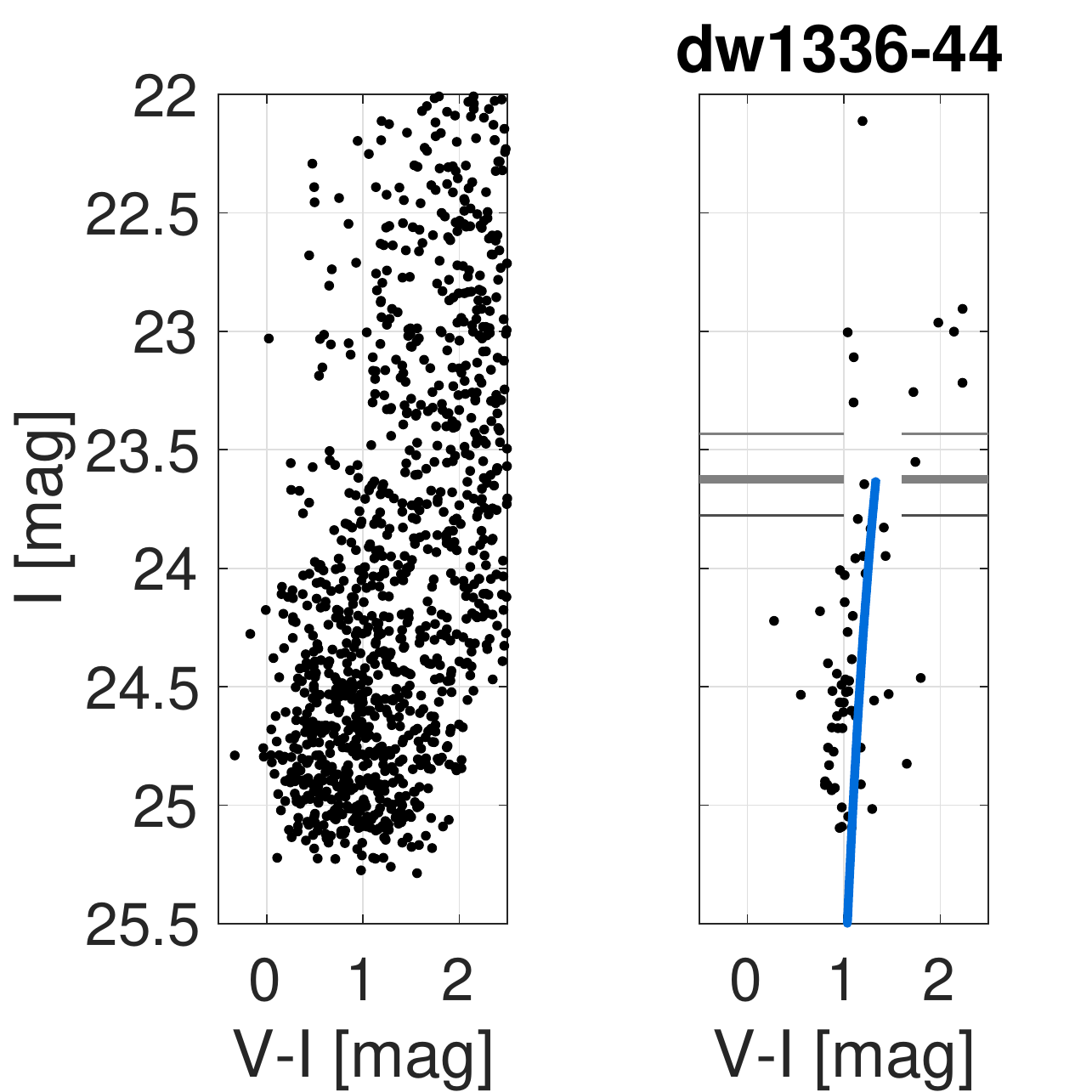}

\includegraphics[width=6cm]{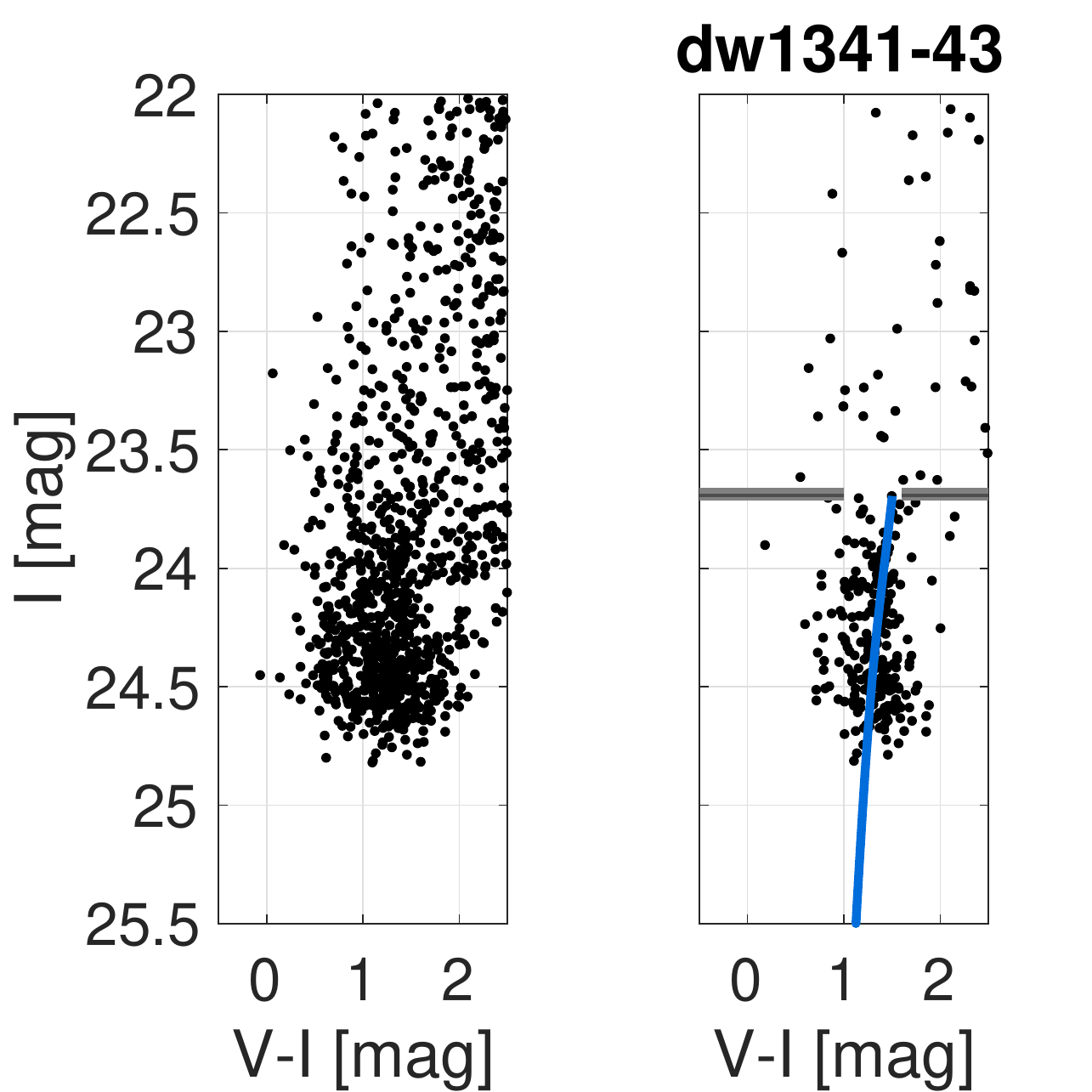}
\includegraphics[width=6cm]{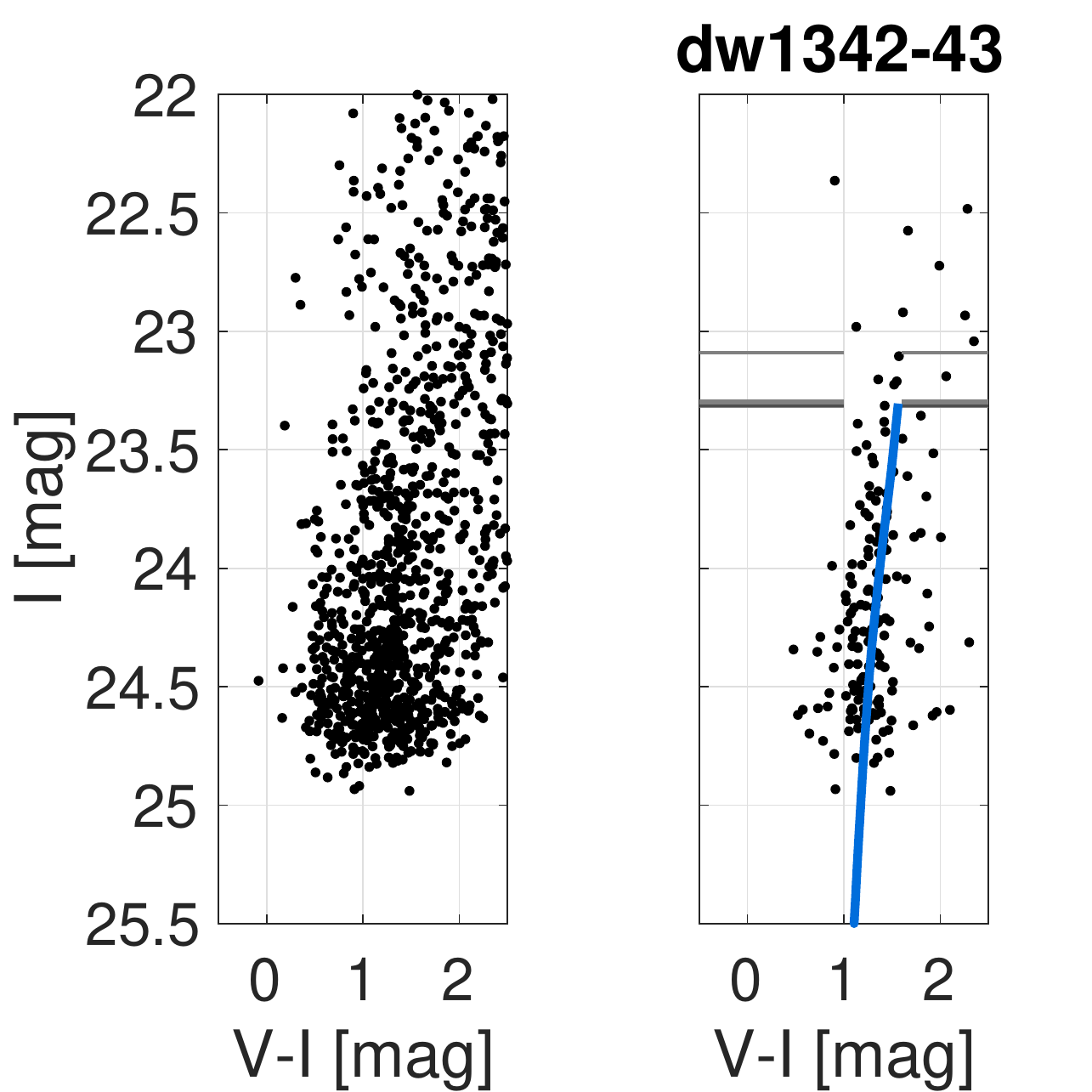}
\includegraphics[width=6cm]{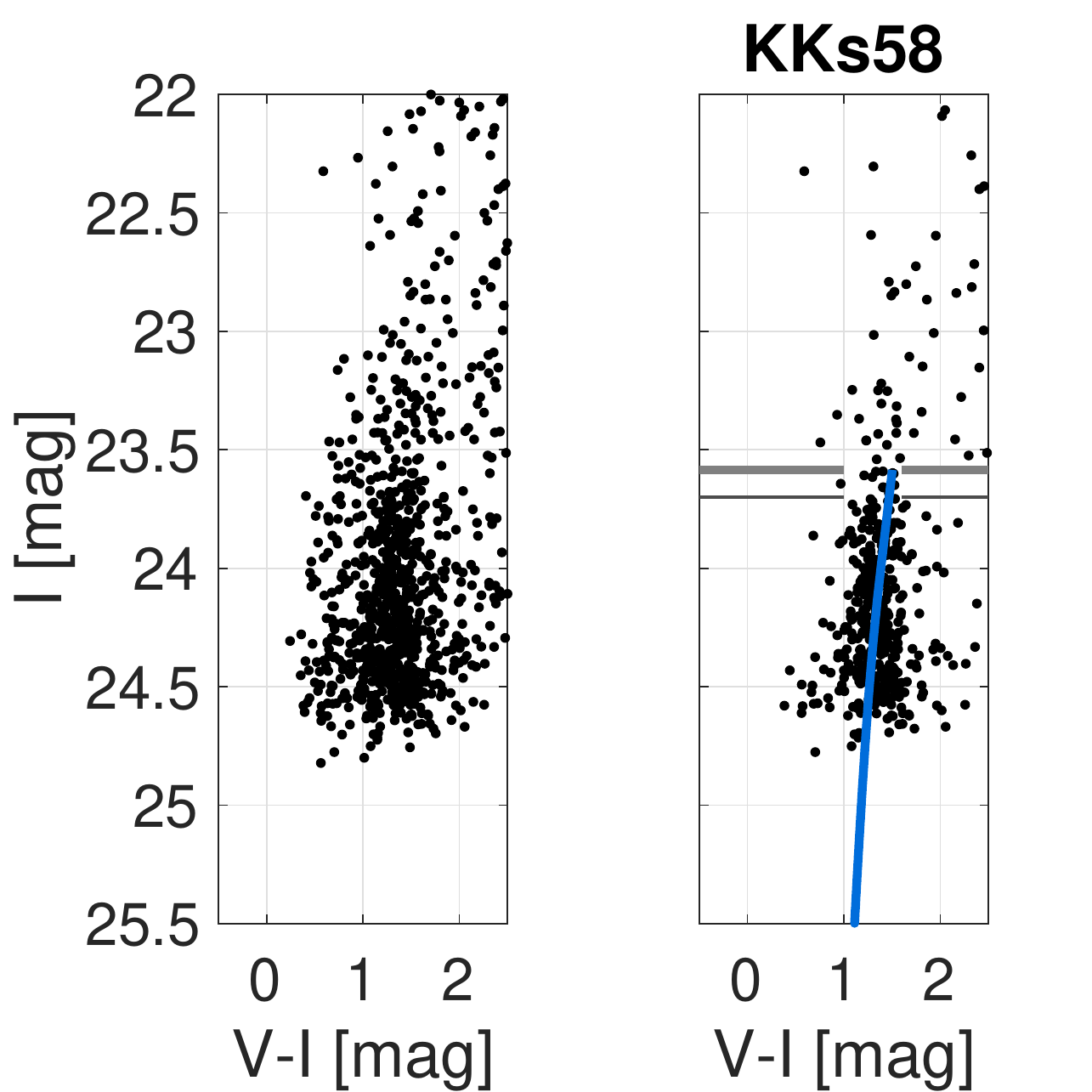}
\caption{Extinction-corrected CMDs for newly studied dwarf galaxies. 
Left: CMD for all stars on the CCD.  Right: Stars within an aperture {of 2.5 times the effective radius}  around the galaxy. The thick {and thin} gray lines indicate the estimated TRGB and the uncertainties, {respectively}. The blue line shows the best fitting isochrone. 
}
\label{cmds}
\end{figure*}

At the 
distance of 4 Mpc {of} the Centaurus group \citep{2013AJ....145..101K}, the dwarf galaxy candidates only cover a small fraction of chip 1 of the CCD. To derive the CMD for the dwarf galaxies, we used a circular aperture of $2.5\times r_{\rm eff}$ (effective radius) around their estimated center to produce the galactic CMDs. The $r_{\rm eff}$ values were taken from integrated light photometry \citep{2000AJ....119..593J,2017A&A...597A...7M}.  The galaxy's {extent} is well approximated with $2.5\times r_{\rm eff}$. Using larger radii 
increase{s} the pollution {by} 
foreground {Milky Way} stars {and unresolved compact background galaxies, which start to be comparable in number to (or even more numerous than) the potential dwarf galaxy members.\ This thus} 
decreases the signal-to-noise ratio. 
We subsequently produced a reference CMD {for each target used} to statistically clean the
{galaxy's} CMD from foreground stars which will unavoidably overlap with the dwarf galaxy. {This reference CMD consists of all stars on Chip 1 which are outside of the galactic aperture radius times $\sqrt{2}$ (to avoid the outmost galaxy stars). The area of the reference CMD is much larger than the galactic aperture because the field of view of Chip 1 is roughly $7\times8$ sq. arcmin, while the targets sizes are typically $<1$ arcmin in radius.}
In Fig.\,\ref{cmds}, we present all stars on the CCD (left) 
and in the galactic aperture (right) for every galaxy for which we were able to resolve the RGB.
The candidates, not being resolved into their individual stars, are discussed below. 
In Fig.\,\ref{cmds}, we also present the detected 
{tip of the RGB (TRGB)} magnitude, as well as the best-fitting isochrone, which are discussed in the next subsection.
It is noteworthy that the quality of the CMDs strongly varies depending {principally on the seeing} conditions during the specific target observations. 

{Every CMD exhibits a continuous  population of stars which can be associated to the RGB. For some galaxies (KKs\,54, dw1322-39, dw1323-40b, and KKs\,58), an extension of the RGB toward brighter magnitudes and redder color is visible, which bears resemblance to a population of AGB stars. If real, they indicate 
{the presence of an} intermediate age stellar population in these dwarf galaxies {due to extended star formation activity}. 
There are also some blue stars with $(V-I)_0<0.7$\,mag in dw1323-40b.
{If associated with dw1313-40b, these blue stars would trace recent star formation, similar}
to the case of dw1335-29 {-- a satellite member of M\,83 --} as suggested by \citet{2017MNRAS.465.5026C}. However, {we find the blue stars to be} almost isotropically distributed {across the galaxy}, indicating that they are probably 
contaminating background galaxies.
}

\begin{figure}[ht]
\centering
\includegraphics[width=6cm]{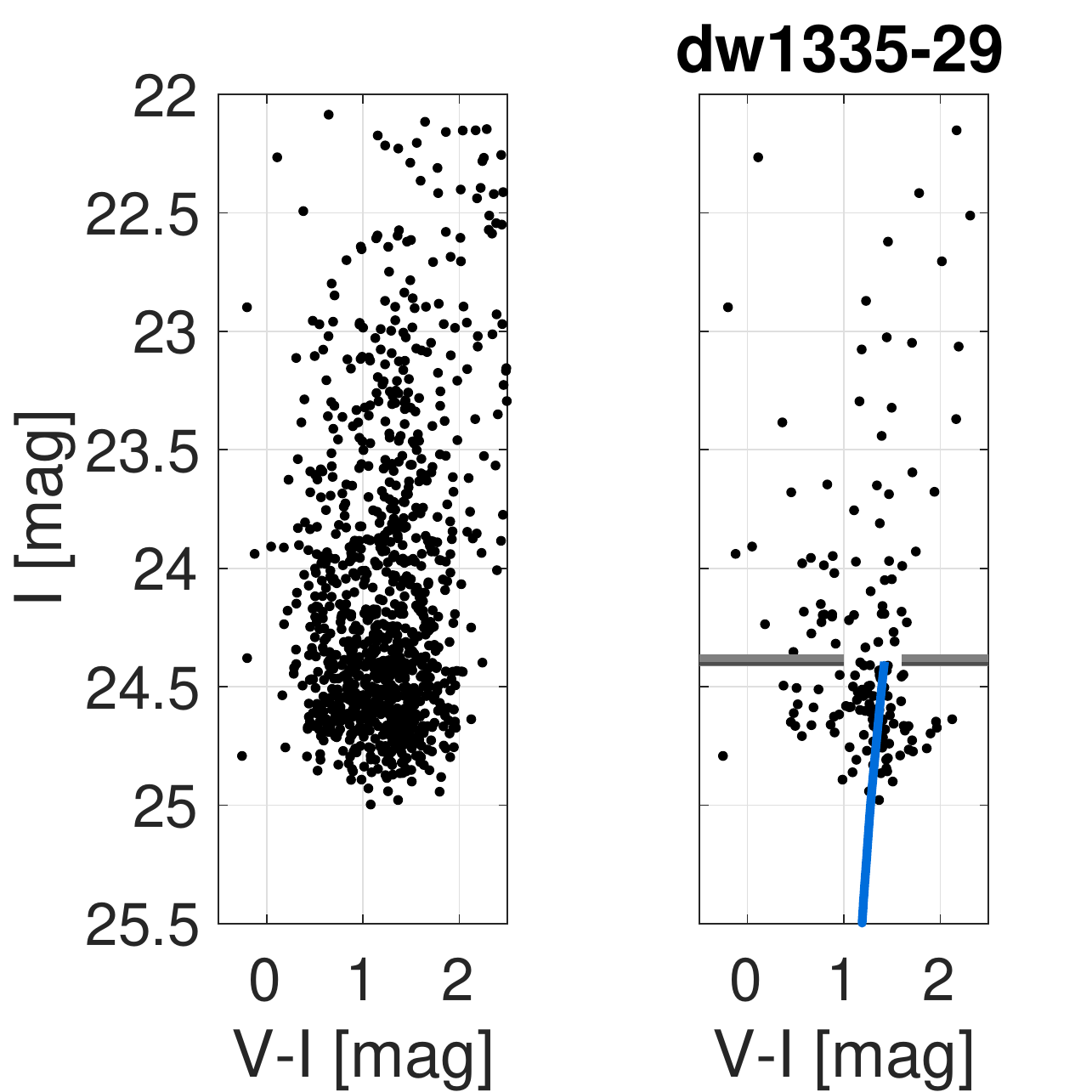}
\includegraphics[width=6cm]{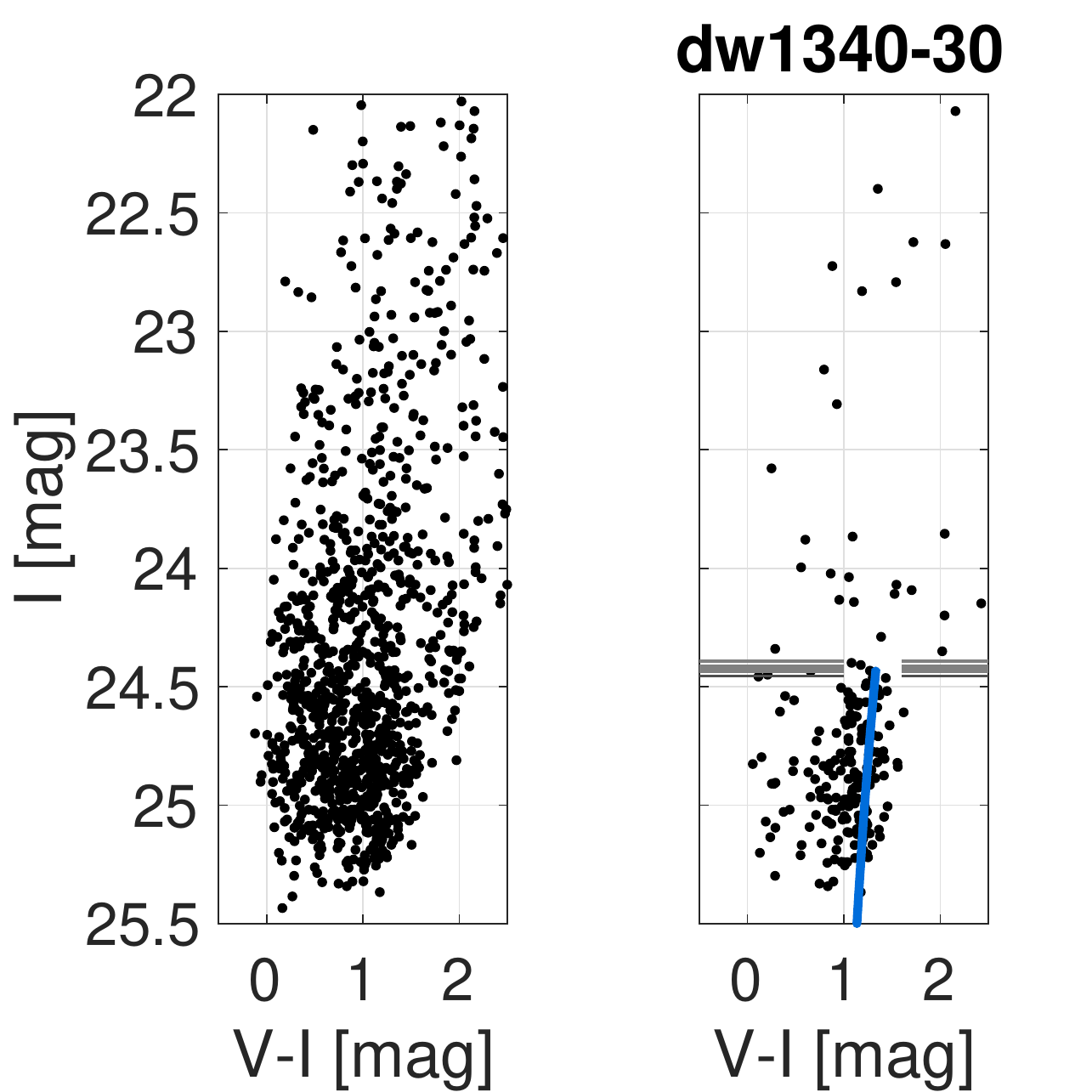}

\caption{Extinction-corrected CMDs for dwarf galaxies from \citet{MuellerTRGB2018}. 
Left: CMDs for all stars on the CCD. Right: Stars within the aperture {(black dots)}, the estimated TRGB {(gray line)}, and the best fitting isochrone {(blue line)}.}
\label{cmds2}
\end{figure}

\subsection{TRGB distance estimation}
In the literature, there are 
{several} methods 
to measure the tip of the red giant branch \citep[TRGB, e.g.,][]{Lee1993,2000A&A...359..601C,2006AJ....132.2729M,2012ApJ...758...11C}. Physically, this sharp {feature}
in the CMD is well understood. On their evolutionary path through the red giant branch, the stars have long exhausted the hydrogen fuel in the core, while still burning hydrogen in the shell. Once the thermal pressure is insufficient to stabilize the star, it contracts until the core  becomes degenerate. Eventually, the temperature reached in the core is hot enough to ignite the helium burning, and an explosive helium flash occurs, ultimately leading to a change in color and temperature. Thus the stars leave the red giant branch. This helium flash is only marginally dependent on the metallicity and is an excellent standard candle to measure the distance of an object \citep{daCostaArmandroff1990,Lee1993}, this has a typical error of 5\% \citep{2015ApJ...802L..25T}. In \citet{MuellerTRGB2018}, we used a Sobel edge detector \citep{Lee1993} {to determine the TRGB}. In the following, we used a different method based on Bayesian considerations \citep{2011ApJ...740...69C,2012ApJ...758...11C}. This was necessary because the Sobel edge detection becomes quite unreliable when the data are noisy, which happens to be the case with {low surface brightness galaxies in general and even more when they are} 
close to the 
{Milky Way} disk ($b=19.4$\,deg). We followed the implementation of \citet{2011ApJ...740...69C,2012ApJ...758...11C}, which employs a Markov Chain Monte Carlo (MCMC) method.
For each star, the probability is calculated for it belonging to a RGB model, a background model, {and} a radial density model. The RGB model uses a simple power law \citep{2006AJ....132.2729M}
\begin{eqnarray*}
\psi=10^{a\,(m-m_{\rm TRGB})}
\end{eqnarray*}
 to describe the luminosity function of stars in the galaxy. 
 {The background contamination is modeled with a polynomial function (BG) fit to the luminosity function of the background stars {on the reference CMD}. From this BG function, together with the area used for the galaxy and reference CMDs, the expected number of background objects within our galactic aperture can be estimated. The fraction of contaminating sources is  given as the contamination factor $c$.} 
 A log likelihood estimation $L$ for a given set of parameters ($a$, $m_{\rm TRGB}$,  and $c$) is calculated by assigning a probability to each star:
\begin{eqnarray*}
M(m)=
 \begin{cases}
  (1-c)\,\psi(m)+ c\,{\rm BG}(m)& {\rm  if}\ m<=m_{\rm TRGB}\\
  c\,{\rm BG}(m)&  {\rm if}\ m>m_{\rm TRGB}\\
\end{cases}
\end{eqnarray*}
\begin{eqnarray*}
N(r)=\exp(-r/r_{\rm eff})
\end{eqnarray*}
\begin{eqnarray*}
L=\sum_{i=1}^N \log M(m(i))\,N(r(i))
\end{eqnarray*}
For this to work, the BG, $\psi$, and the exponential density profile need to be probability density 
{distributions, which }
need to be normalized. Now the MCMC scheme comes into play: We propose a new set of parameters ($a_{\rm proposed}$, $m_{\rm TRGB, proposed}$, and $c_{\rm proposed}$) by randomizing their values drawn from a normal distribution on top of the current values ($a_{\rm current}$, $m_{\rm TRGB, current}$,and $c_{\rm current}$). The standard deviation is chosen such that the new values can fully cover the  space we expect them to be. 
{Next,} the two likelihoods $L_{\rm proposed}$ and  $L_{\rm current}$ are compared with each other.\ If their ratio is less than $r$, which is drawn from a uniform distribution between zero and one, the proposed set of parameters is accepted and added to the chain. If not, the current value is added. This is repeated 100,000 times. Subsequently, the chain will produce a posterior distribution  for ($a$, $m_{TRGB}$, and  $c$), where the mode of the posterior distribution of $m_{TRGB}$ will be our estimated TRGB magnitude. The uncertainties are given by the 68\% confidence interval. The posterior distribution will not follow a normal distribution, meaning that this error includes the interval with $\pm34$\% around the mode.

The calibration of the TRGB magnitudes is done in accordance to \citet{2004A&A...424..199B} and \citet{2007ApJ...661..815R}:
\begin{eqnarray*}
 M^{TRGB}_I&=&-4.05 +0.217[(V-I)_0-1.6],\\
 (V-I)_0&=&0.581 \rm{[Fe/H]}^2 + 2.472 \rm{[Fe/H]}+4.013.
 \end{eqnarray*}
To estimate the mean metallicities [Fe/H] we fit theoretical BaSTI isochrones \citep{2004ApJ...612..168P}, {using an age of 10\,Gyr for the isochrones}. {BaSTI isochrones are available at  $\rm{[Fe/H]}$ of -2.267, -1.790, -1.488, and -1.266\,dex steps, which is sufficient for the given accuracy in the color.}
{The selection of the best-fitting isochrone was done by minimizing the distances between the color of the stars and the color of the different isochrones, respectively.}
The isochrones are presented in the CMDs (Figs.\,\ref{cmds} and \ref{cmds2}). The properties of the dwarfs are presented in Table \ref{properties}.

As a consistency check, we applied the MCMC method to the two already confirmed M\,83 dwarfs, dw1335-29 and dw1340-30 from \citet{MuellerTRGB2018}, where we used a Sobel Edge detection to find the position of the TRGB{, employing deep VLT+FORS2 $VI$ photometry}. The TRGB magnitudes were estimated to be 24.43$\pm$0.1\,mag for both galaxies \citep{MuellerTRGB2018}. Here we derived 24.40$^{+0.01}_{-0.02}$\,mag and 24.43$^{+0.03}_{-0.03}$\,mag, respectively, which are 
{fully} consistent with the previous estimates. For one of the dwarfs, dw1335-29, \citet{2017MNRAS.465.5026C} measured a distance of $5.01^{+0.74}_{-0.22}$\,Mpc independently with HST imaging, which is consistent with our previous estimate of 5.03$\pm$0.24\,Mpc \citep{MuellerTRGB2018}, and our current distance estimate of 4.96$^{+0.02}_{-0.05}$\,Mpc. For dw1340-30, we get a new distance estimate of 5.06$^{+0.07}_{-0.08}$\,Mpc.


\subsection{Metallicity spread}
\label{metalspread_subsection}
\begin{figure}[ht]
\centering
\includegraphics[width=9cm]{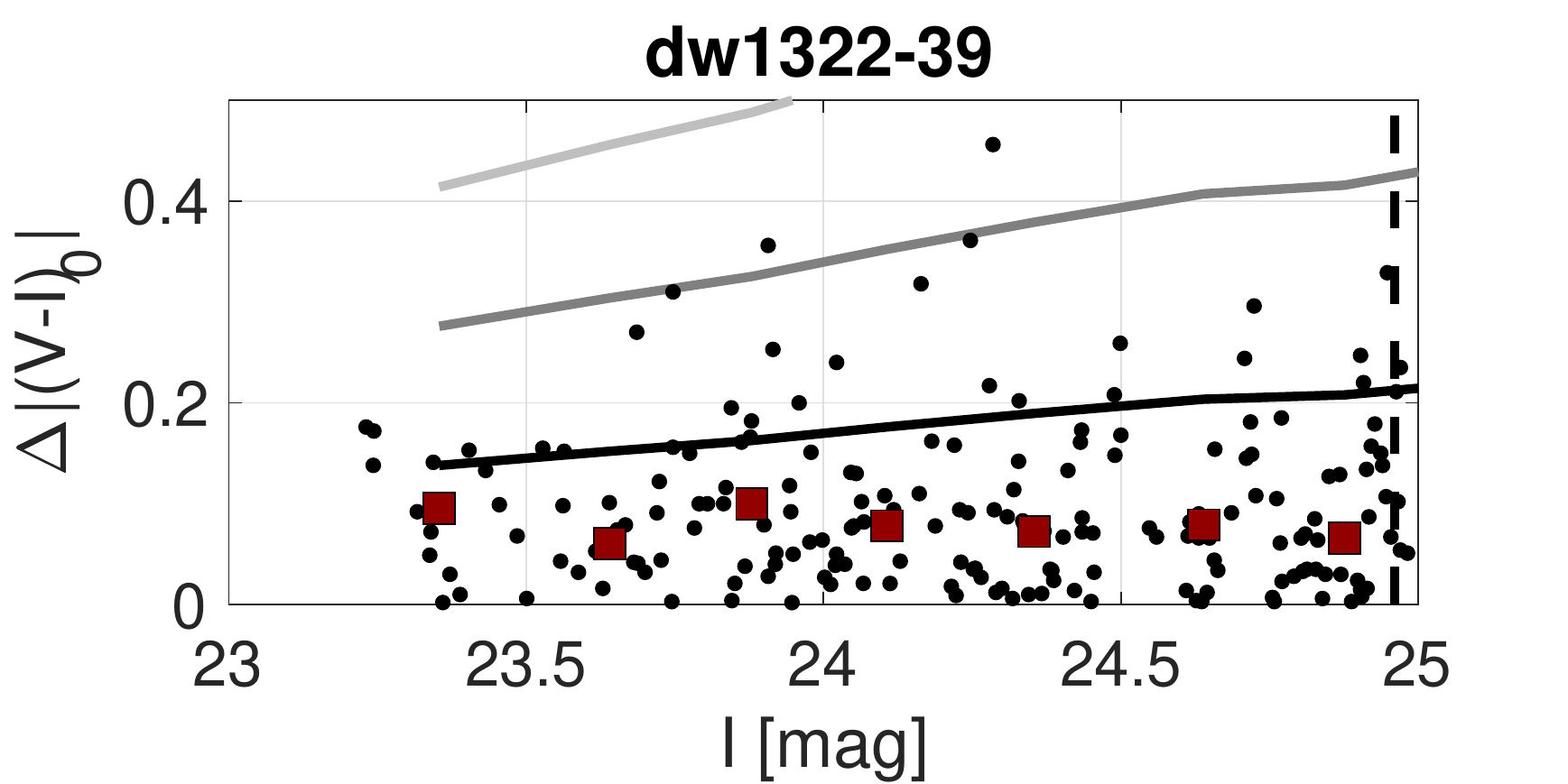}
\caption{Measured $(V-I)_0$ spread {as function of $I$-band magnitude} for {stars observed in} dw1322-39 (black dots) {in comparison with} expected uncertainty derived from our artificial star tests. {The red squares correspond to the mean values binned over 0.25\,mag steps.  The solid lines correspond to the 1, 2, and 3$\sigma$ levels, the dotted line to the 75\% completeness limit. {The remaining galaxies are shown in Fig.\,\ref{app:spread} in the Appendix.}}
}
\label{metalspread}
\end{figure}

The majority of the stars in the CMDs studied here follow, within a certain spread, the best-fitting isochrone, see for example the CMD of KKs\,58 in Fig\,\ref{cmds}. This spread could potentially indicate a complex stellar population {and spread in metallicity}, as in the globular cluster $ega$ Centauri \citep{2014ApJ...791..107V} or the Sculptor dwarf spheroidal for example \citep{2005MNRAS.359..985B}. Before such conclusions can be drawn, the error budget has to be understood {since it can induce} a spread in the $(V-I)_0$ {due to limited photometric precision}. For that purpose, we used our artificial star tests to see how a perfect population of stars is scattered due to measurement uncertainties. The scatter is mainly driven by two components: the systematic error and the standard deviation at given magnitude (see Fig.\,\ref{artTests}). To assess the expected spread at given $V$ and $I$ band magnitudes, we calculated: 
$$\sigma=\sqrt{I_{err}^2+V_{err}(I)^2 + I_{std}^2 + V_{std}(I)^2},$$
by taking $I_{err}$, $V_{err}$,  $I_{std}$, and $V_{std}$ directly from Fig.\,\ref{artTests} and interpolated where necessary. We then measured the spread of $(V-I)_0$ for the stars in the CMD along the isochrone, binned over 0.25\,mag steps. {Only stars within a color range between 0.9 and 1.8\,mag were considered.} The spread in this magnitude bin is then compared to the expected spread of photometric uncertainty, coming from our artificial star experiment. If the measured trend is lower, then there is no significant metallicity spread detected. If it is larger, the spread could be of physical nature. For every galaxy, we created a plot like Fig.\,\ref{metalspread}, where we indicate the measured spread, as well as the 1, 2, and 3$\sigma$ limits from the artificial star data. We show all plots in the Appendix (Fig.\,\ref{app:spread}). {For all but one galaxy, the measured spread is systematically below the $1\sigma$ line, indicating that there is either no metallicity spread, or the uncertainties from the photometric pipeline are too large to detect such a spread. Only for dw1329-45 is there a marginal signal, however, this galaxy is affected by light from a very close-by star (see Fig.\,\ref{targetsVLT}). Because our artificial star experiments take advantage of the full CCD, the uncertainty in the close proximity of this star could be underestimated.}
We therefore conclude that with our photometry, we have 
not detected a 
metallicity spread
within the studied dwarf galaxies. 
The resolved dwarf galaxies are most likely akin to old gas-poor dwarf Spheroidal population around the Milky Way and the Andromeda galaxy. {However, we remark that while for the Milky Way dwarf galaxy, Carina, the color spread in the RGB is rather small, multiple sequences of star formation have been measured \citep{2010PASP..122..651B}, showing that the absence of a significant color spread in the RGB doesn't rule out a more complex underlying stellar population (i.e.,\ a combination of different ages and metallicities can result in a relatively thin RGB).}

\subsection{Unresolved targets}
\begin{figure*}[ht]
\centering
\includegraphics[width=6cm]{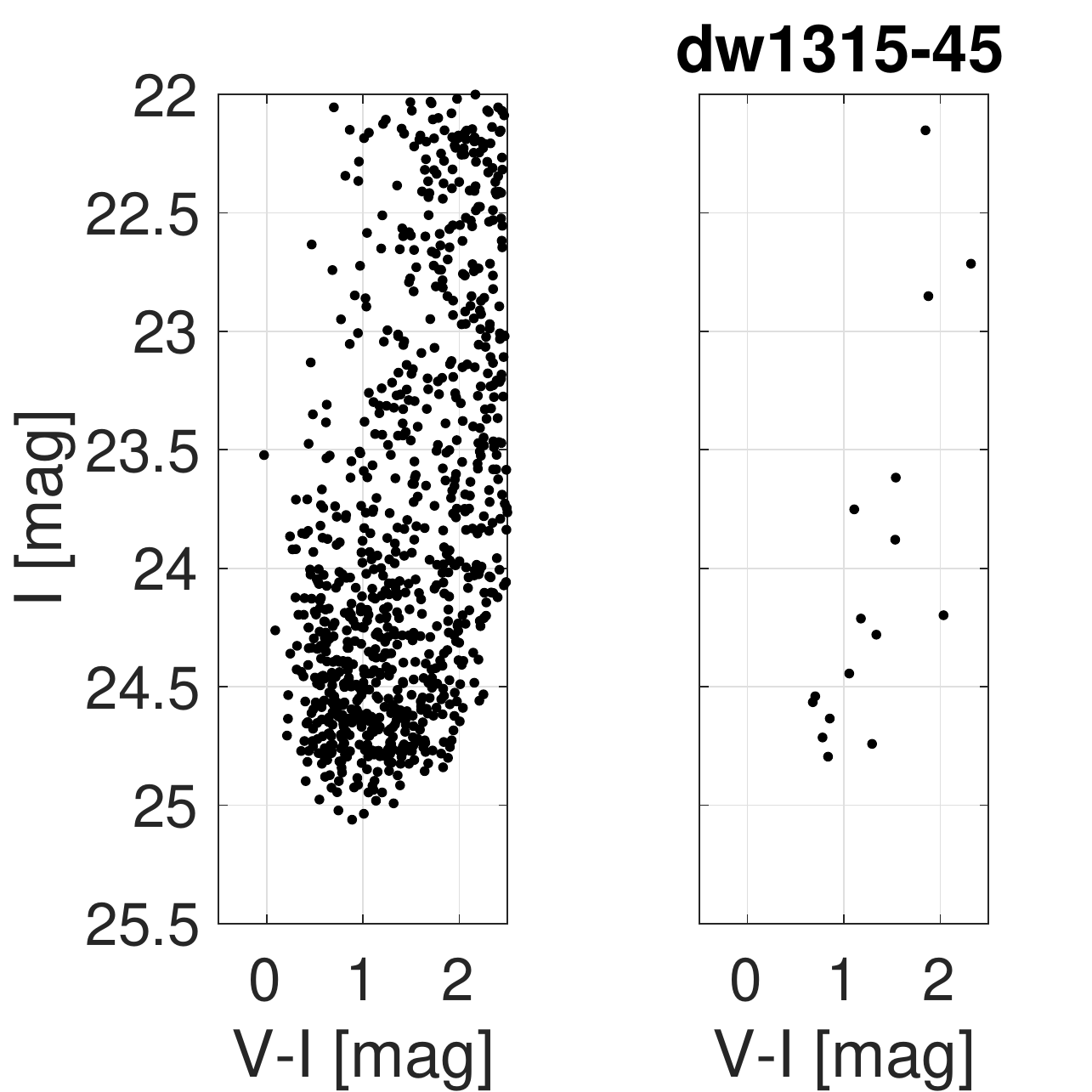}
\includegraphics[width=6cm]{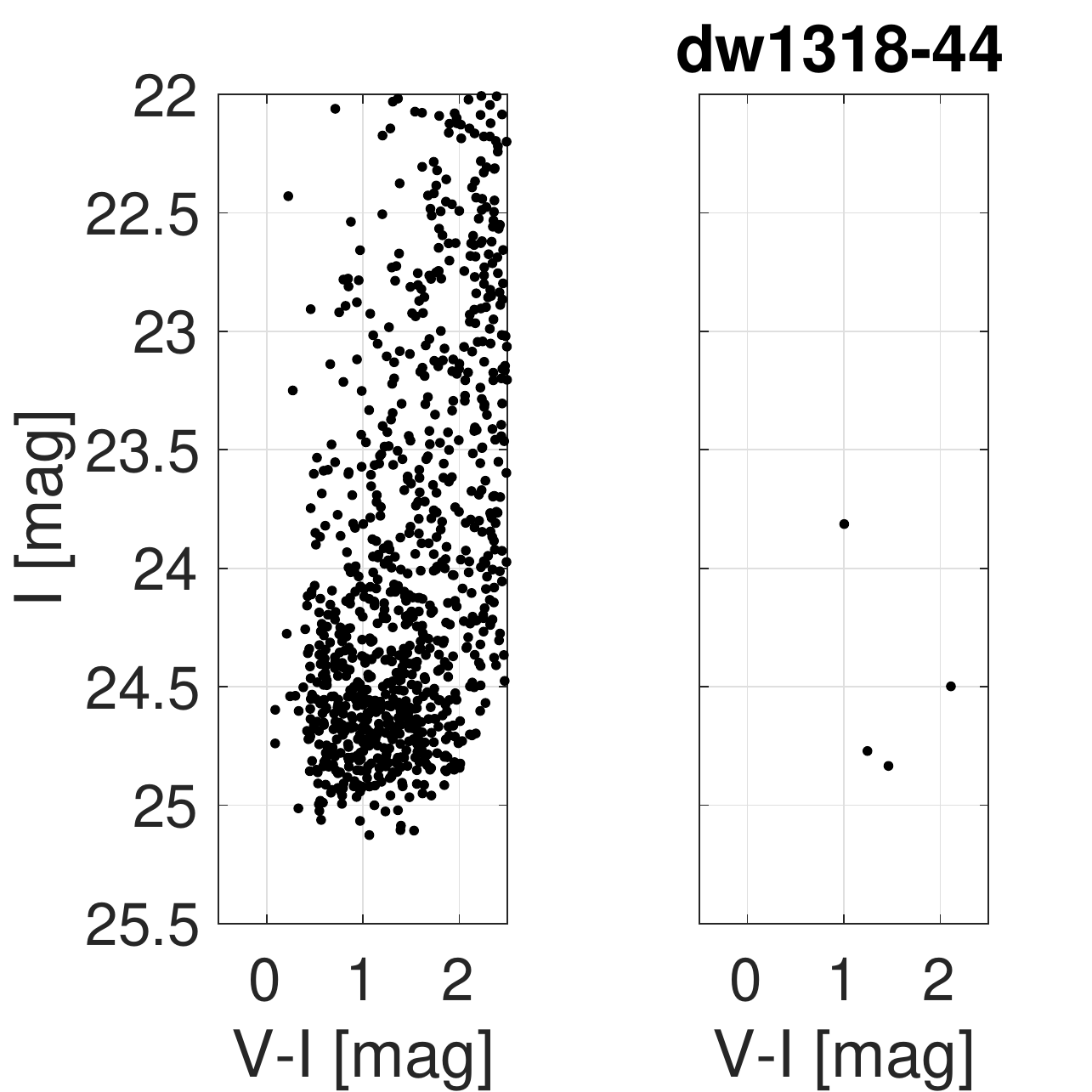}
\includegraphics[width=6cm]{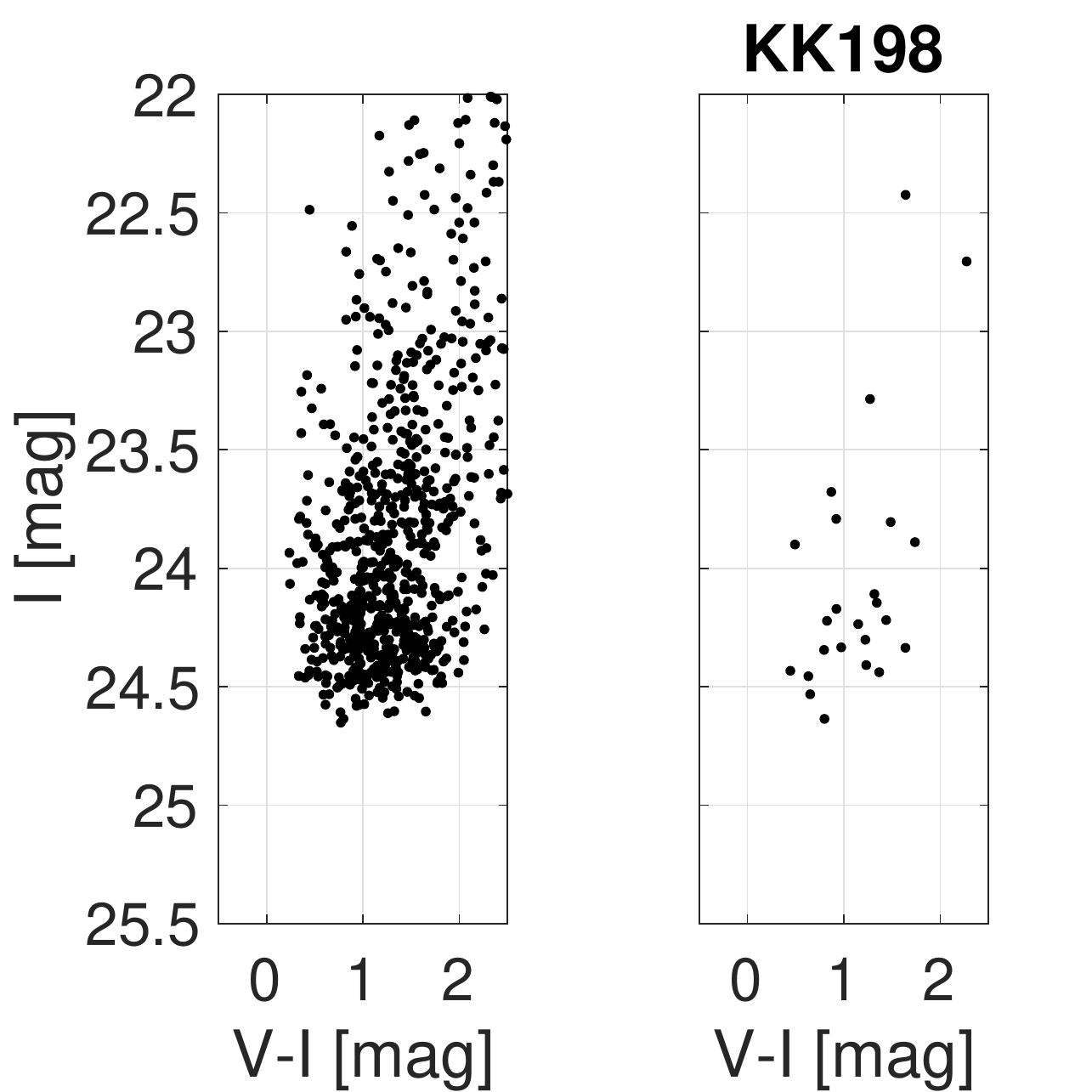}

\includegraphics[width=6cm]{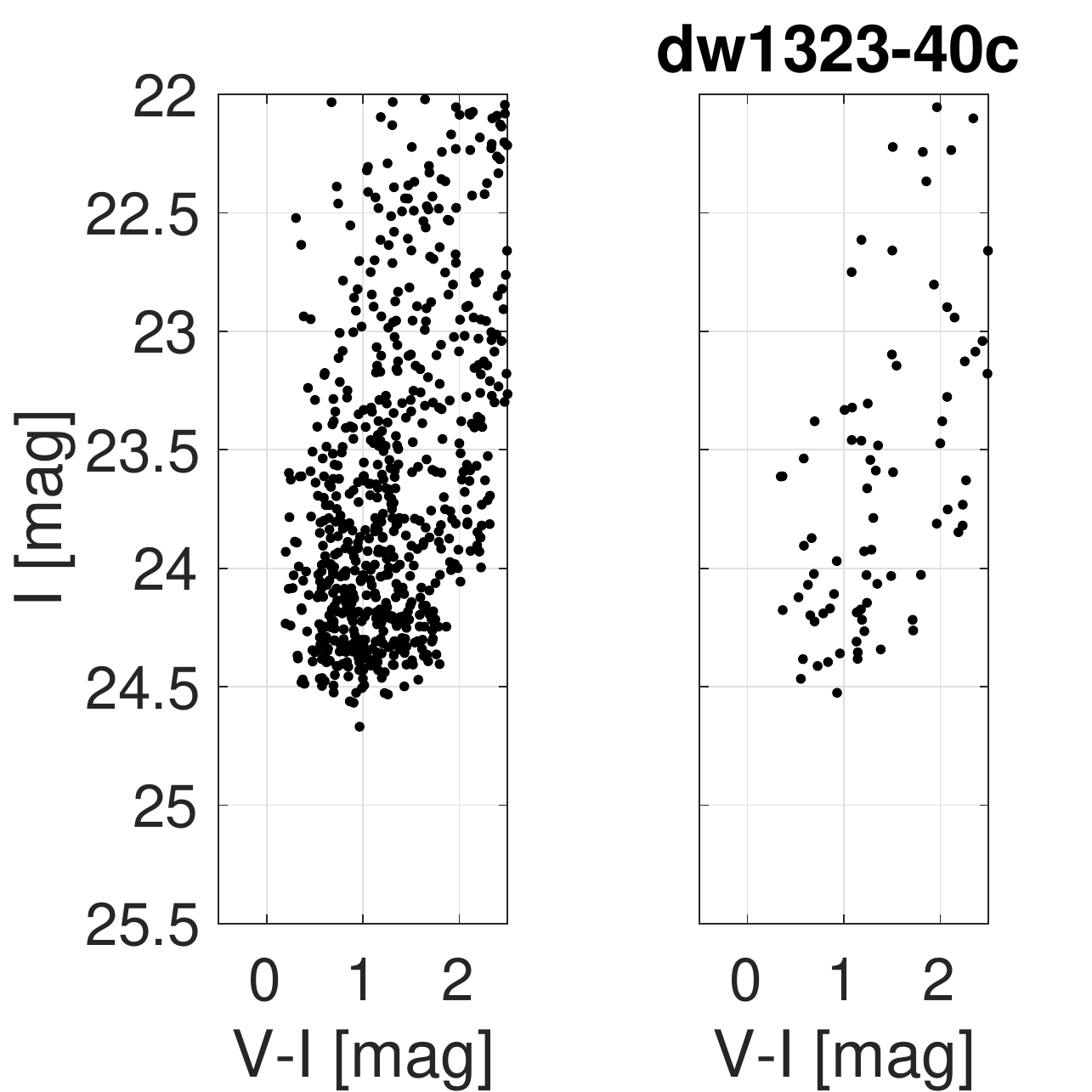}
\includegraphics[width=6cm]{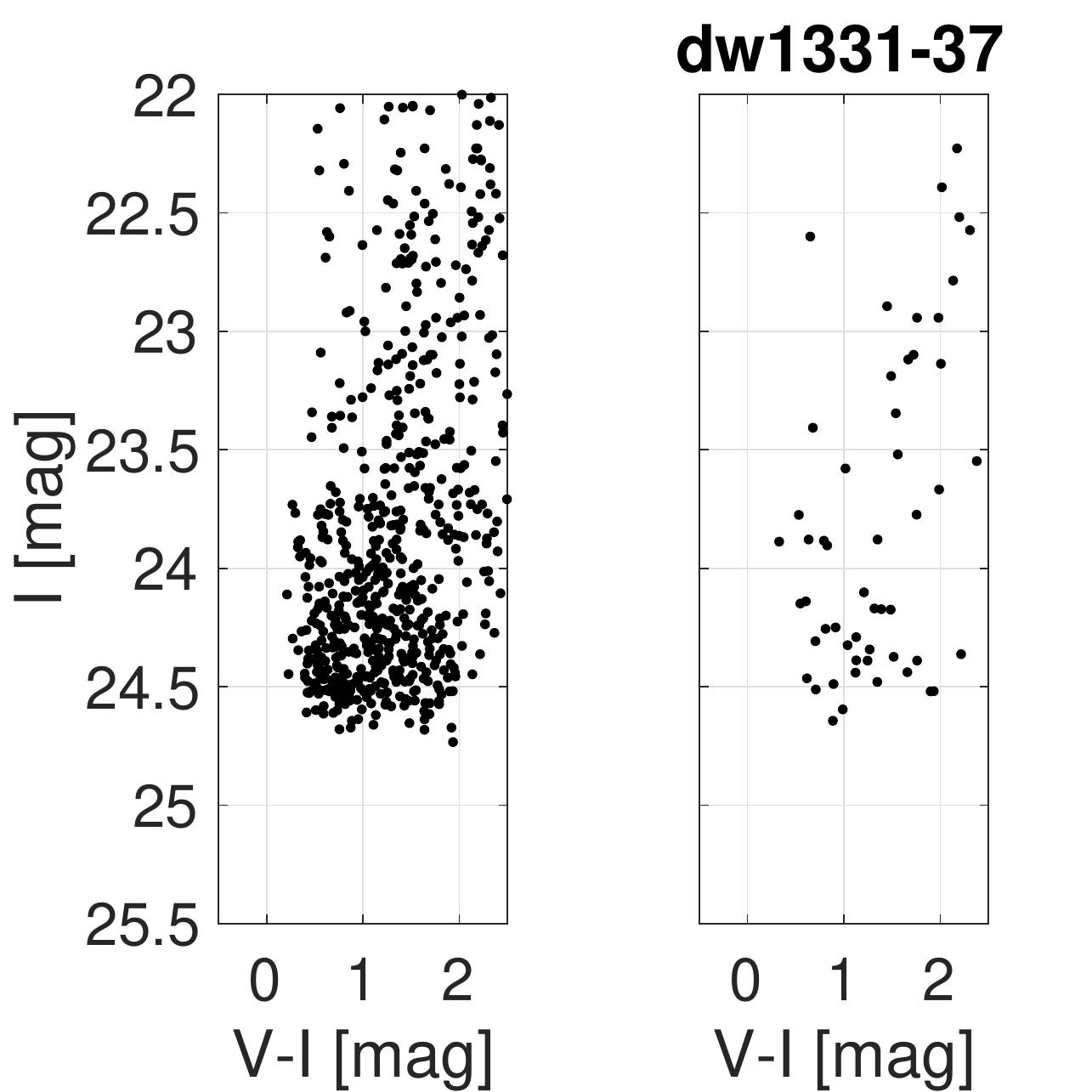}
\includegraphics[width=6cm]{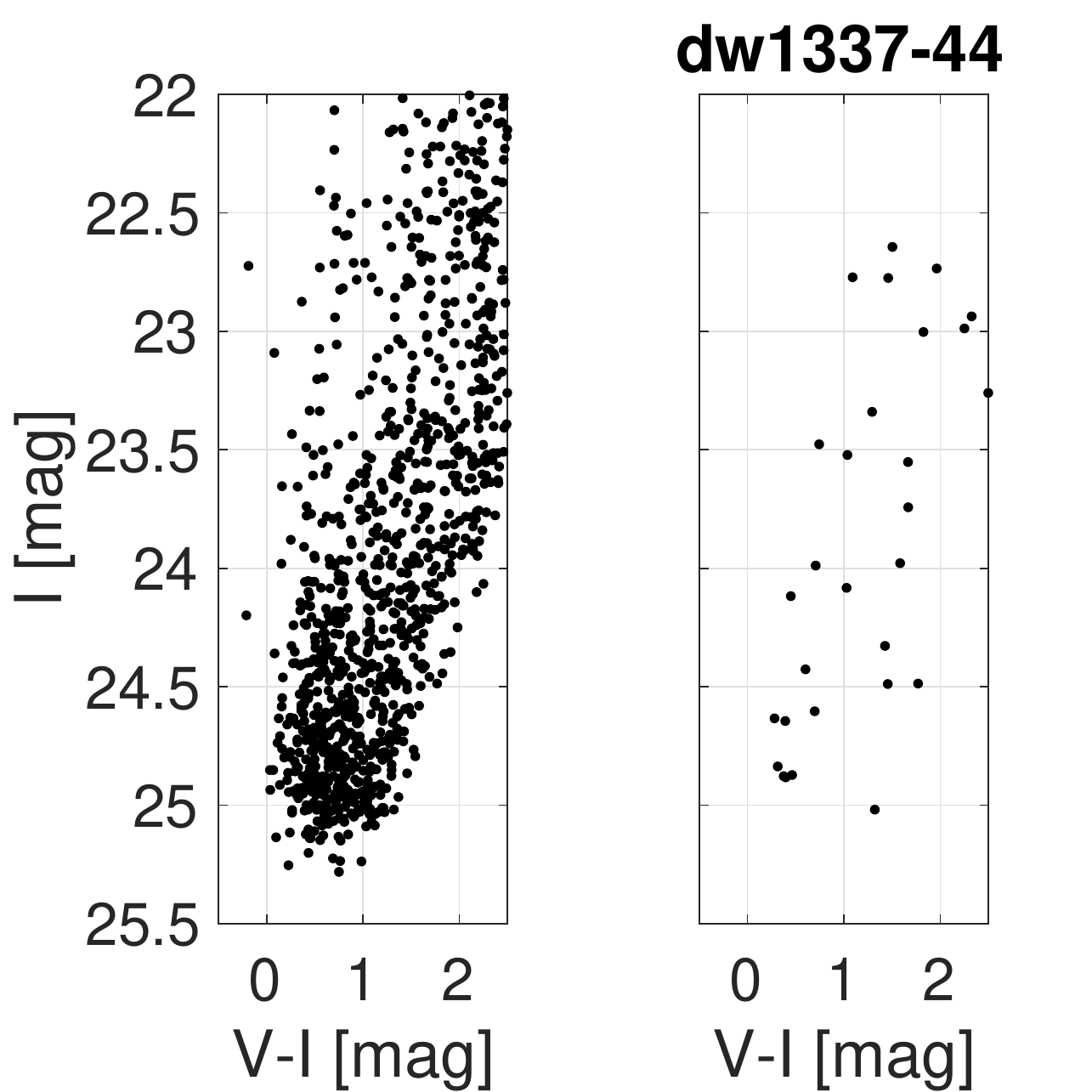}
\caption{Extinction-corrected color-magnitude diagrams for unresolved targets.
Left: CMD for all stars on the CCD. Right: Stars within the aperture.}
\label{cmds_false}
\end{figure*}

\begin{figure*}[ht]
\includegraphics[width=18cm]{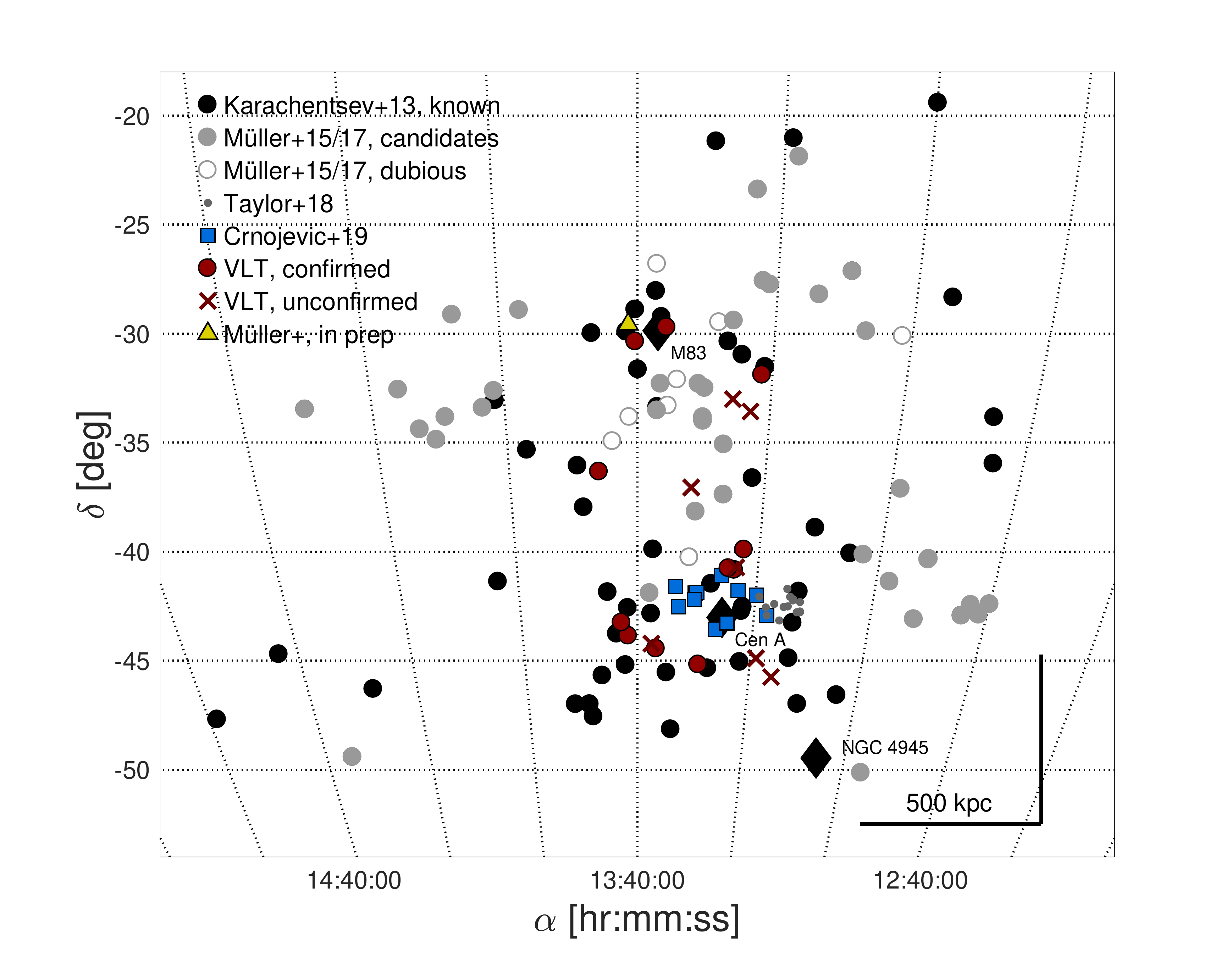}
\caption{{On-sky distribution of galaxies in field of Cen\,A/M\,83. The large black diamonds correspond to M\,83 (top), Cen\,A (middle), and NGC\,4945 (bottom). The black rulers indicate 500\,kpc at the distance and declination of Cen\,A. The black dots are confirmed dwarf galaxies in this field \citep{2013AJ....145..101K}, the gray dots are the remaining candidates in the field \citep{2015A&A...583A..79M,2017A&A...597A...7M,2018ApJ...867L..15T}.  With the gray circles, we denote dubious candidates (probably Galactic cirrus patches), based on their morphology, size, and faintness. The red dots correspond to confirmed dwarf galaxies employing the VLT (\citealt{MuellerTRGB2018}, and this work), the blue squares to the recently confirmed dwarfs with HST \citep{2019ApJ...872...80C}. The red crosses are the candidates which we ruled out as dwarf galaxies in the immediate vicinity of Cen\,A/M\,83 (\citealt{MuellerTRGB2018}, and this work).  {The yellow triangle indicates a newly discovered dwarf galaxy of Cen\,A, confirmed using SBF distance measurements (M\"uller et al, in preparation, see Section \ref{sec:planarity}).}
Our Cen\,A survey ends at roughly $\delta=-45$\,deg.
}
}
\label{field}
\end{figure*}

From our 15 targets, we were able to confirm nine to be members of Cen\,A based on their resolved stellar population. For six targets, however, we could not 
identify clear overdensity of stars {either} spatially, 
{or as RGB sequence} in CMDs, see Fig.\,\ref{cmds_false} for the corresponding CMDs.
One candidate, dw1337-44, is clearly a patch of spurious galactic cirrus. Another candidate, KK198, is actually a LSB background spiral galaxy. Even with our deep imaging, the spiral arms are barely visible, indicating that indeed such objects will contaminate any dwarf galaxy candidate list based on shallow imaging campaigns. The candidate dw1315-45 appears to be a background galaxy
in projection close to a galaxy at redshift $z=0.08$ ($D=360$\,Mpc). Putting dw1314-45 at this distance, it would have an effective radius of $r_{\rm eff}=17$\,kpc. 
Spectroscopy will be needed to unravel the true nature of this object, but it is reasonable to assume that this object lies closer than 360\,Mpc. For dw1323-40c, we found no feature what-so-ever, meaning that the detection in \citet{2017A&A...597A...7M} happened to be due to noise. Lastly, two objects, dw1318-44 and dw1331-37, are clearly visible as extended LSB objects and resemble dwarfs in their morphology. They may be in the outer {distant} part of the Centaurus aggregate, which is out of reach from our detection limit. {Their 25 percent completeness limit is 24.46\,mag and 24.72\,mag in the $I$-band, corresponding to a lower distance limit of 5.2\,Mpc and 5.8\,Mpc, respectively.}
{In Fig.\,\ref{field}, we show the on-sky distribution of all galaxies and remaining candidates in the field around Cen\,A/M\,83.}

\section{Properties of the new dwarf galaxies}
\label{sec:properties}
{Having derived the distances of the dwarf galaxies, it is possible to {investigate their location on the} scaling relations {defined by} Local Group dwarfs. Furthermore, we searched for 
tidal signatures in the {stellar} distribution.}
\subsection{Scaling relations}
\begin{figure*}[ht]
\includegraphics[width=9cm]{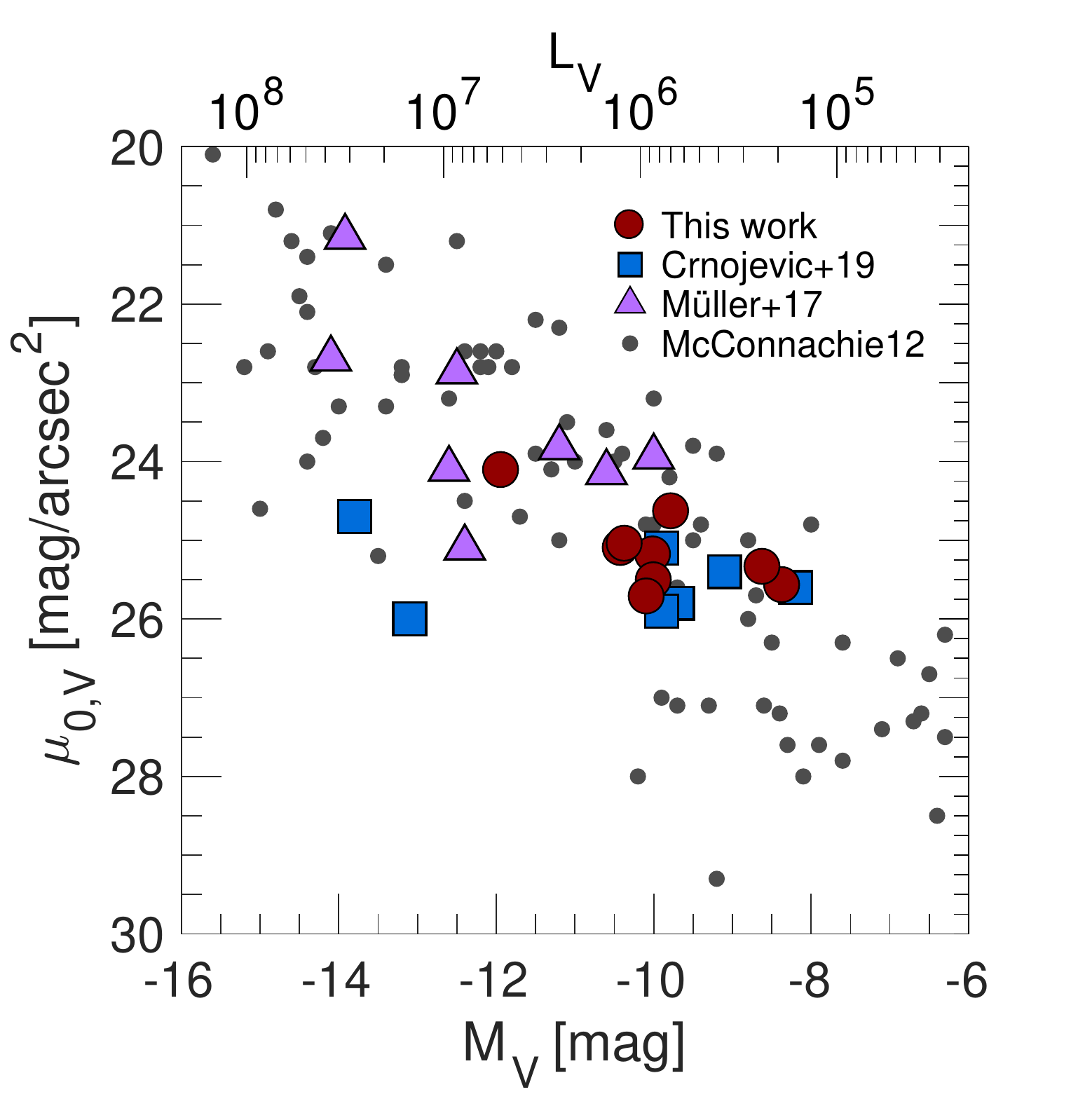}
\includegraphics[width=9cm]{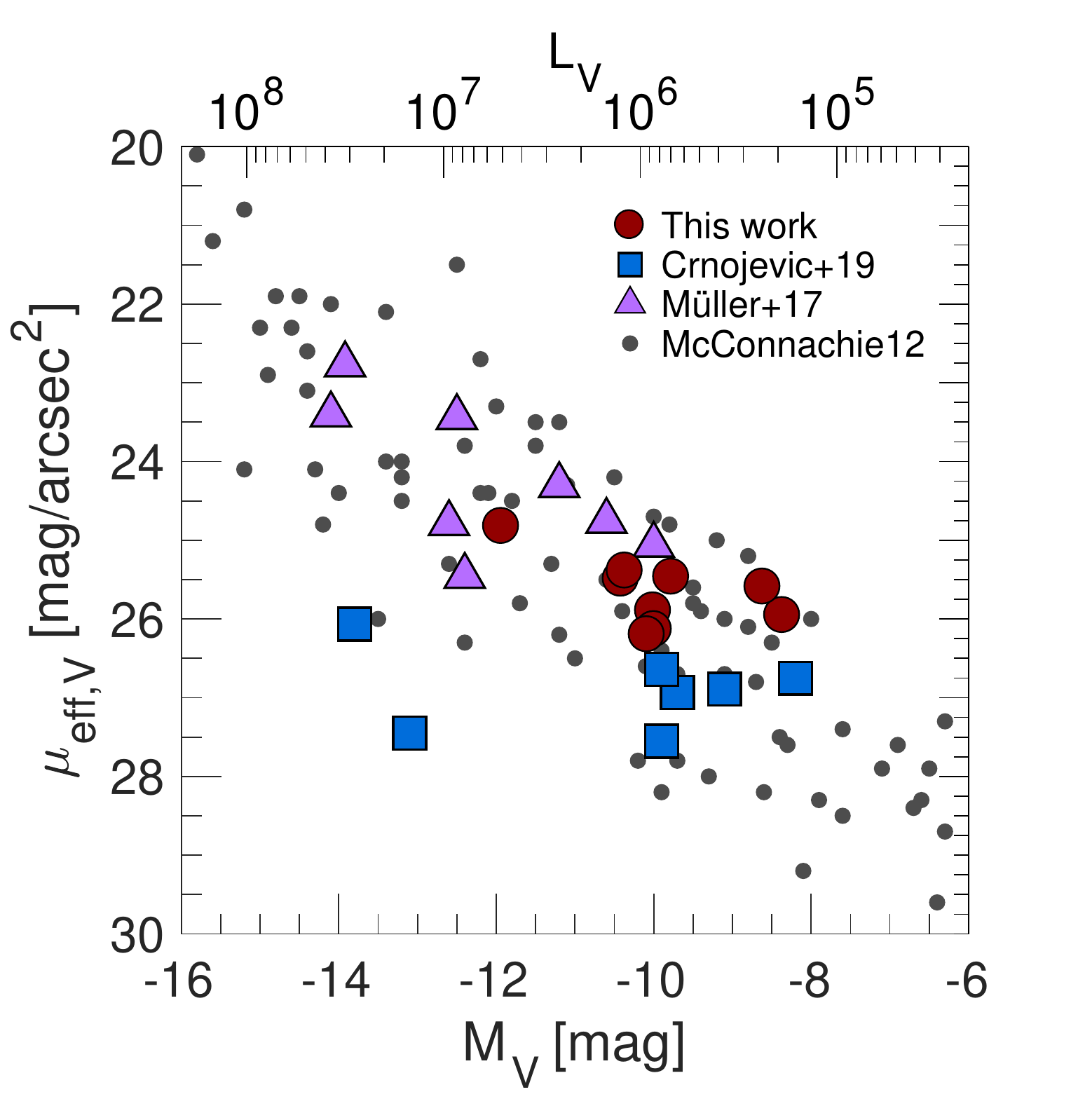}

\includegraphics[width=9cm]{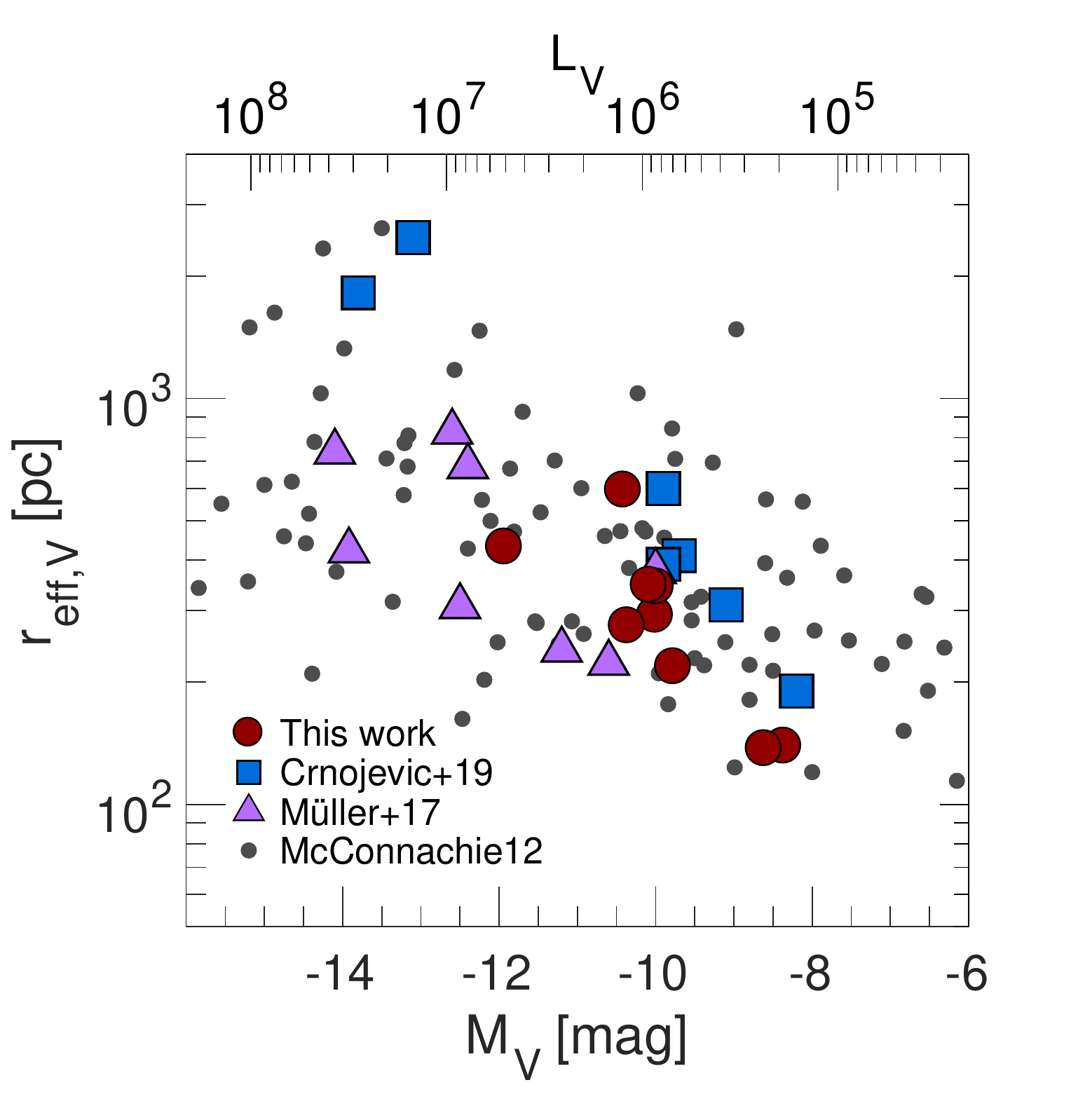}
\includegraphics[width=9cm]{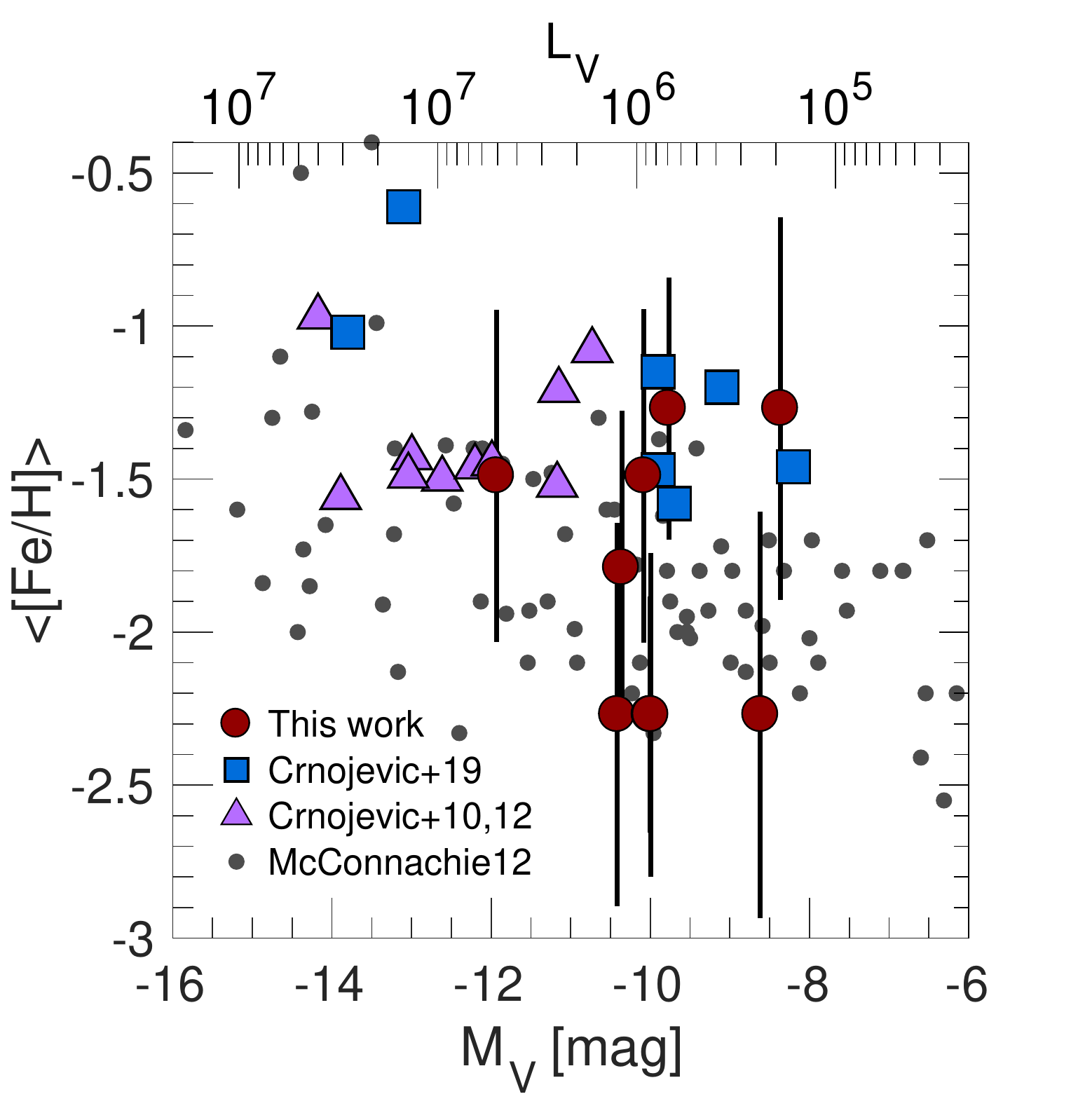}
\caption{{Scaling relations of dwarf galaxies: central surface brightness (top-left), effective surface brightness (top-right), effective radius (bottom-left), and mean metallicity (bottom-right) vs. absolute luminosity. We show Local Group dwarfs (gray dots, \citealt{2012AJ....144....4M}), previously studied Cen\,A dwarfs (blue squares and purple triangles, \citealt{2010A&A...516A..85C,2012A&A...541A.131C,2019ApJ...872...80C,2017A&A...597A...7M}), and our newly confirmed members (red circles).}}
\label{relations}
\end{figure*}

The dwarf galaxies in the Local Group follow several scaling relations \citep{2008ApJ...684.1075M,2013ApJ...779..102K}, such as the central surface-brightness to absolute luminosity relation or the effective radius to absolute luminosity relation (see also \citealt{2012AJ....144....4M}). These relations are used to estimate the membership of a dwarf galaxy candidate before follow-up observations are conducted \citep[e.g., ][]{2017A&A...597A...7M} by assuming a distance which corresponds to the putative host galaxy. Once the distance has been measured, no assumptions have to be made and the properties of the dwarfs can be compared to those {in} the Local Group. {We did not rederive the structural parameters on our FORS2 images, we rather used the values derived from integrated light photometry \citep{2000AJ....119..593J,2017A&A...597A...7M}. This is due to the difficulty of deriving structural parameters via the resolved stellar population (however, see \citealt{2008ApJ...684.1075M,2014ApJ...793L...7S}), especially when only a handful of stars are available.}
In Fig.\,\ref{relations}, we present several of the scaling relations for our newly confirmed dwarf galaxies and compare them to other known Local Group {\citep{2012AJ....144....4M}} and Centaurus A dwarf galaxies {\citep{2010A&A...516A..85C,2012A&A...541A.131C,2019ApJ...872...80C,2017A&A...597A...7M}}. It is apparent that the Centaurus A dwarf galaxies follow the relations as spanned by the Local Group dwarfs. In Fig.\,\ref{relations}, we also show the mean metallicity estimated from the CMD as a function of the luminosity. The errors are estimated by converting the mean color 
0.5\,mag below the TRGB into a 
{mean} metallicity via \citet{2007ApJ...661..815R}:
\begin{eqnarray*}
{\rm [Fe/H]}= - 12.64 + 12.6\,(V -I )_0 - 3.3\,(V-I )_0^2.
\end{eqnarray*}
{This metallicity value is fully consistent with the one obtained by fitting stellar evolutionary isochrones from the BaSTI database \citep{2004ApJ...612..168P} and assuming a 10 Gyr old population.} {In Table \ref{properties}, we further present the ellipticity ($e=1-b/a$) and position angle ($PA$) of the dwarf galaxies, measured with MTO \citep{teeninga2013bi}, which is a software program to detect astronomical sources. The ellipticities range between 0.1 and 0.6, which is consistent with the bulk of Local Group dwarfs \citep{2012AJ....144....4M}.}

\begin{table*}[ht]
\caption{Properties of newly confirmed dwarf galaxies.}
\centering                          
\small
\setlength{\tabcolsep}{2pt} 
\begin{tabular}{l c c c c c c c c c}        
\hline\hline                 
 & KKs54 &dw1322-39 & dw1323-40b & dw1323-40  & dw1329-45 & dw1336-44 & dw1341-43 & dw1342-43 & KKs58 \\    
\hline      \\[-2mm]                  
RA (J2000)  & 13:21:31.8 & 13:22:31.8 & 13:23:55.6 & 13:24:53.9  & 13:29:09.9 & 13:36:49.0 & 13:41:36.6 & 13:42:43.8 &  13:46:00.4\\ \addlinespace[0.05cm]
DEC (J2000)  & $-$31:53:09.8 & $-$39:54:21.6 & $-$40:50:09.1 & $-$40:45:38.7 & $-$45:10:33.7 &$-$44:26:55.5 & $-$43:51:19.6 & $-$43:15:21.1 & $-$36:19:42.4\\ \addlinespace[0.05cm]
$I_{TRGB}$ (mag) & 
$23.78^{+0.01}_{-0.12}$ & 
$23.26^{+0.01}_{-0.06}$ & 
$23.87^{+0.13}_{-0.62}$ & 
$23.78^{+0.00}_{-0.19}$ & 
$23.31^{+0.07}_{-0.11}$ & 
$23.63^{+0.15}_{-0.19}$ &  
$23.70^{+0.00}_{-0.03}$ &   
$23.31^{+0.01}_{-0.22}$ &  
$23.59^{+0.12}_{-0.01}$ \\ \addlinespace[0.05cm]
$[$Fe/H$]$ (dex) & 
$-2.27$ & 
$-2.27$ & 
$-2.27$ & 
$-1.79$ &
$-1.27$ &
$-2.27$ &  
$-1.49$ & 
$-1.27$ & 
$-1.49$\\ \addlinespace[0.05cm]
$M_I^{TRGB}$ (mag) & 
$-4.09$ & 
$-4.09$ &
$-4.09$ &
$-4.08$ & 
$-4.00$ & 
$-4.09$ & 
$-4.05$ & 
$-4.00$ & 
$-4.05$\\ \addlinespace[0.05cm]
$(m-M)_0$ (mag) &  
$27.87^{+0.01}_{-0.12}$ &  
$27.35^{+0.01}_{-0.06}$ &
$27.96^{+0.13}_{-0.62}$ & 
$27.86^{+0.00}_{-0.19}$ & 
$27.31^{+0.07}_{-0.11}$ &  
$27.72^{+0.15}_{-0.19}$ & 
$27.74^{+0.00}_{-0.03}$ &
$27.31^{+0.01}_{-0.22}$ & 
$27.63^{+0.12}_{-0.01}$\\ \addlinespace[0.05cm]
Distance (Mpc) & 
$3.75^{+0.01}_{-0.21}$ & 
$2.95^{+0.02}_{-0.08}$ &
$3.91^{+0.24}_{-0.97}$ &
$3.73^{+0.01}_{-0.31}$ &  
$2.90^{+0.10}_{-0.14}$ &   
$3.50^{+0.25}_{-0.30}$ & 
$3.53^{+0.00}_{-0.04}$ &  
$2.90^{+0.01}_{-0.27}$ &   
$3.36^{+0.18}_{-0.02}$ \\ \addlinespace[0.05cm]
$A_V, A_I$ (mag)& 0.176, 0.097 & 0.244, 0.134 & 0.334, 0.183 & 0.306, 0.168  & 0.252, 0.138 & 0.338, 0.186 & 0.255, 0.140 & 0.218, 0.120 & 0.169, 0.093\\ \addlinespace[0.05cm]
$M_{V}$ (mag) & 
$-10.41^{+0.01}_{-0.12}$ & 
$-10.03^{+0.01}_{-0.06}$ & 
$-9.99^{+0.13}_{-0.62}$ & 
$-10.36^{+0.00}_{-0.19}$ & 
$-8.36^{+0.07}_{-0.11}$ & 
$-8.61^{+0.15}_{-0.19}$ & 
$-10.08^{+0.00}_{-0.03}$ &
$-9.77^{+0.01}_{-0.22}$ & 
$-11.93^{+0.12}_{-0.01}$\\ \addlinespace[0.05cm]
$L_{V}$ ($10^6$ M$_\odot$) & 
$1.25^{+0.01}_{-0.14}$ & 
$0.85^{+0.01}_{-0.05}$ & 
$0.85^{+0.11}_{-0.37}$ &
$1.19^{+0.00}_{-0.19}$ & 
$0.19^{+0.01}_{-0.02}$ & 
$0.24^{+0.04}_{-0.04}$ & 
$0.92^{+0.00}_{-0.02}$ &  
$0.69^{+0.01}_{-0.12}$ & 
$5.06^{+0.57}_{-0.05}$ \\ \addlinespace[0.05cm]
$r_{eff,r}$ (pc) & 
$594^{+1}_{-17}$ & 
$296^{+2}_{-8}$ & 
$343^{+20}_{-80}$ &
$275^{+0}_{-23}$ & 
$139^{+5}_{-7}$ & 
$137^{+10}_{-12}$ & 
$346^{+0}_{-4}$ & 
$218^{+1}_{-21}$ & 
$430^{+23}_{-2}$\\ \addlinespace[0.05cm]
{Ellipticity} ($1-b/a$) &
0.21& 
0.50&
0.64&
0.10&
--&
0.33&
0.08&
0.28&
0.14 \\ \addlinespace[0.05cm]
$PA$ {(north to east)} &
96.8&
120.8&
168.1&
18.8&
-- &
81.0&
0.3&
67.4&
38.5 \\ \addlinespace[0.05cm]

\hline 
\end{tabular}
\label{properties} 
\tablefoot{The properties for $M_{V}$, $L_{V}$, and $r_{eff,r,pc}$ were derived using the distances estimate here and the published structural parameters from \citet{2000AJ....119..593J} for KKs54 and KKs58, and from \citet{2017A&A...597A...7M} for the rest. {Due to its closeness to a bright star, we are not able to derive $e$ and $PA$ for dw1329-45.}}
\end{table*}

\subsection{Tidal features}
The dwarf galaxies studied here are {likely bound members}
of Cen\,A \citep{2006AJ....132.2424W}
{and tidal forces may be important.}
One may wonder if 
{tidal disturbances are}
detectable in the on-sky distribution of the observed RGB stars. For that, we created a {stellar} map for each detected point source between I=23.0 mag and I=25.5 mag. 
Additionally, we used the best fitting isochrone to color-code the stars corresponding to a mask of the dwarf galaxy's red giant branch in red {(having a color between 0.9 and 1.8 mag)}, and the remaining stars are in blue. These {stellar} maps are then binned into $100\times100$ {pixels} boxes, which are color-coded according to the {RGB} stellar density within the box. 
{The centers of the galaxies were estimated with a k-mean algorithm as explained in \citet{MuellerTRGB2018}.}  We inspected each {stellar} map overlayed with its binned density map and searched for the following two distinct features: (a) {asymmetries in the overall stellar distribution},
and (b) overdensities of potential RGB stars around the dwarf galaxy.

For dw1323-40b, we indeed detected some overdensities of stars at the southern outskirt of the galaxy. The stellar density in this region is almost as high as in the center. By studying the stacked image itself, it is apparent that this dwarf galaxy is elongated along this direction {and is highly elliptical with $e=0.64$}. However, with the depth of the photometry no conclusive statement can be made about it being truly tidally distorted or not; the overabundance could well be a small-number fluctuation {and/or an effect of the large ellipticity}. 

For dw1336-44, we find a discrepancy between the center of the galaxy derived by the 
{RGB} stars only and all the stars within the galactic aperture. But this can be explained by the few stars found within the aperture ($\sim 60$ stars).
In the field outside of the galactic apertures of the studied dwarf galaxies, there are no significant stellar overdensities. 

To summarize, we find no convincing evidence that the dwarfs are tidally disturbed based on the {probable member star distribution}
on the CCD chips. This is not unexpected, as there was also no visible evidence in the integrated-light images of our DECam survey \citep{2017A&A...597A...7M}. In contrast, the dwarf galaxies Cen\,A-MM-dw3 \citep{2016ApJ...823...19C} and KK\,208 \citep{2002A&A...385...21K} are visibly disturbed in the integrated-light images, see for example Fig.\,10 in \citet{MuellerTRGB2018} for the latter. {These two galaxies are also, on average, 
closer to the large host (Cen A and M83) than our targets. 
}


\begin{figure}[ht]
\centering
\includegraphics[width=9cm]{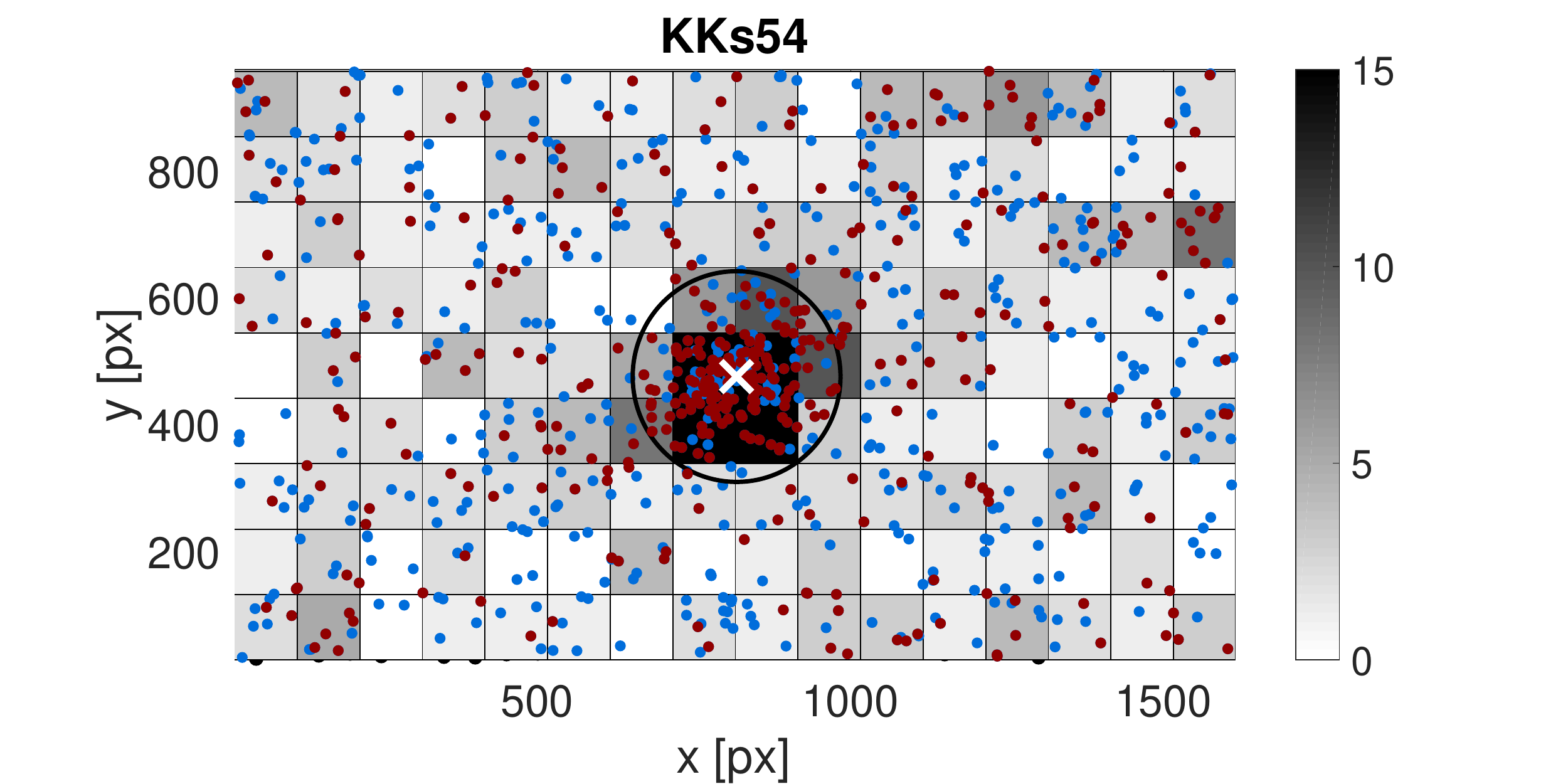}
\caption{Stellar map of one confirmed dwarf galaxy. Red indicates stars {within} the RGB mask according to the best-fitting isochrone. Blue indicates the remaining stars. The circle {shows} the aperture {used to construct the CMD.}
{The remaining maps are shown in Fig.\,\ref{app:maps} and \ref{app:maps2} in the Appendix.}
}
\label{maps}
\end{figure}

\section{The luminosity function of Cen\,A {satellite system}}
\label{LFcena}

\begin{table}[ht]
\caption{{Dwarf galaxies within projected radius of 200\,kpc around Cen\,A.}}
\centering                          
\small
\begin{tabular}{l c c c r}        
\hline\hline                 
Name & $\alpha$ & $\delta$ &   $D$ &  $M_V$\\    
 & deg (J2000.0) & deg (J2000.0)  & Mpc & mag \\  
\hline      \\[-2mm]                  
KK189             &             198.1875        &-41.8319&      4.23&   -11.2\\
ESO269-066        &             198.2875        &-44.8900&      3.75&   -14.1\\
NGC5011C          &             198.2958        &-43.2656&      3.73&   -13.9\\
CenA-MM-Dw11            &       199.4550        &-42.9269&      3.52&   -9.4\\
CenA-MM-Dw5       &             199.9667        &-41.9936&      3.61&   -8.2\\
KK196             &             200.4458        &-45.0633       &3.96&  -12.5\\
KK197             &             200.5042        &-42.5356&      3.84&   -12.6\\
KKs55           &               200.5500        &-42.7308&      3.85&   -12.4\\
CenA-MM-Dw10       &            200.6214&       -39.8839&       3.27&   -7.8 \\
dw1322-39       &               200.6558        &-39.9084&      2.95&   -10.0\\
CenA-MM-Dw4       &             200.7583        &-41.7861&      4.09&   -9.9\\
dw1323-40b       &              201.0000        &-40.8367&      3.91&   -9.9\\
dw1323-40       &               201.2421        &-40.7622&      3.73&   -10.4\\
CenA-MM-Dw6       &             201.4875        &-41.0942&      4.04&   -9.1\\
CenA-MM-Dw7       &             201.6167        &-43.5567&      4.11&   -7.8\\
ESO324-024       &              201.9042        &-41.4806&      3.78&   -15.5\\
KK203             &             201.8667        &-45.3525&      3.78&   -10.5\\
dw1329-45       &               202.3121        &-45.1767&      2.90&   -8.4\\
CenA-MM-Dw2       &             202.4875        &-41.8731&      4.14&   -9.7\\
CenA-MM-Dw1       &             202.5583        &-41.8933&      3.91&   -13.8\\
CenA-MM-Dw3       &             202.5875        &-42.19255&     3.88&   -13.1\\
CenA-MM-Dw9       &             203.2542        &-42.5300&      3.81&   -9.1\\
CenA-MM-Dw8       &             203.3917        &-41.6078&      3.47&   -9.7\\
dw1336-44       &               204.2033        &-43.8578&      3.50&   -8.6\\
NGC5237          &              204.4083        &-42.8475 &     3.33&   -15.3\\
KKs57           &               205.4083        &-42.5819&      3.83&   -10.6\\
dw1341-43       &               205.4221        &-44.4485&      3.53&   -10.1\\
dw1342-43       &               205.7029        &-43.8561&      2.90&   -9.8\\
KK213            &              205.8958        &-43.7692&      3.77&   -10.0\\ 
\hline 
\end{tabular}
\label{table:lf} 
\tablefoot{{The TRGB distances are compiled in the LV catalog \citep{2013AJ....145..101K} and are from the HST programs of \citet{2007AJ....133..504K}, \citet{2019ApJ...872...80C}, and the VLT data presented here. 
The photometry comes from \citet{2000AJ....119..593J}, \citet{2017A&A...597A...7M}, and \citet{2019ApJ...872...80C}.}}
\end{table}

A crucial test in small-scale cosmology is the number of predicted dark matter subhalos in comparison to the observed number of dwarf galaxies. Around the Milky Way, the number of predicted subhalos {from dark matter-only simulations} exceeded the observations by a factor of more than 100 \citep{1999ApJ...524L..19M}. This has since been revised by taking baryons into account with {recent simulations} 
matching {the} observations \citep[e.g., ][]{2016MNRAS.457.1931S,2018MNRAS.478..548S}{, leading to good agreement between the observations and cosmology in the Local Group}. \citet{2018ApJ...863..152S}  claimed a dearth of satellite galaxies around M\,94 {and find only two satellites in deep Subaru Hyper Suprime Cam imaging within 100\,kpc. This}  is
incompatible with state-of-the-art cosmological simulations. {However, in the more extended vicinity ($\sim$500\,kpc), 14 probable satellites with known radial
velocities scatter around M\,94 \citep{2014AJ....148...50K}. Hence, it is still to be seen how the total number of satellites around M\,94 compares to cosmological simulations}. 

Around Cen\,A, we established a good census of dwarf galaxies within 200\,kpc {(see Table \ref{table:lf})}, which {we now compare}
to cosmological simulations. {The completeness of the dwarf galaxy catalog is $M_V=-10$\,mag, based on artificial galaxy tests \citep{2017A&A...597A...7M}. {At fainter luminosity, incompleteness kicks in, meaning that we underestimated the total number of satellites.}}
We used the IllustrisTNG Project \citep{2018MNRAS.475..676S}, 
specifically the simulation TNG100-1 with a box size of 110.7 Mpc and a dark matter particle mass $m_\mathrm{DM} = 7.5 \times 10^6\,M_\sun$. We used the publicly available $z=0$ galaxy catalogs \citep{2018arXiv181205609N}.

Our Cen\,A analogs were selected to be the primaries of halos within a mass range of $4.0\times10^{12} \leq M_\mathrm{200} \leq 1.2\times10^{13}$ \citep{2007AJ....133..504K,2006AJ....132.2424W,2007AJ....134..494W,2010AJ....139.1871W}, where $M_\mathrm{200}$\ is the total mass within a sphere that has a mean density of 200 times the critical density of the Universe. To ensure a comparable isolation to Cen\,A, all hosts that have another halo with $M_\mathrm{200} \geq 0.5\times10^{12}$ within a distance of 1.4\,Mpc were discarded {(representing the separation between Cen\,A and M\,83)}. This left us with 237 hosts. {We mock-observed these systems}
from random directions by placing them at a distance of 3.8\,Mpc and selecting all luminous galaxies that {(1)} lie within a cone of opening angle $3.5^\circ$ around the host galaxy, {(2)} are no closer than $0.5^\circ$\ to the host, and {(3)} have a distance of less than 800\,kpc from the host, mimicking the selection volume of our observed galaxies. The $V$-band magnitudes $M_\mathrm{V}$\ of the selected ``satellites,'' as provided by the IllustrisTNG catalog, are then compared to the magnitudes of the observed satellite system.

\begin{figure*}[ht]
\includegraphics[width=8.31cm]{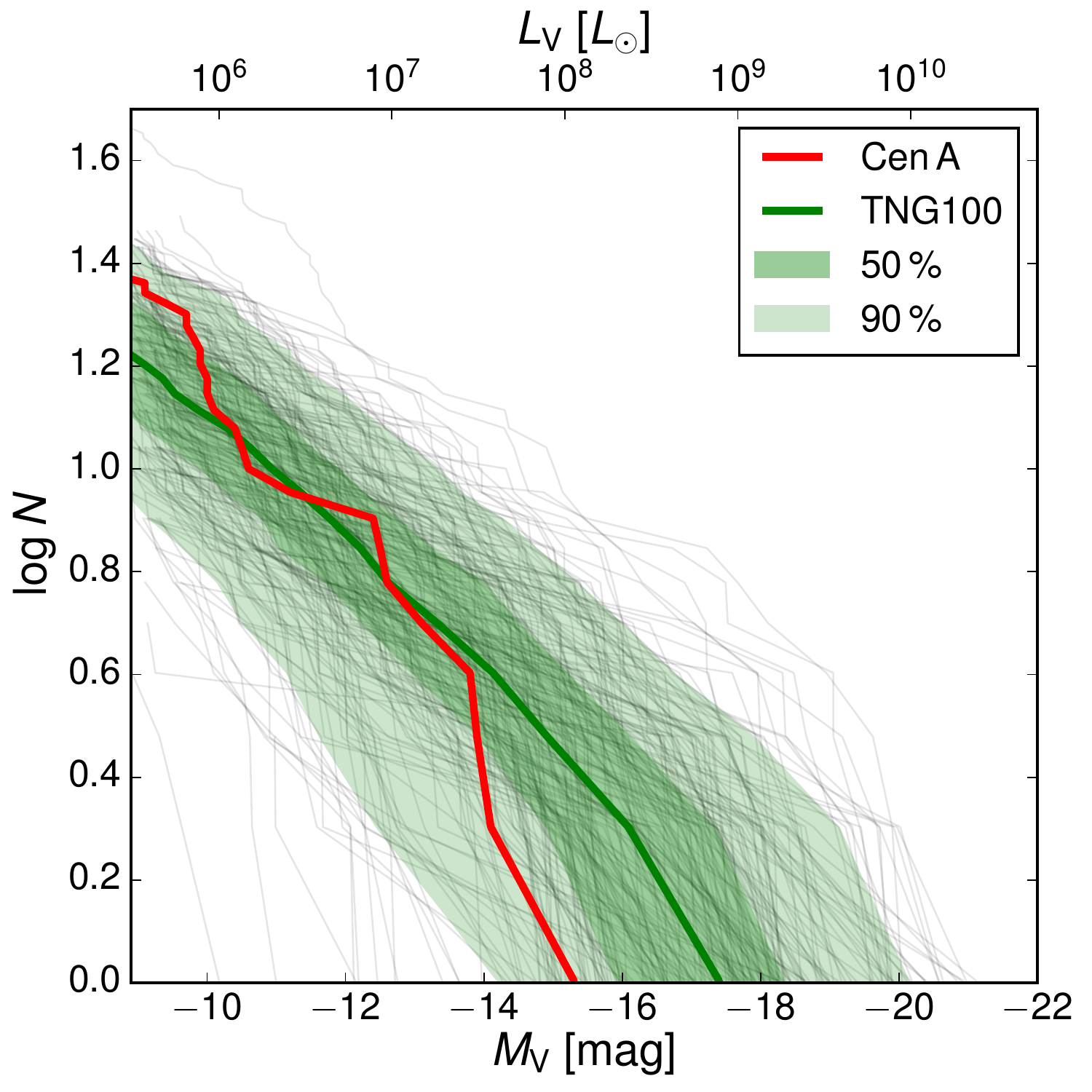}
\includegraphics[width=8.31cm]{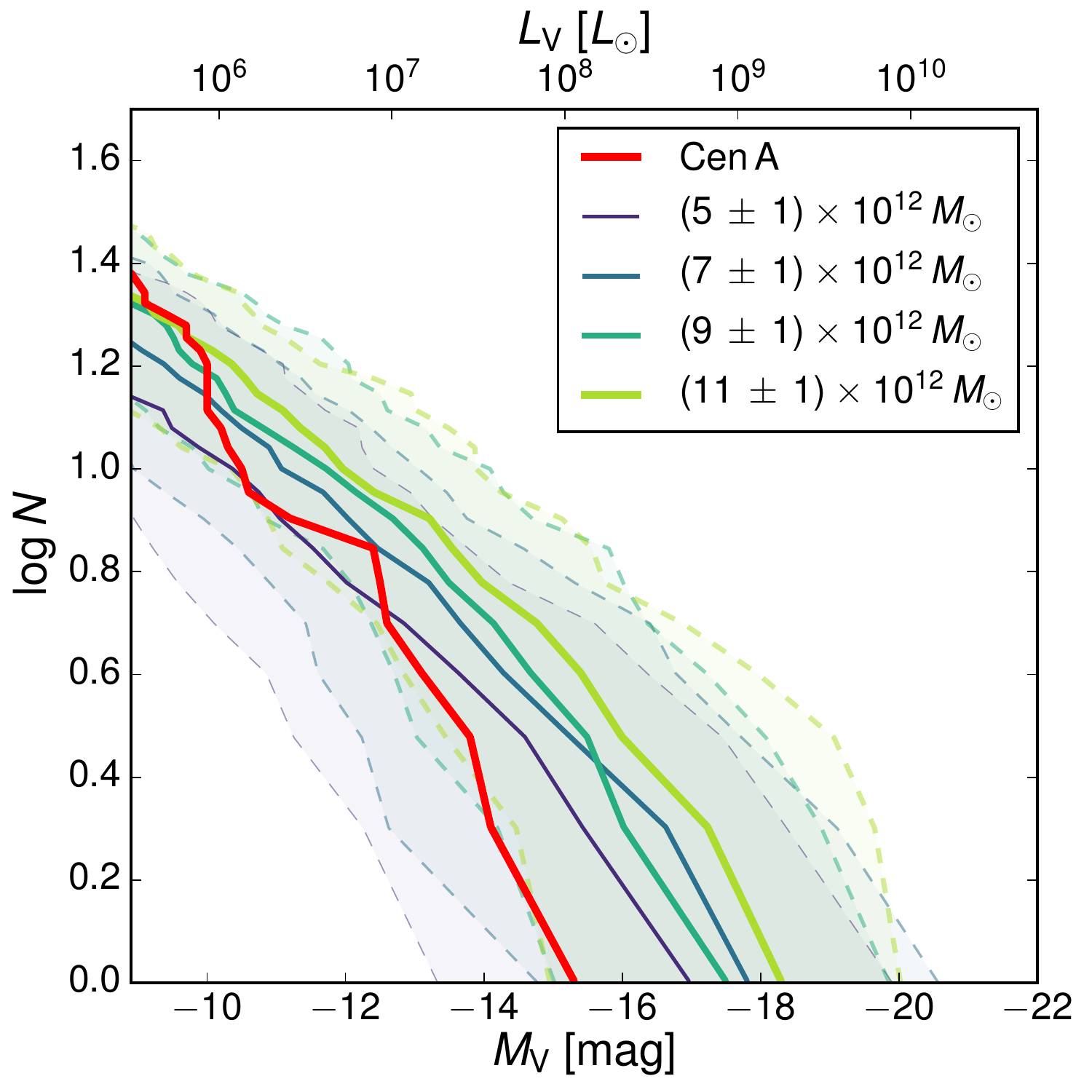}
\caption{
{
Comparison between the observed and the simulated luminosity functions. Left panel: The luminosity function of Cen\,A within 200\,kpc (red line) compared to the Illustris TNG100-1 simulation. The gray lines correspond to individual simulated Cen\,A analogs, the green line to the median, and the green {shaded areas} to the 90\% and 50\% intervals. Right panel: Same {as the left panel}, but now the simulated systems are binned {based on} their host halo mass $M_\mathrm{200}$, with solid lines indicating their mean relation and dashed lines their 90\% intervals.
}
}

\label{LF}
\end{figure*}

In Figure\,\ref{LF}, we show the {observed} luminosity function of the Cen\,A {satellite system} and the analogs from the simulations. Interestingly, there seems to be a lack of bright satellites. {However, in absolute numbers this constitutes only a lack of two to three satellites with
$M_{\rm V} < -15$, 
and this is furthermore still within the 90\% interval defined by the simulated analogs. The right panel indicates that within the considered range, the lower host halo masses of $M_{\rm 200} = 4$\ to $6 \times 10^{12}\,M_\sun$ provide a better match to the observed luminosity function {(down to $M_V\approx-10$\,mag).\ Nevertheless,}  for all halo masses there is agreement to within the 90 percent interval such that no strong conclusions on the halo mass of Cen\,A can be drawn.} On the faint end of the luminosity function ($M_{\rm V} > -10$), the number of observed dwarfs is actually overshot (but still within the 90\% interval). {However, this might be a sign of lacking convergence in the simulations due to insufficient resolution to reliably model the smaller satellite galaxies.}

\section{The spatial structure of the {Centaurus} group}
\label{spatial}
\begin{figure*}[ht]
\includegraphics[width=18cm]{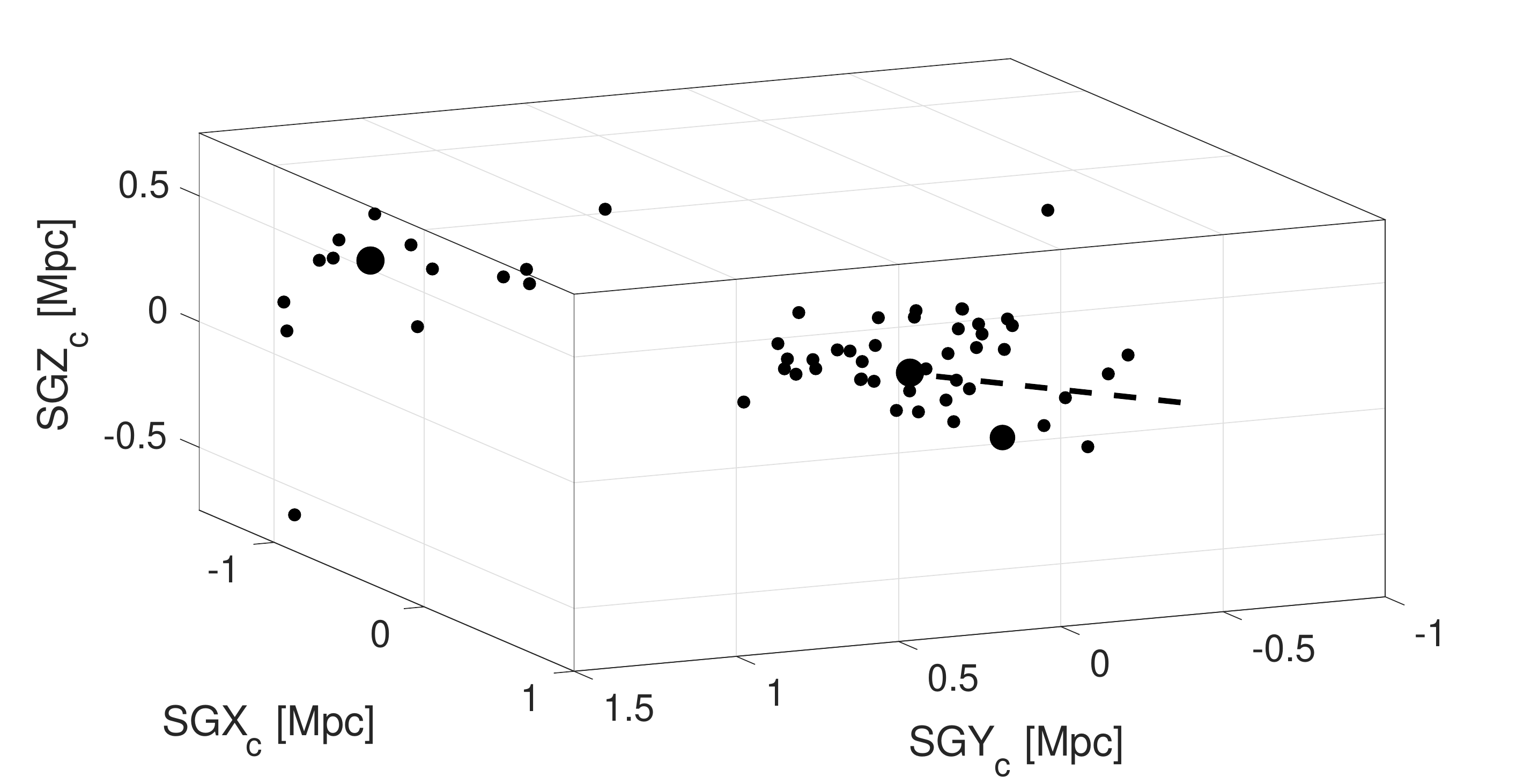}
\caption{3D distribution of galaxies in Centaurus association, {using} supergalactic coordinates centered on Cen\,A. Two subgroups {centered} around M\,83 ({large dot on the} left) and Cen\,A ({large dot on the} right) {are clearly visible}. In the Cen\,A subgroup, NGC\,4945 is also indicated {with a large dot}. The dotted line indicates our line-of-sight toward Cen\,A, originating from the $SGY_cSGZ_c$ plane. The literature distance values are taken from \citet{2015ApJ...802L..25T}, \citet{MuellerTRGB2018}, and \citet{2019ApJ...872...80C}.
}
\label{3d}
\end{figure*}

Having derived {accurate} distances, we {can}
study the {3D} distribution of the galaxies in the Centaurus association. In Fig.\,\ref{3d}, we show all confirmed galaxies in the Centaurus region with TRGB distances. There are the following two {main} concentrations: a well populated one around Cen\,A and a less populated one around M\,83. Apart from these concentrations, there are only a handful of galaxies (i.e., four to five) in the field around the two main galaxies. Based on the sky distribution of the dwarf galaxy candidates from our DECam survey, we suspected a bridge of dwarf galaxies {similar to the ones} discovered in the Local Group \citep{2013MNRAS.435.1928P}.
Such a connection is not apparent here, with the dwarf galaxies clustering around the major galaxies. {However,} several dwarf galaxy candidates still lack distance measurements, especially in the southern region of M\,83, {as is apparent in Fig.\,\ref{field}}.

\subsection{{Pairs of satellites}}

\citet{2014ApJ...795L..35C} announced the first discovery of a pair of satellites outside of the Local Group, the dwarf galaxies Cen\,A-MM-dw1 and Cen\,A-MM-dw2, which are satellites of Cen\,A. Their projected separation is only 3\,arcmin (= 3.9\,kpc at Cen\,A's distance), and they shared the same distance estimates. But with their updated distances \citep{2019ApJ...872...80C}, Cen\,A-MM-dw2 seems to be located 200\,kpc further away than what was previously assumed. \citet{2019ApJ...872...80C} still argue that the probability for a chance alignment is negligible and, therefore, that  they are indeed a pair of satellites given the distance uncertainties. We tested this assessment by making a simple Monte Carlo simulation. In every run we created two satellites at a random position in a sphere of 0.4\,Mpc radius (i.e., the virial radius).\ We then calculated their true separation as well as their projected separation in the $xy$-plane (approximating the sky). After performing this 100,000 times, we counted the number of times we found a pair of satellites with an $xy$ projection of 4\,kpc or less. In total, 0.01 percent of the random realizations appear to be a pair of satellites by chance, given their on-sky projection. 
 This shows that in principle, we can detect pairs of satellites, given that their on-sky separation is small and their distance estimates are similar. 
 
 We calculated the on-sky distances between the dwarf galaxy satellites as well as the 3D distances. {We} found that {one} other potential pair, KK\,197 and KKs\,55, have a projection separation of 12 arcmin (= 13 kpc at a distance of 3.85\,Mpc) and a 3D separation of 17\,kpc.  
 {There are  two other galaxies, Cen\,A-MM-dw1 and Cen\,A-MM-dw3, with a projection separation of 18 arcmin (= 20 kpc at a distance of 3.89\,Mpc) and a 3D separation of 36\,kpc. However, Cen\,A-MM-dw3 is a tidally disrupted dwarf extending over several hundred kpc, which complicates the analysis. We therefore excluded these two galaxies from the list of pairs.}
 The chance alignments of { KK\,197 and KKs\,55} on-sky projections is with $\sim$0.1 percent, which is still quite low.\ Furthermore, adding the 3D separation will make it vanish. The question remains of at which separation do we count two galaxies as pairs. Let us assume that the virial mass of a typical dwarf galaxy is around $10^{10}$\,M$_\odot$ \citep{2017ARA&A..55..343B}, which leads us to a virial radius of 45\,kpc (using a Hubble constant of 68\,km s$^{-1}$ Mpc$^{-1}$). Thus, with their measured separations, the two galaxies could indeed be identified as a physical pair. KK\,197 is 2.5 times brighter than KKs\,55 \citep{2017A&A...597A...7M}, that is 2.5 times more massive assuming the same mass-to-light ratio for both galaxies. 
 Assuming that these {four} dwarf galaxies are indeed {two} physical pairs, the abundance of pairs of galaxies within the virial radius of Cen\,A  is approximately 14  percent{, that is 4 out of 28 dwarf galaxies}. However, we note that to confirm these candidate  pairs, they should share a similar systemic velocity.
 
\subsection{{Planarity of the satellite distribution}}
\label{sec:planarity}
\begin{figure}[ht]
\includegraphics[width=9cm]{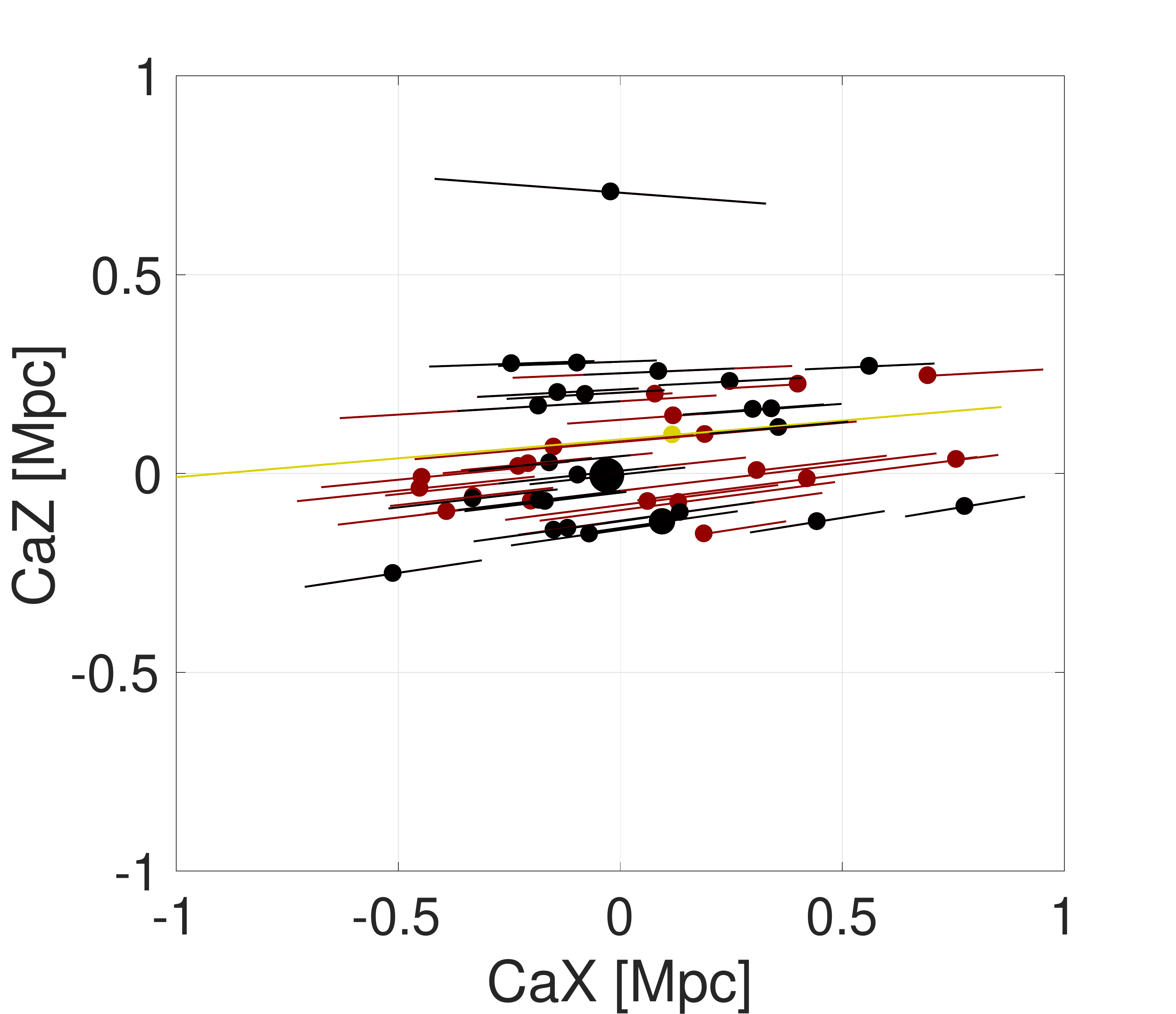}
\caption{Edge-on view of Centaurus group {in Centaurus A centered coordinate system according to \citet{2015ApJ...802L..25T}, where CaZ corresponds to normal of plane}. The two large dots correspond to Cen\,A and NGC\,4945, the small ones to the dwarf galaxies with measured distances (black for the previously known dwarfs, red for the new dwarfs from this study and from \citealt{2019ApJ...872...80C}, {and yellow for the new dwarf of M\"uller et al. (in preparation), based on SBF}). The lines indicate the distance uncertainties. 
}
\label{tullyPlot}
\end{figure}

{
Let us now proceed to the question {of} the planes-of-satellites around Cen\,A. In \citet{2016A&A...595A.119M}, we have studied the significance of the two planes {proposed by \citet{2015ApJ...802L..25T}, using} our newly detected dwarf candidates \citep{2017A&A...597A...7M} and the newly established members by \citet{2016ApJ...823...19C}. We found that two distinct {planes}
are no longer supported by the new data, with dwarfs filling up the gap between the two planes. The distribution is consistent with a single normal distribution within $2\sigma$. Recently, \citet{2019ApJ...872...80C} revised the distances to Cen\,A dwarf galaxies 
{from their 2016 ground-based study using}
superior HST imaging. Additionally, they found two new dwarf members. These satellites, together with the dwarfs presented here, doubles the data available to study the 3D distribution {(with respect to the sample used by \citealt{2015ApJ...802L..25T})}.
{One more dwarf galaxy belonging to Cen\,A was discovered in very deep DECam images taken from the NOAO archive (M\"uller et al, in preparation). Its distance of $3.90^{+1.43}_{-0.86}$\,Mpc has been estimated by employing surface brightness fluctuations (SBF, \citealt{1988AJ.....96..807T}). At first, this galaxy was expected to be a member of M\,83 due to its close on-sky projection of $\sim$1\,deg (see also Fig\,\ref{field}), but its distance rather suggests that it is a member of Cen\,A (even though the uncertainties in SBF are quite large).}
In Fig.\,\ref{tullyPlot}, we show the edge-on view of the Centaurus planes, as it was defined in \citet{2015ApJ...802L..25T}. We indicate the new data presented here, from \citet{2019ApJ...872...80C}, {as well as the new dwarf measured with SBF (M\"uller et al, in preparation)}.
It is evident that the overall structure is flattened. 
If we assume that the two planes hypothesis is correct, then the addition of more galaxies should increase the signal-to-noise, which is not apparent here. Rather, the {new} dwarfs seem to be distributed between the two proposed planes, as their projected on-sky position already suggested. 
{In particular,} the dwarfs can only reside along the indicated lines, {given by} the {distance} uncertainty along our line-of-sight.\ This means that there is no confusion between being a member of the flattened structure or being outside of it. Only one confirmed dwarf galaxy, PGC051659, resides outside {this structure}. {However,} several dwarf galaxy candidates in the outskirts {of Cen\,A} still lack distance measurements \citep{2017A&A...597A...7M,2018ApJ...867L..15T}.

{Taking all 45 galaxies with accurately estimated distances within 1\,Mpc of Cen\,A into account {and excluding the dwarf with the
SBF distance estimate}, we measured a minor-axis rms height of 160\,kpc and a minor-to-major axis ratio of $c/a=0.50$ using the tensor of inertia method (ToI, e.g., \citealt{2015ApJ...815...19P}). The intermediate-to-major axis ratio is $b/a = 0.71$. Removing the outlying galaxy PGC051659, the height was estimated to be 133\,kpc with axis ratios of $c/a=0.41$ (i.e., a semi major axis of 327\,kpc) and $b/a = 0.69$. The normal  of this sample (i.e., the minor axis) is given as $n_{\rm CASP}=[-0.250, -0.270,  +0.930]$\ in supergalactic coordinates, and it is perpendicular to the line-of-sight to within $4^\circ$. The 3D $rms$ radius for all 45 galaxies is 404\,kpc, which agrees well with the virial radius of Cen\,A. The long axis of the satellite distribution is aligned to within $8^\circ$\ with the line-of-sight to Cen\,A. This begs the question of whether the found flattening could arise from the direction of the measurement uncertainties. To test this, we assumed a typical 5\% uncertainty along the line-of-sight for every galaxy. Then, the true major $a_{\rm intr}$\ axis length is given by $a_{\rm intr}^2 =  a_{obs}^2 - a_{\rm err}^2$, where $a_{\rm obs}$ is the observed major axis length and $a_{\rm err}$ is the length arising from measurement uncertainties. With the 5\% uncertainty at the distance of 3.7\,Mpc to Cen\,A, this gives an intrinsic major axis length of $a_{\rm intr}=260$\,kpc and a $c/a_{\rm intr} = 0.62$. This indicates that the measurement uncertainties are indeed an important contribution, but they do not dominate the result. An additional caveat we face in this analysis is the obstruction of the Milky Way disk. In our dwarf galaxy survey \citep{2015A&A...583A..79M,2017A&A...597A...7M}, we limited our survey area at $\delta\approx-45$\,deg due to the increasing contamination of Galactic cirrus and the number of stars; however, several of the plane members \citep{2015ApJ...802L..25T} are outside of this area at roughly $\delta\approx-50$\,deg. The survey footprint thus limits the extent along the plane direction and thus biases toward less extreme flattening {because it could potentially cut-off a part of the plane. In other words, the structure could be more extended than what we observe here.} The SCABS survey footprint \citep{2016arXiv160807285T,2017MNRAS.469.3444T,2018ApJ...867L..15T} should overlap with this region and shed new light on the distribution of satellites there. {In addition, the new dwarf galaxy, confirmed with SBF (M\"uller et al, in preparation), lies in a region of the sky where further dwarf candidates await follow-up measurements. In principle, some of them could lie in the plane-of-satellites as this recently discovered dwarf galaxy. Therefore, a final conclusion about the overall flattening cannot be drawn as long as we still lack distance information for many of the dwarf candidates.}
}

}

\section{Summary and conclusion}
The $\Lambda$CDM small-scale crisis in the Local Group motivates the study of satellite systems in other groups of galaxies. One of our closest neighbors, Cen\,A, provides a unique opportunity for such 
studies because it is close enough to accurately determine the distance of its satellite members. In \citet{2015A&A...583A..79M,2017A&A...597A...7M}, we found 58 possible dwarf galaxy candidates in a 500 square degree field around Cen\,A. Here, we present deep $VI$ band follow-up {observations} with FORS2 at the VLT for 15 of these candidates and confirm nine to be members of Cen\,A thanks to their resolved stellar populations. For the confirmed members, we employed a Bayesian approach to accurately measure the tip of the red giant branch and derive distances with a typical accuracy of 5\%. We also fit theoretical isochrones to the color-magnitude diagrams to estimate the mean metallicity of each dwarf. 
As expected, the confirmed satellites follow the scaling relations (luminosity-surface brightness, luminosity-effective radius, and luminosity-metallicity) defined by the Local Group dwarfs. {Our photometric precision is not sufficient to detect metallicity spread in individual galaxies.}
{
This could either mean that there is simply no metallicity spread, which is an indication of a complex star formation history, or that the uncertainties are too large to detect them.} {We have further searched for evidence of tidal disruptions in our dwarf galaxies, but again we find no significant signs. 
For some galaxies, there is a hint of an asymptotic giant branch, indicating an intermediate age stellar population. Future observations  using near-infrared data could confirm or refute the presence of intermediate age AGB in these galaxies {(see e.g.,\ \citealt{2006A&A...448..983R,2011A&A...530A..58C})}. We have also looked for the presence of a young star population but find no convincing candidates.

Out of the 15 targets, six could not be resolved. Out of these, two resemble dwarf galaxies in morphology and could be either in the farther extent of the Centaurus halo (e.g., at 6\,Mpc) or be dwarf galaxies in the background. It could be possible to measure their distances with the surface brightness fluctuation method to get an idea where these objects reside. 
{One unresolved candidate was a patch of cirrus},  one a low-surface brightness background galaxy, and another one probably a ghost in the DECam images. The remaining unresolved candidate is an extended, diffuse object and is projected very close to a spiral galaxy at redshift $z=0.08$. It could  be some tidal ejecta of this background galaxy, or closer to us.}

{We now have a complete census of dwarf galaxies up to a projected distance of 200\,kpc from Cen\,A to a brightness limit of $V=-10$\,mag, which allows us to study the galaxy luminosity function {and to compare it to} the $\Lambda$CDM standard model of cosmology. 
For this purpose, we selected Cen\,A analogs within the IllustrisTNG cosmological simulation and counted the abundance of luminous dark matter subhalos and compared it to the measured luminosity function of Cen\,A. Interestingly, there seems to be a lack of bright, as well as an overabundance of faint, dwarfs. The former is still within the 90\% interval defined by the Cen\,A analogs and the latter could be attributed to unreliable modeling in the smaller satellites due to the resolution of the simulation. Overall, the observed luminosity function is within the 90\% interval and thus agrees with the cosmological $\Lambda$CDM simulation, when modeling baryonic feedback.}

{Finally, we looked at the spatial distribution of the {dwarf galaxies around Cen\,A}. We find two pairs of dwarf galaxies, which could be a physical pair based on their on-sky and 3D separation. 
In \citet{2016A&A...595A.119M}, we made an analysis based on the two proposed planes-of-satellites by \citet{2015ApJ...802L..25T} and have found that this interpretation seems to vanish by taking all our candidates, based on their on-sky positions and the newly discovered dwarfs by \citet{2016ApJ...823...19C}, into account.
Since then, the distances for the dwarfs in \citet{2016ApJ...823...19C} have been revised by \citet{2019ApJ...872...80C} using superior Hubble Space Telescope data. We studied the overall distribution of the previously known satellites together with the dwarf galaxies presented here and confirm this assessment. We find that overall, there is a flattening in the distribution of satellites { of $c/a=0.5$, with the major axis being oriented approximately along the line-of-sight.} The distance uncertainties are too small to explain this elongation {but certainly contribute to it}. 
Only one dwarf galaxy seems to reside outside of the flattened structure. {Removing this outlier, we measured a $rms$ height of 133\,kpc and a semi major axis length of 327\,kpc.}
{However, many dwarf galaxy candidates still lack accurate distance and velocity measurements, so it is too early to draw a final conclusion about the overall distribution of satellites in this galaxy group.} }

\label{summary}
\begin{acknowledgements} {We thank the referee for the constructive report, which helped to clarify and improve the manuscript.}
O.M. is grateful to the Swiss National Science Foundation for financial support. O.M. also thanks Peter Stetson for the fast support solving a DAOPHOT related problem, and Morgan Fouseneau for interesting discussions concerning the statistical subtraction of foreground stars. 
\end{acknowledgements}

\bibliographystyle{aa}\bibliography{aanda}

\begin{appendix}
\renewcommand\thefigure{A.\arabic{figure}}

\begin{figure*}[ht]
\includegraphics[width=6cm]{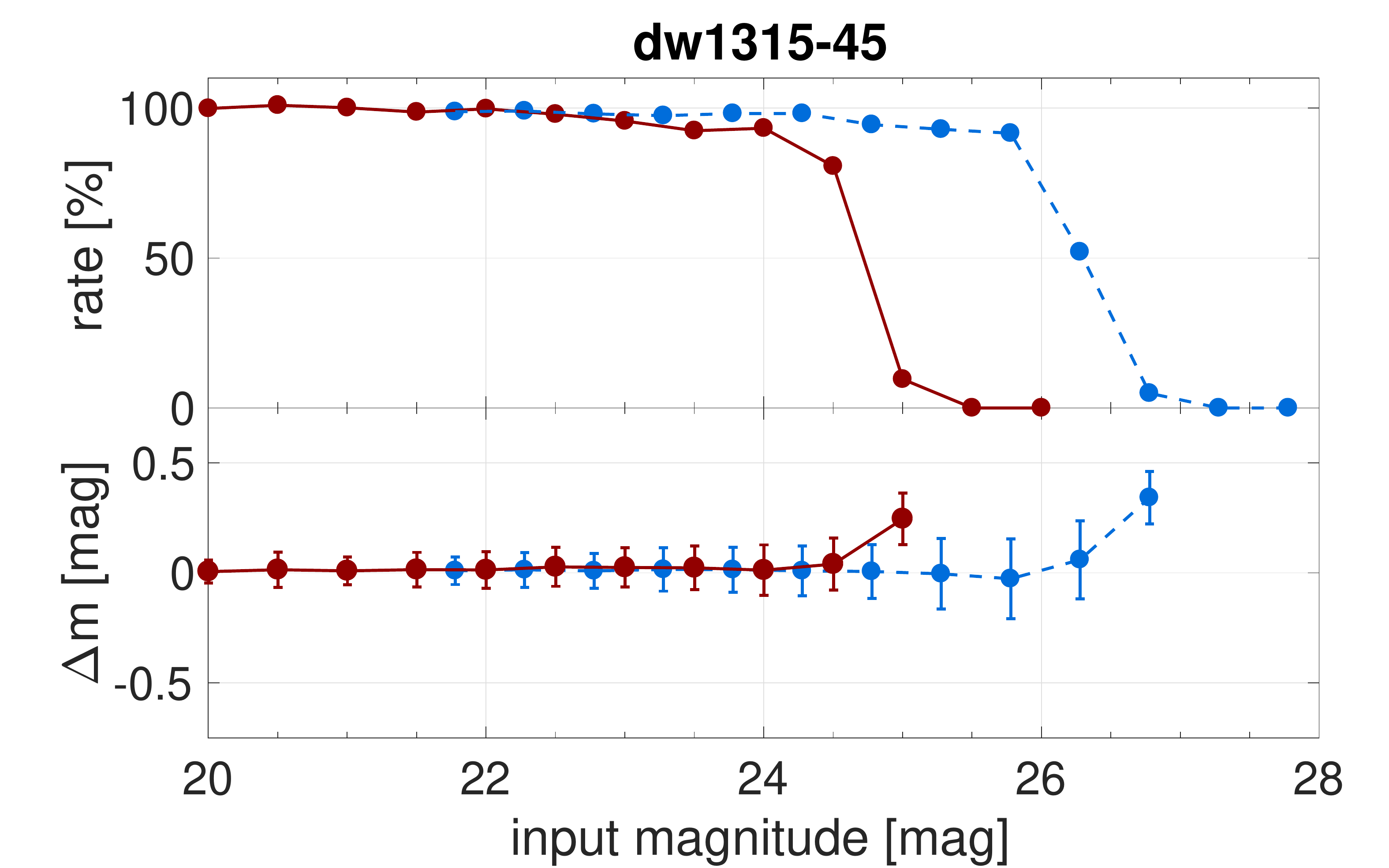}
\includegraphics[width=6cm]{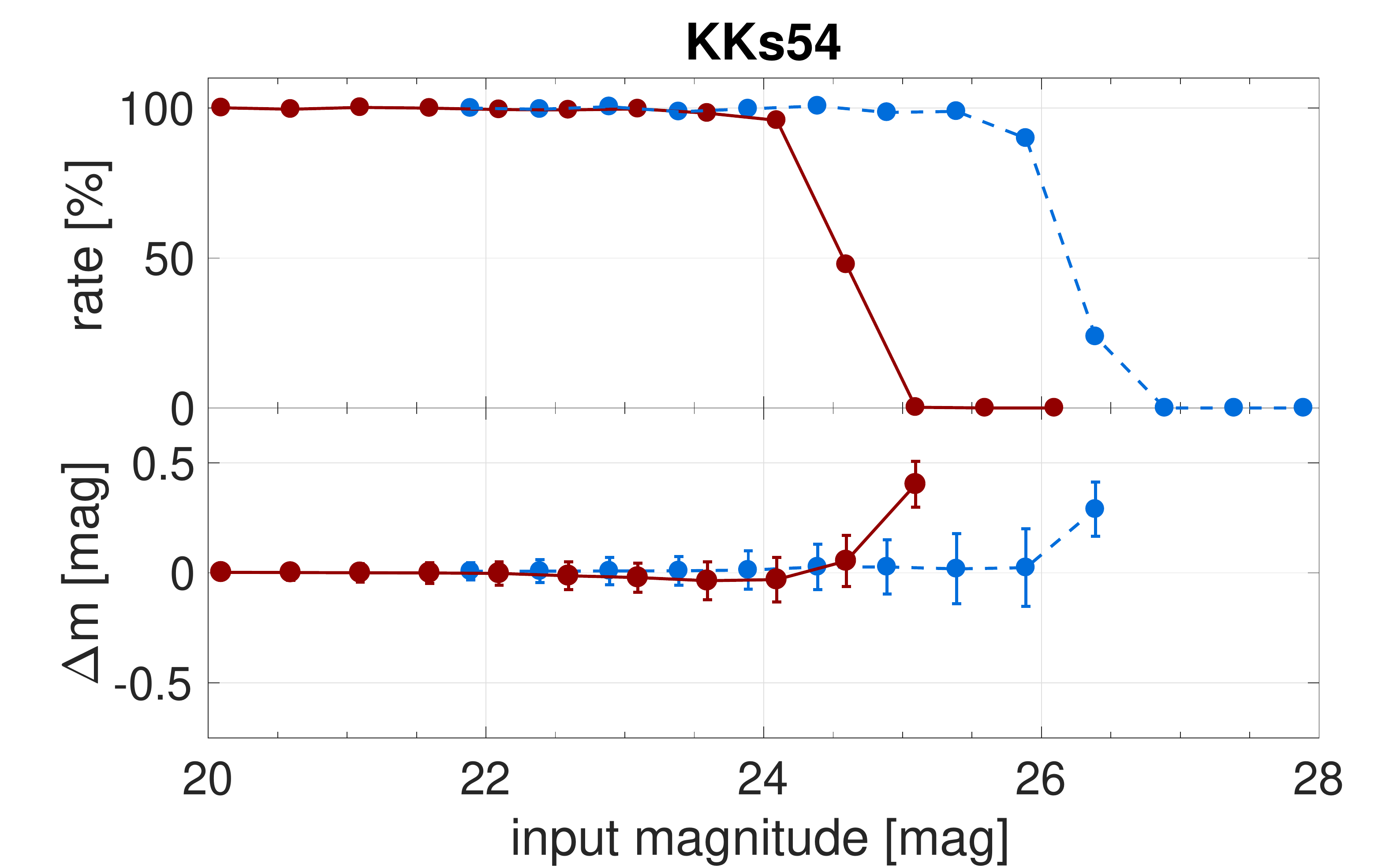}
\includegraphics[width=6cm]{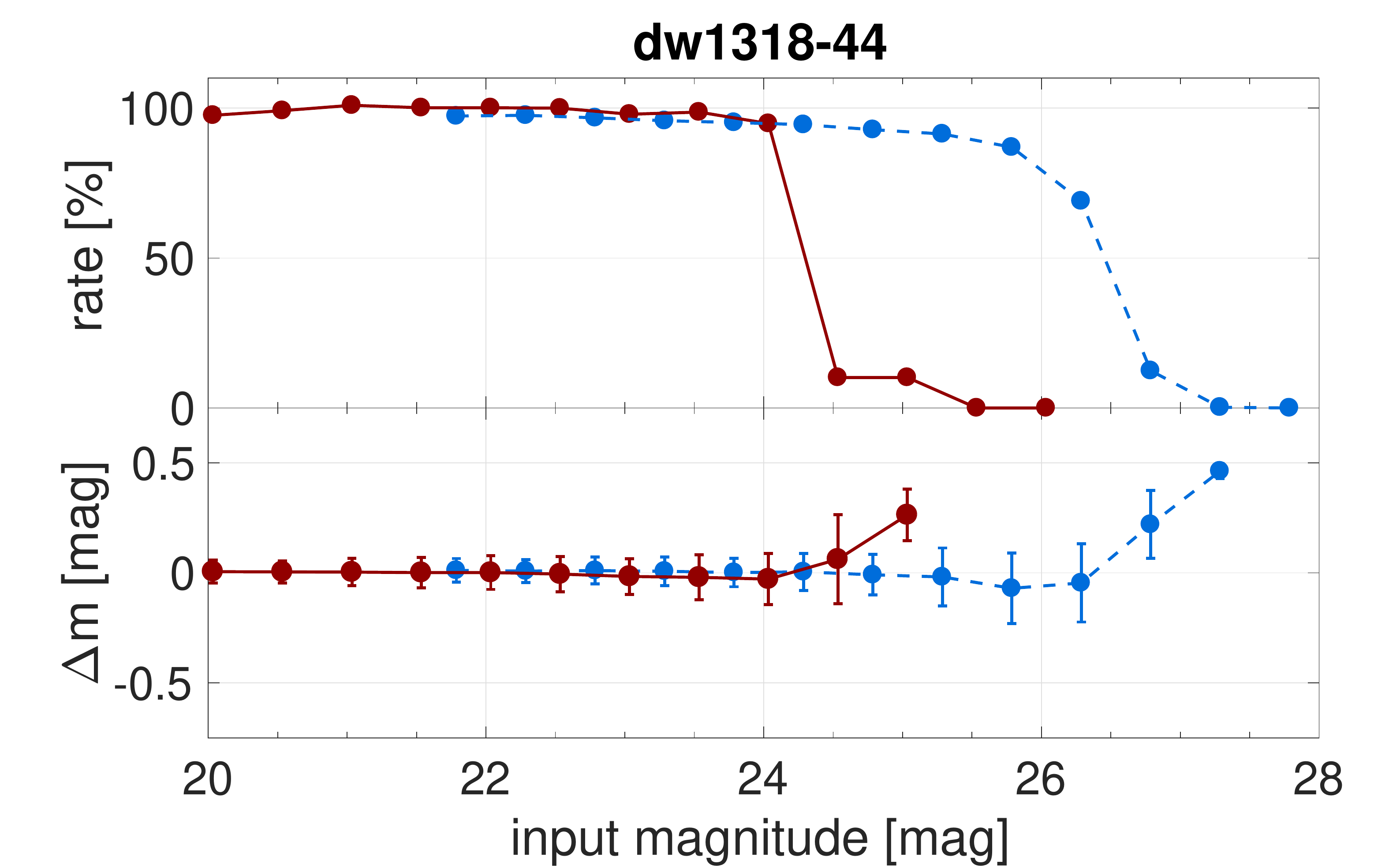}

\includegraphics[width=6cm]{complet_dw1322-39.pdf}
\includegraphics[width=6cm]{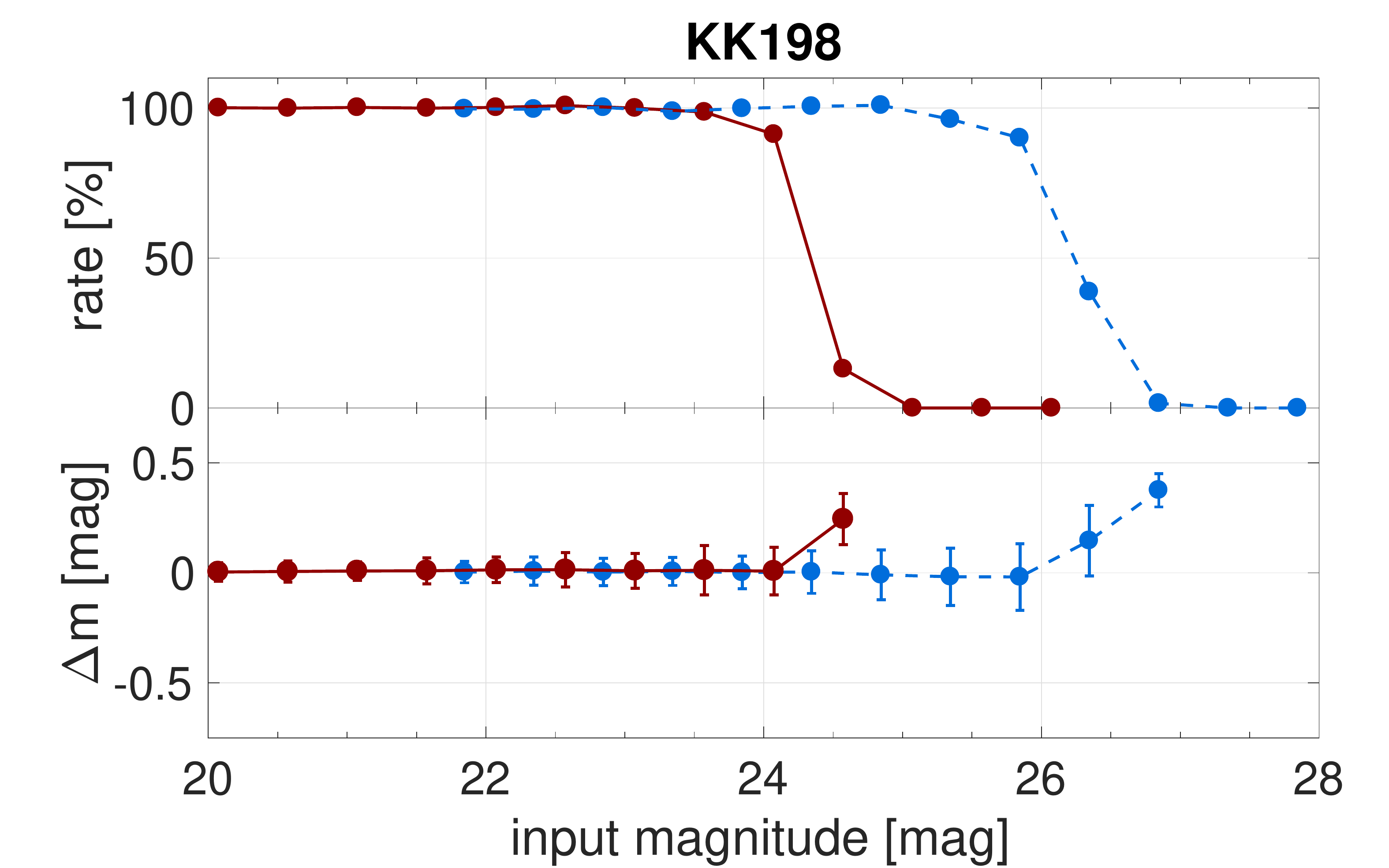}
\includegraphics[width=6cm]{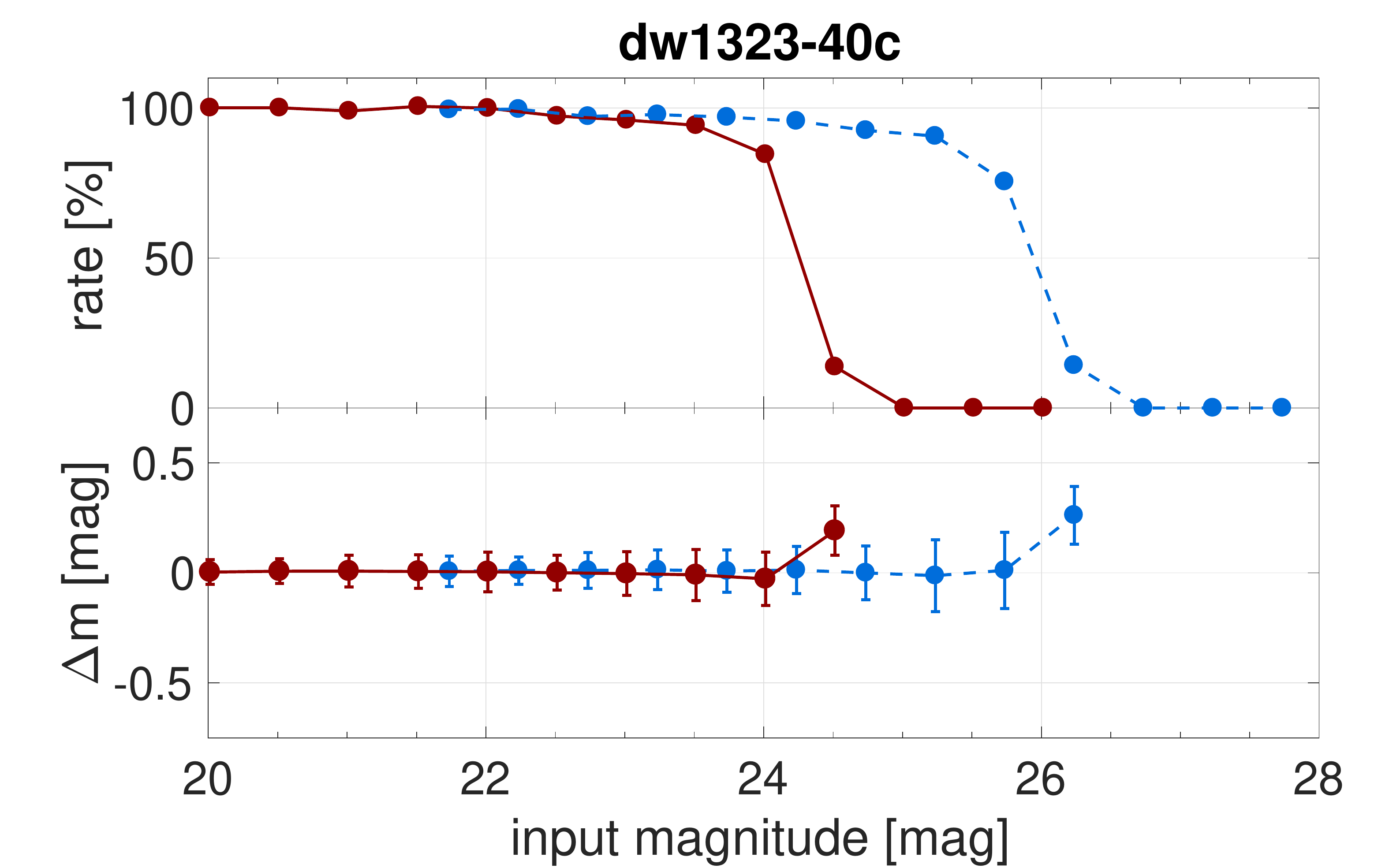}

\includegraphics[width=6cm]{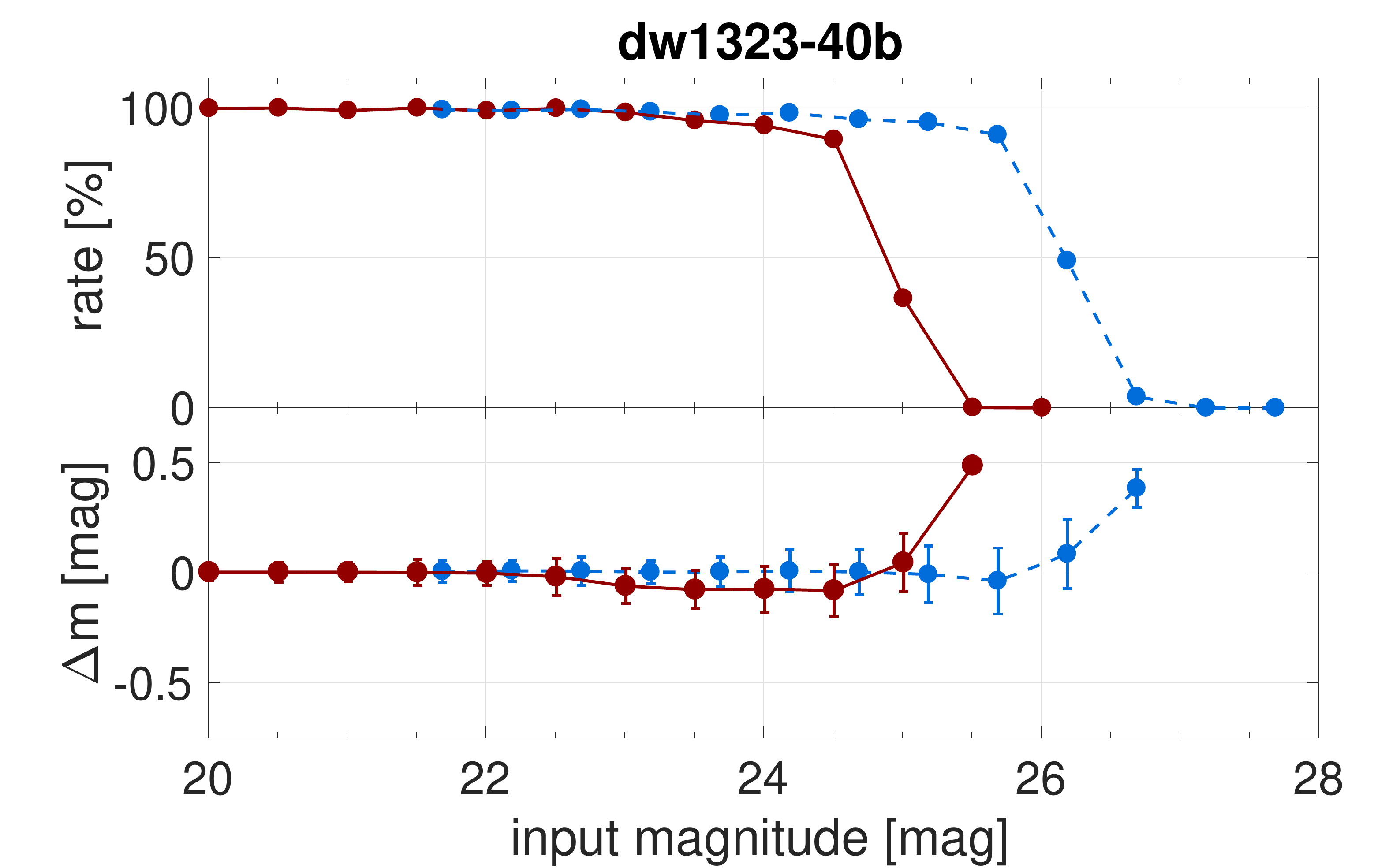}
\includegraphics[width=6cm]{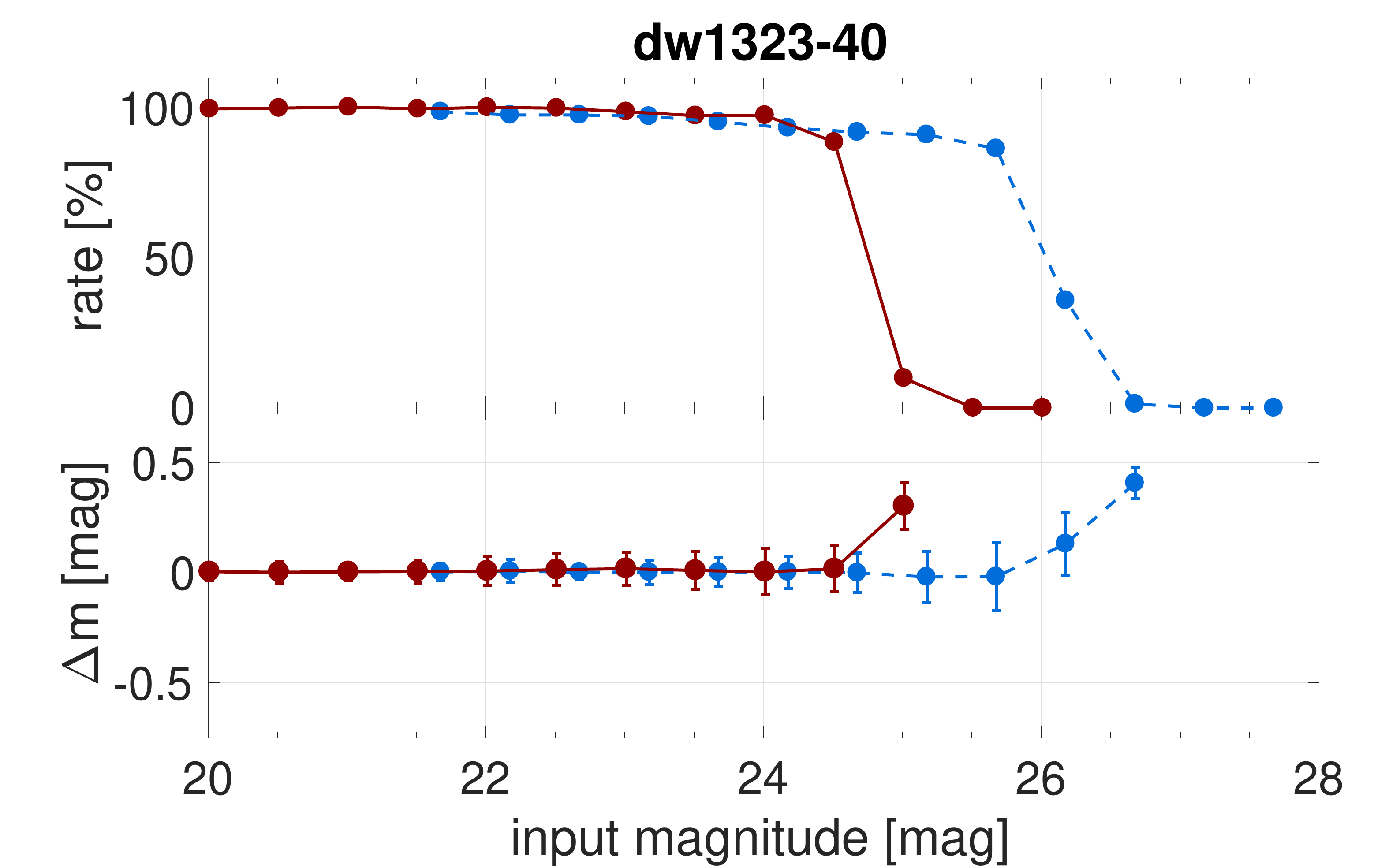}
\includegraphics[width=6cm]{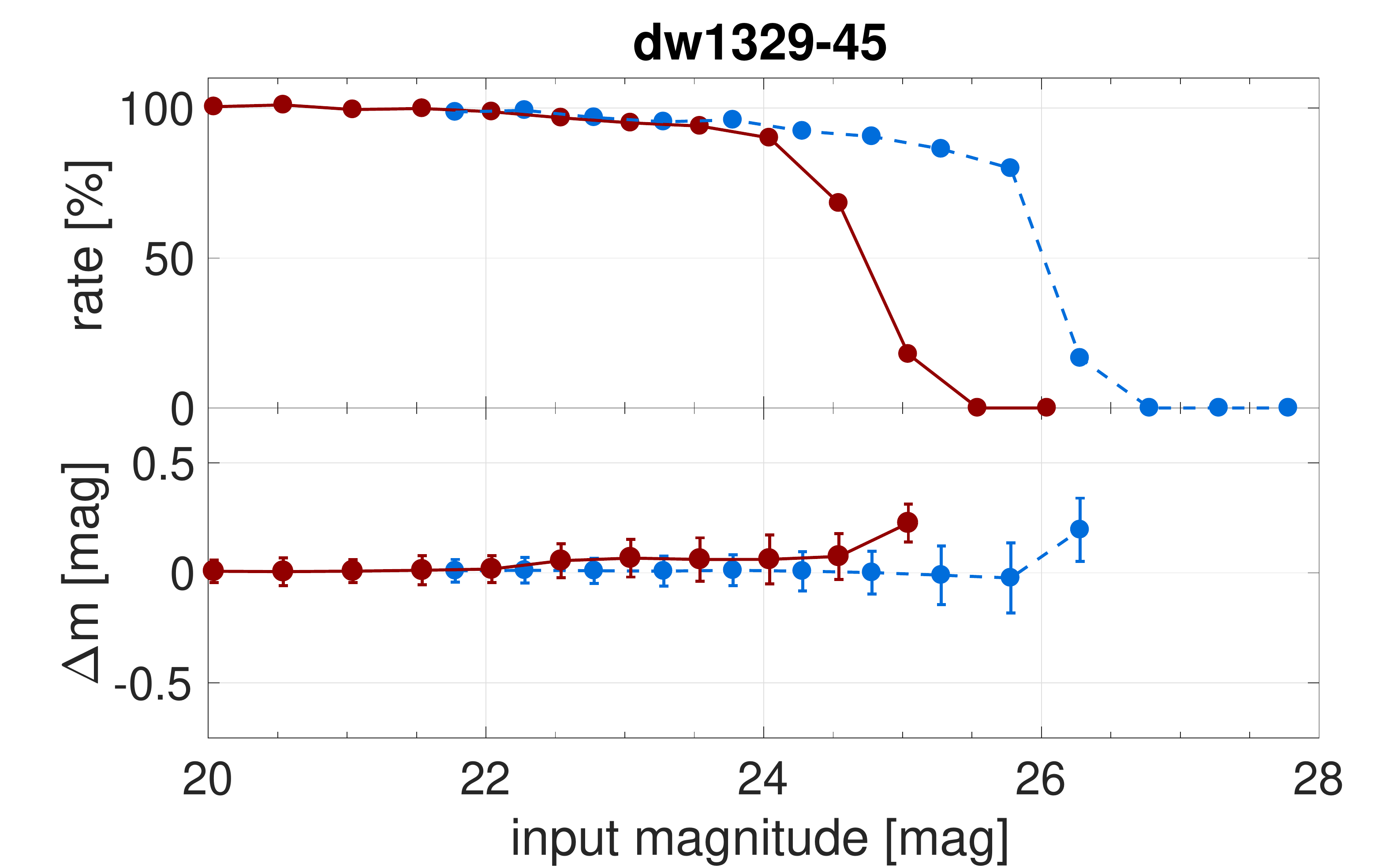}

\includegraphics[width=6cm]{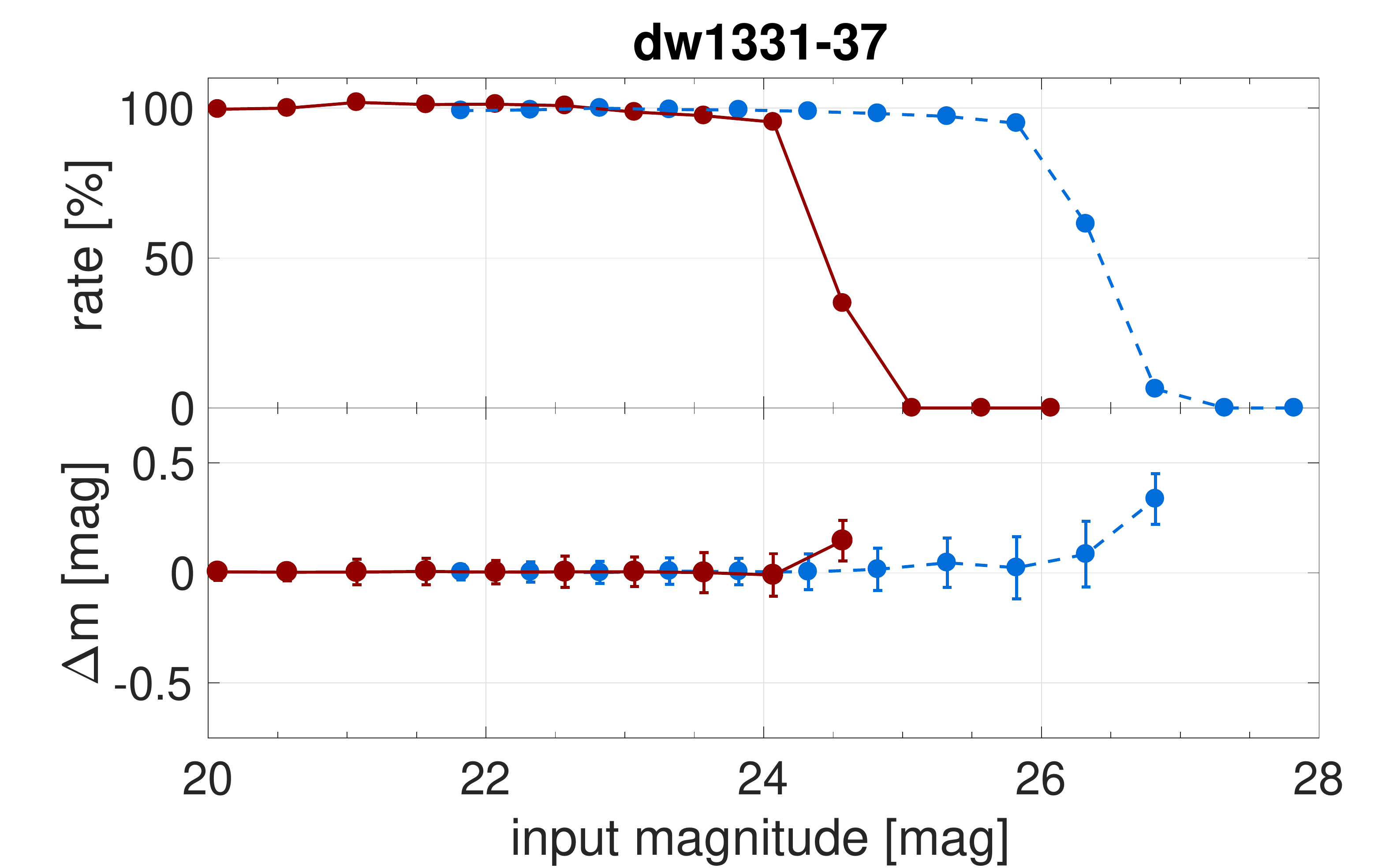}
\includegraphics[width=6cm]{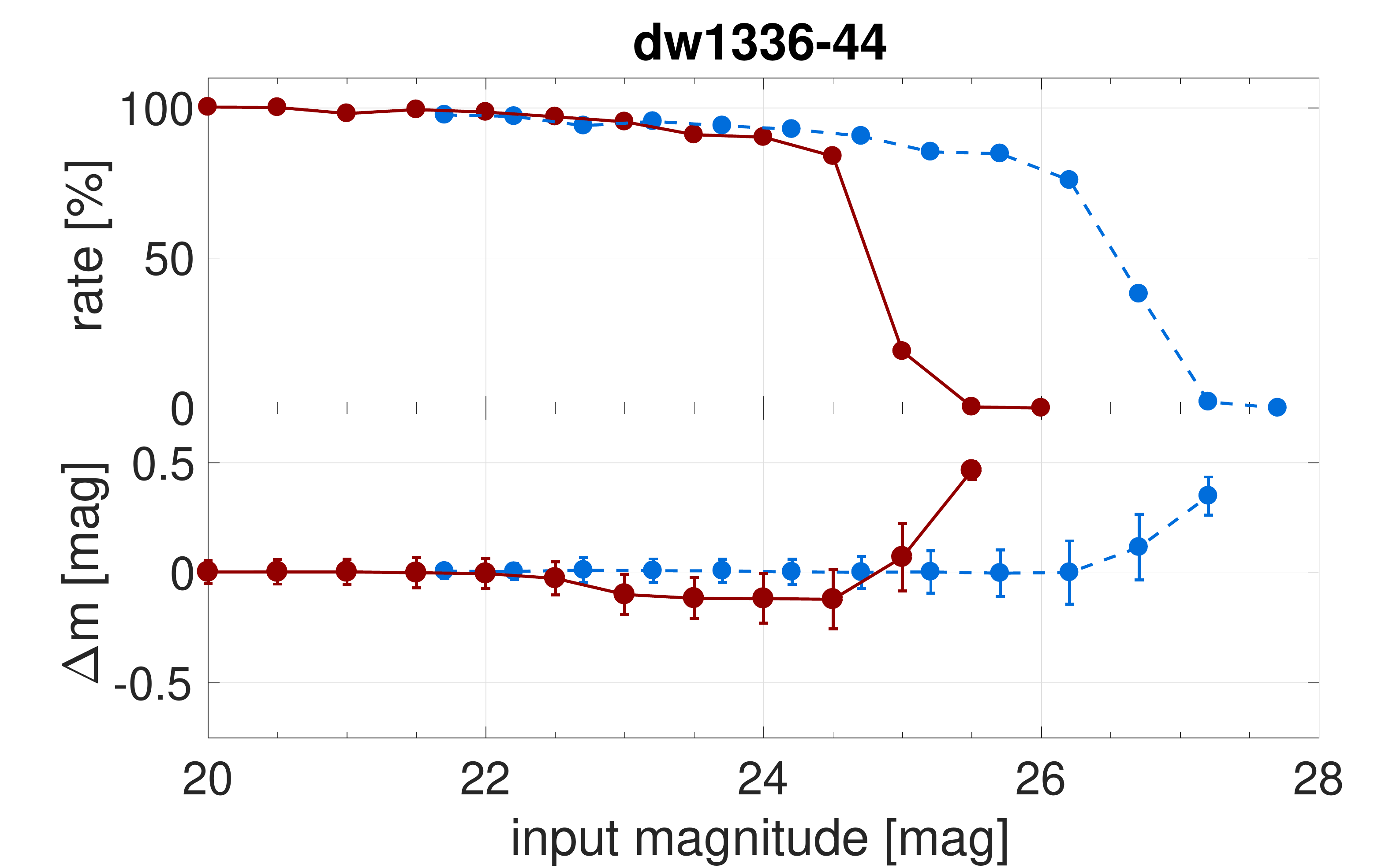}
\includegraphics[width=6cm]{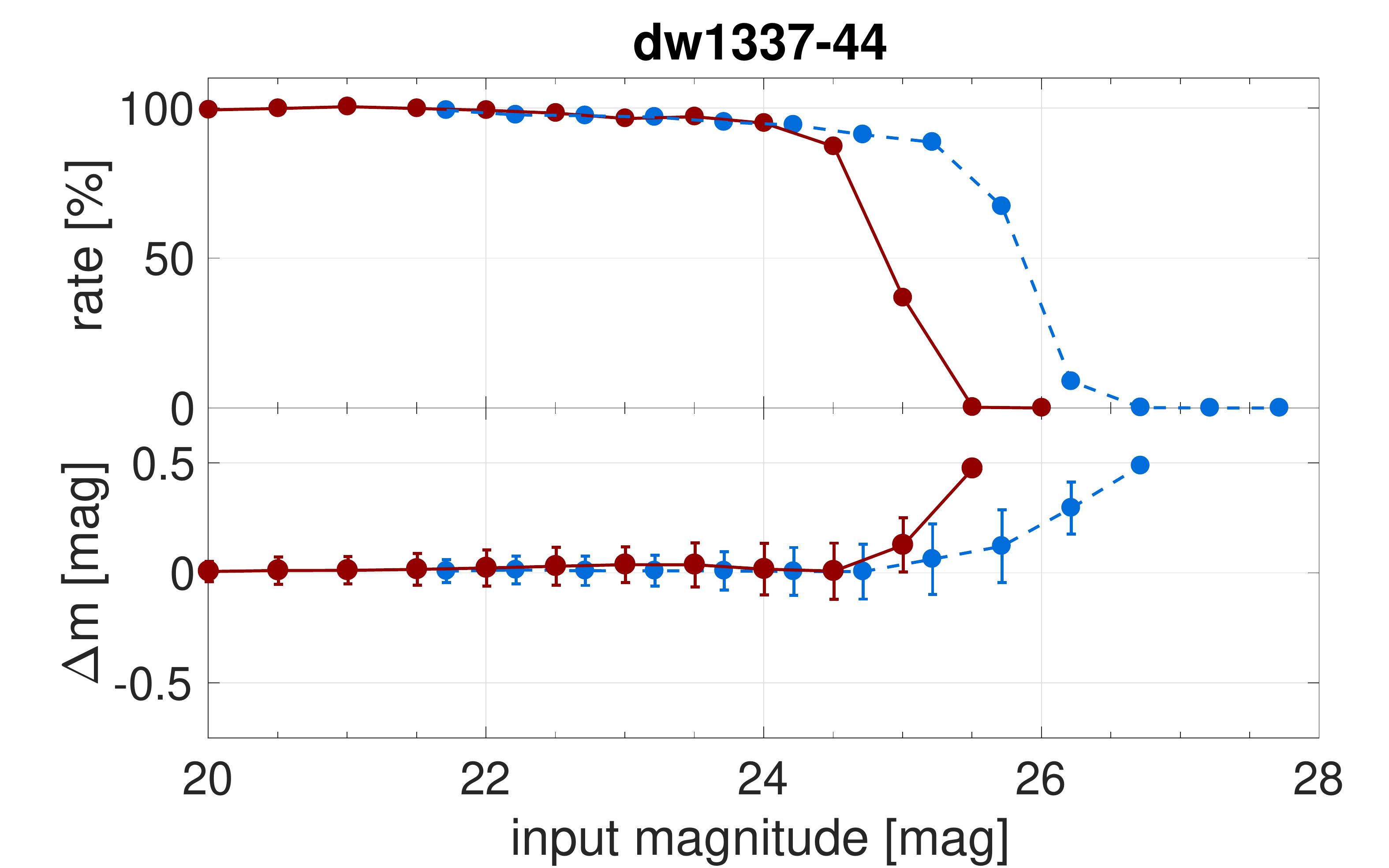}

\includegraphics[width=6cm]{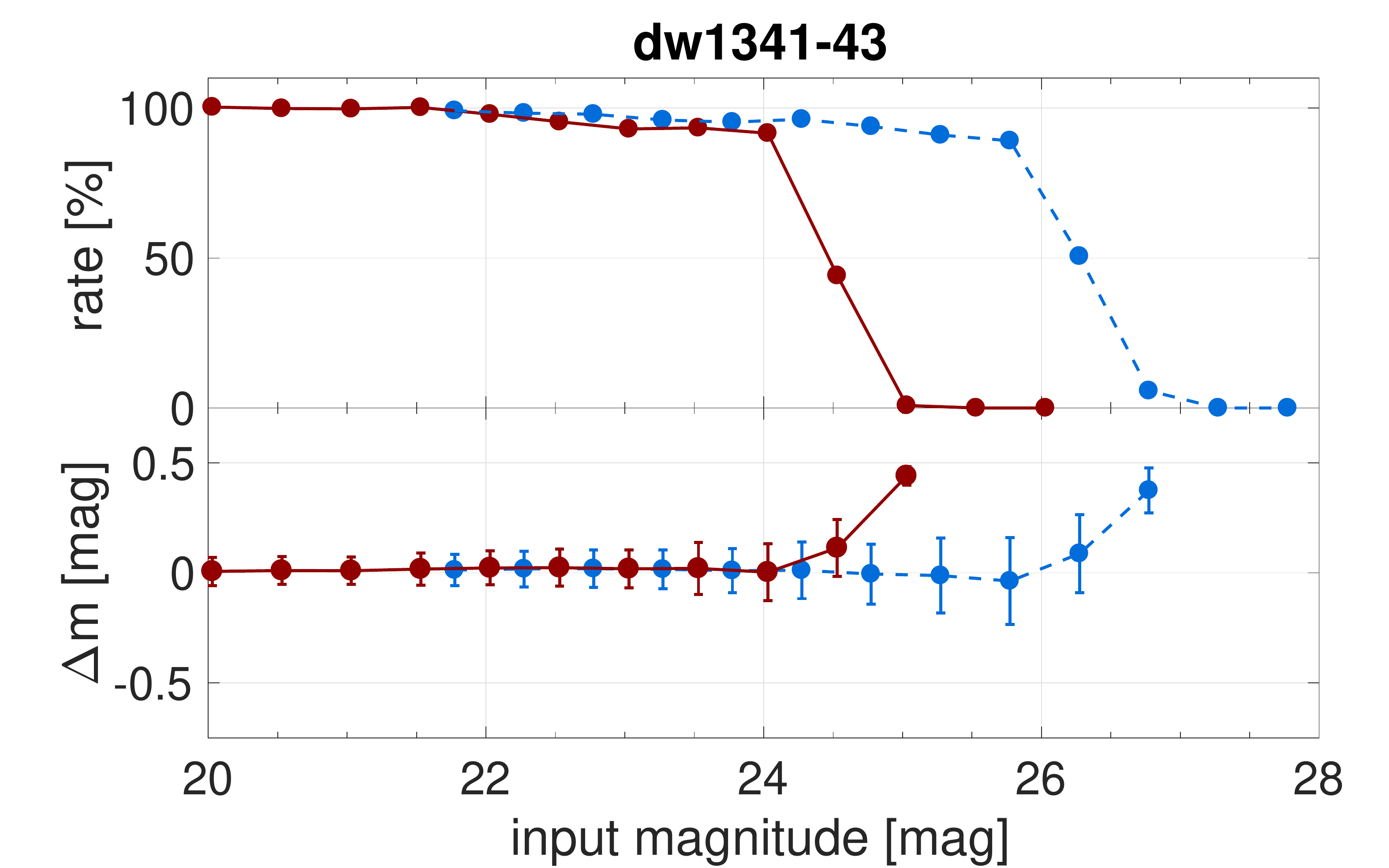}
\includegraphics[width=6cm]{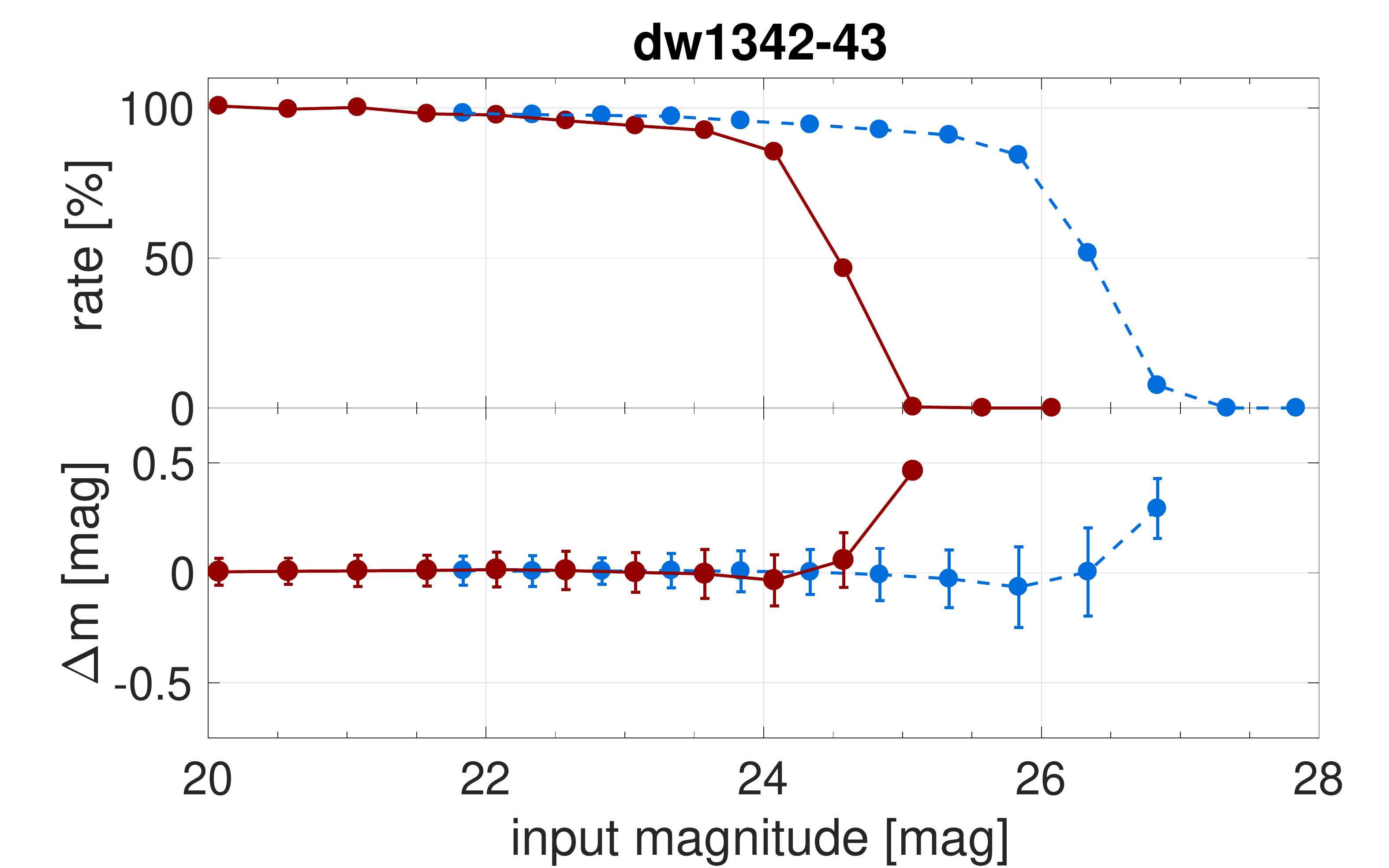}
\includegraphics[width=6cm]{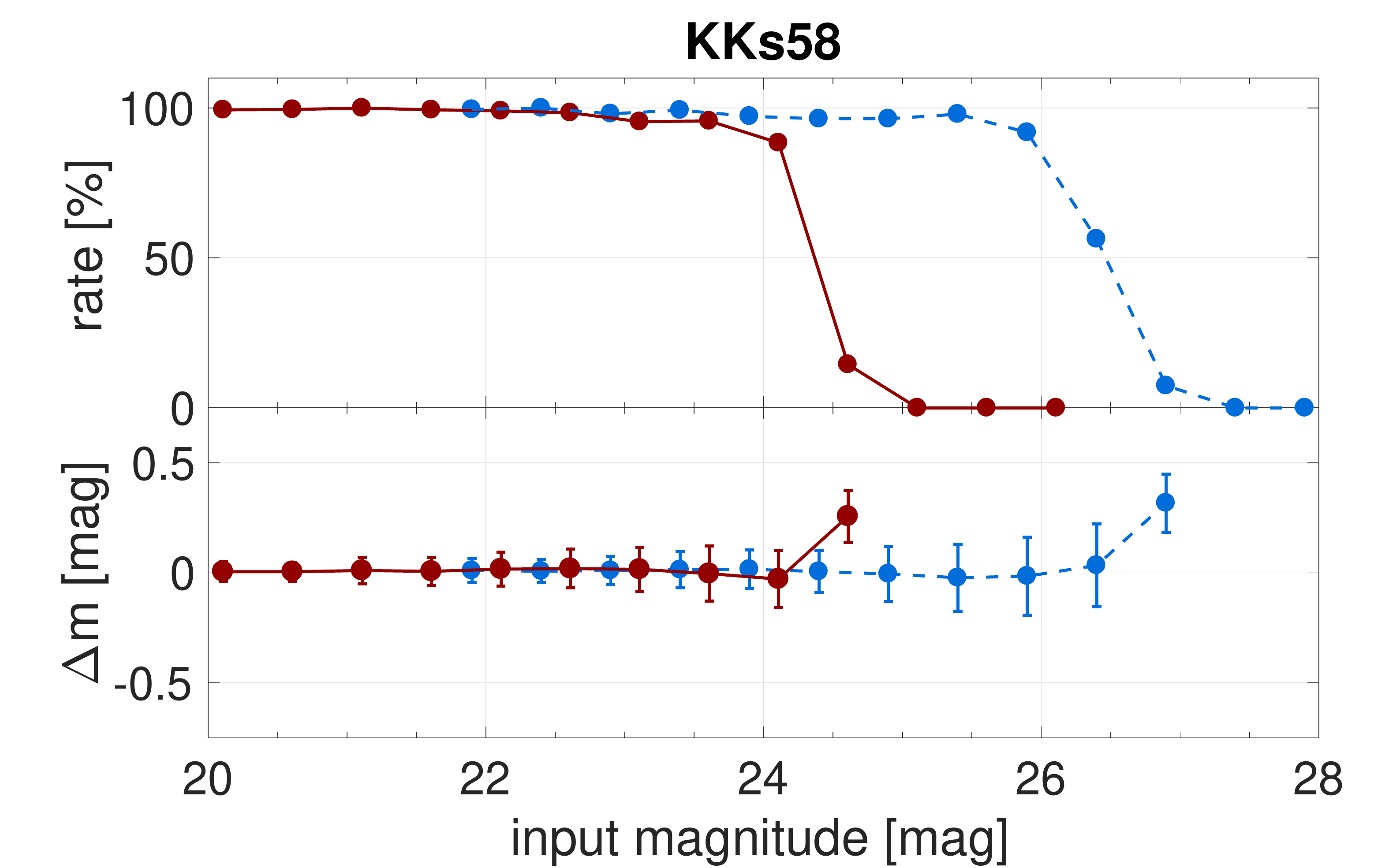}
\caption{{Results of our artificial star experiments for $I$ (red) and $V$ (dashed blue) bands.} Upper {panel}: The recovery 
fraction of the artificial stars induced into the science frames as a function of the input magnitude.
Bottom {panel}: The difference between the input
and measured  magnitude as a function of input magnitude.  
}
\label{app:artTests}
\end{figure*}

\begin{figure*}[ht]
\centering
\includegraphics[width=6cm]{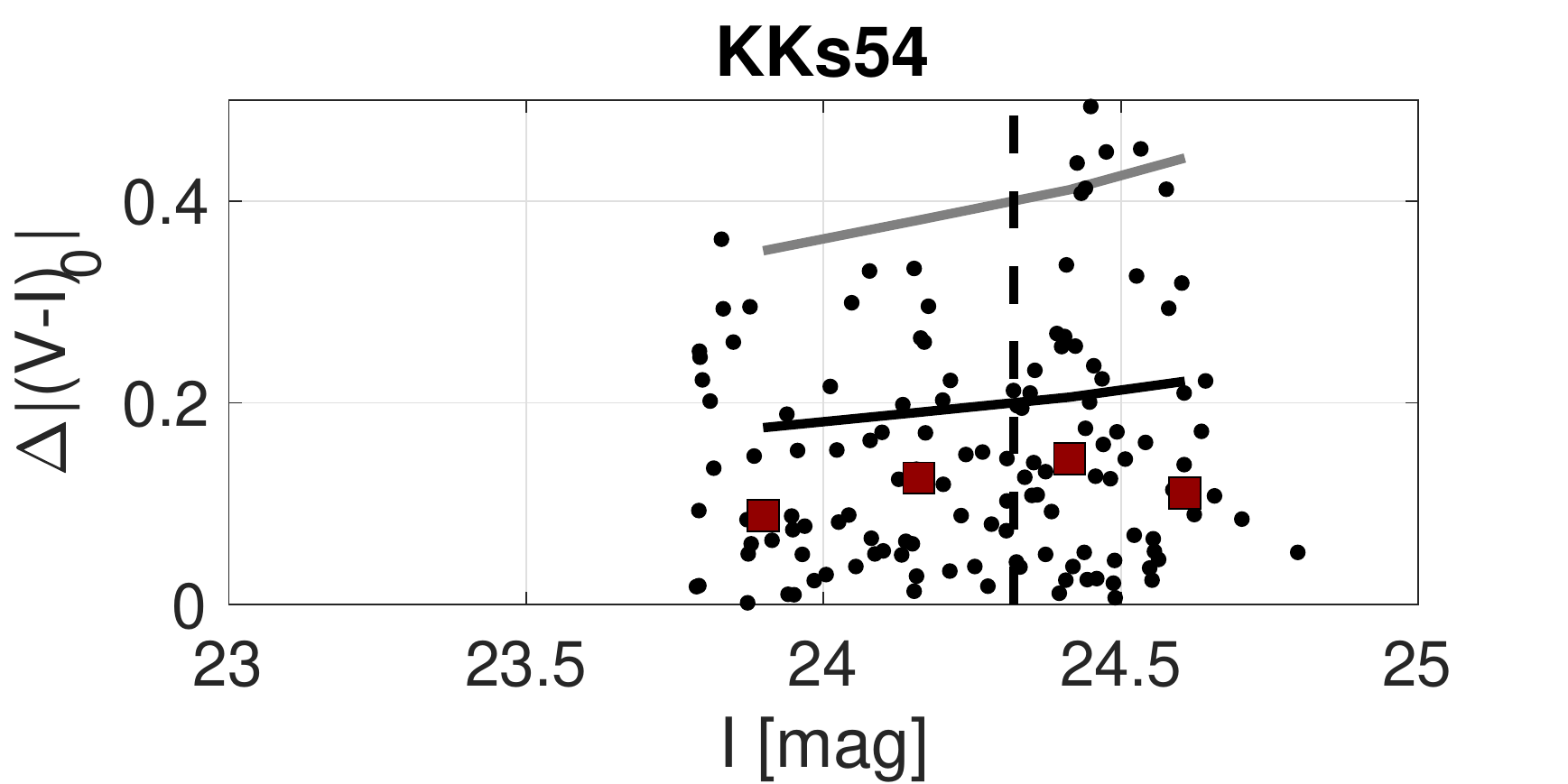}
\includegraphics[width=6cm]{metallicityspread_dw1322-39.pdf}
\includegraphics[width=6cm]{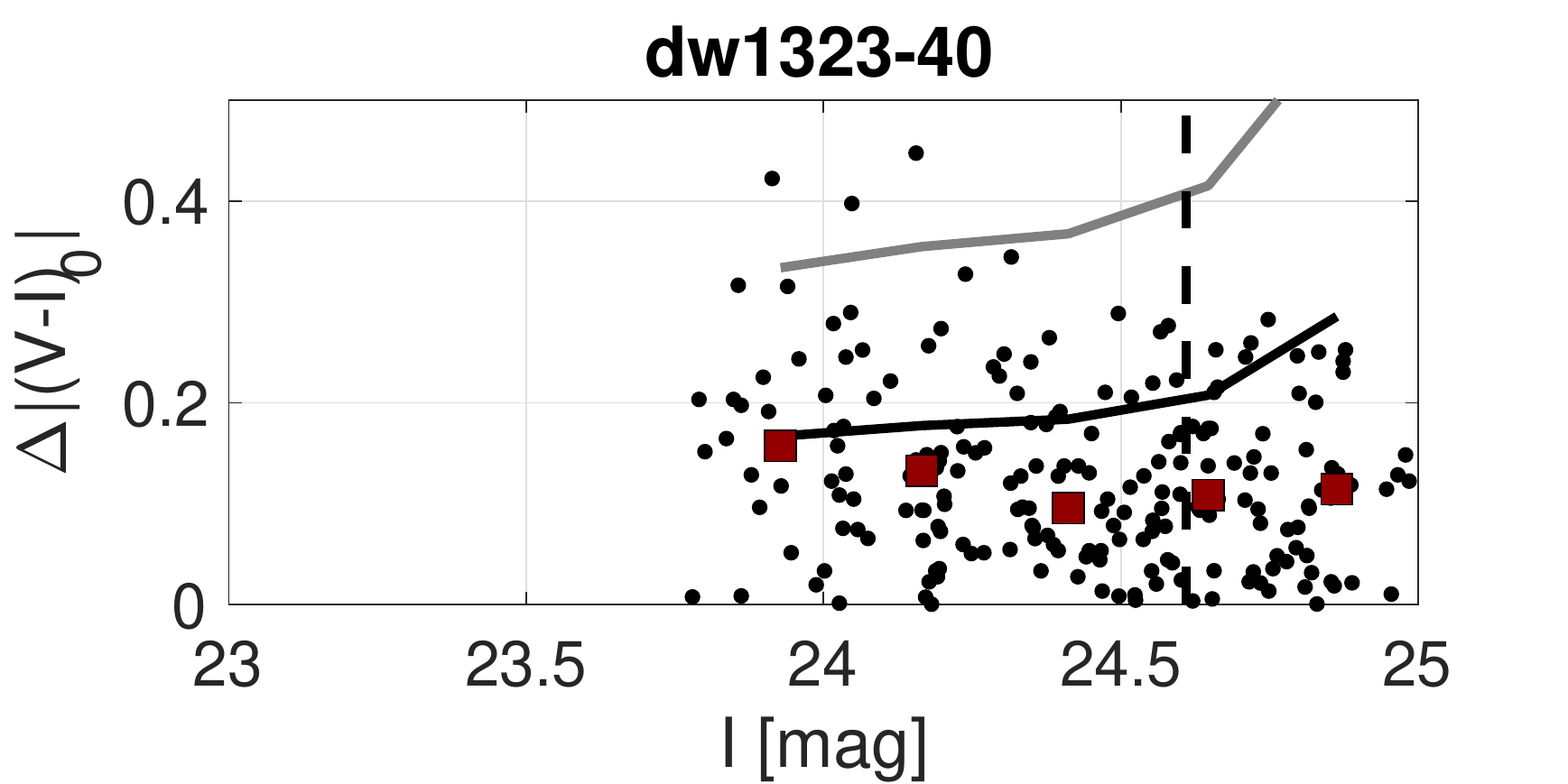}

\includegraphics[width=6cm]{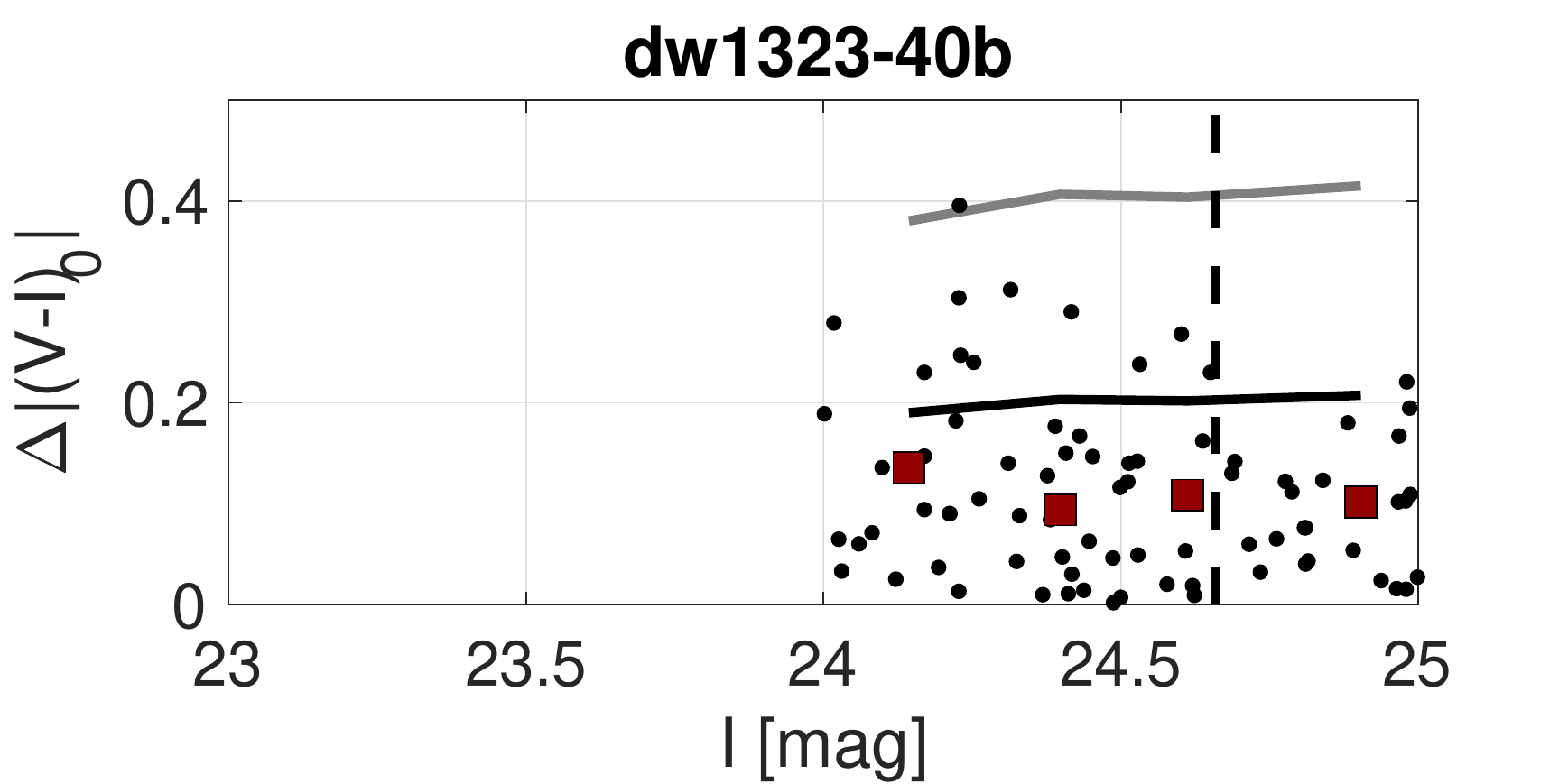}
\includegraphics[width=6cm]{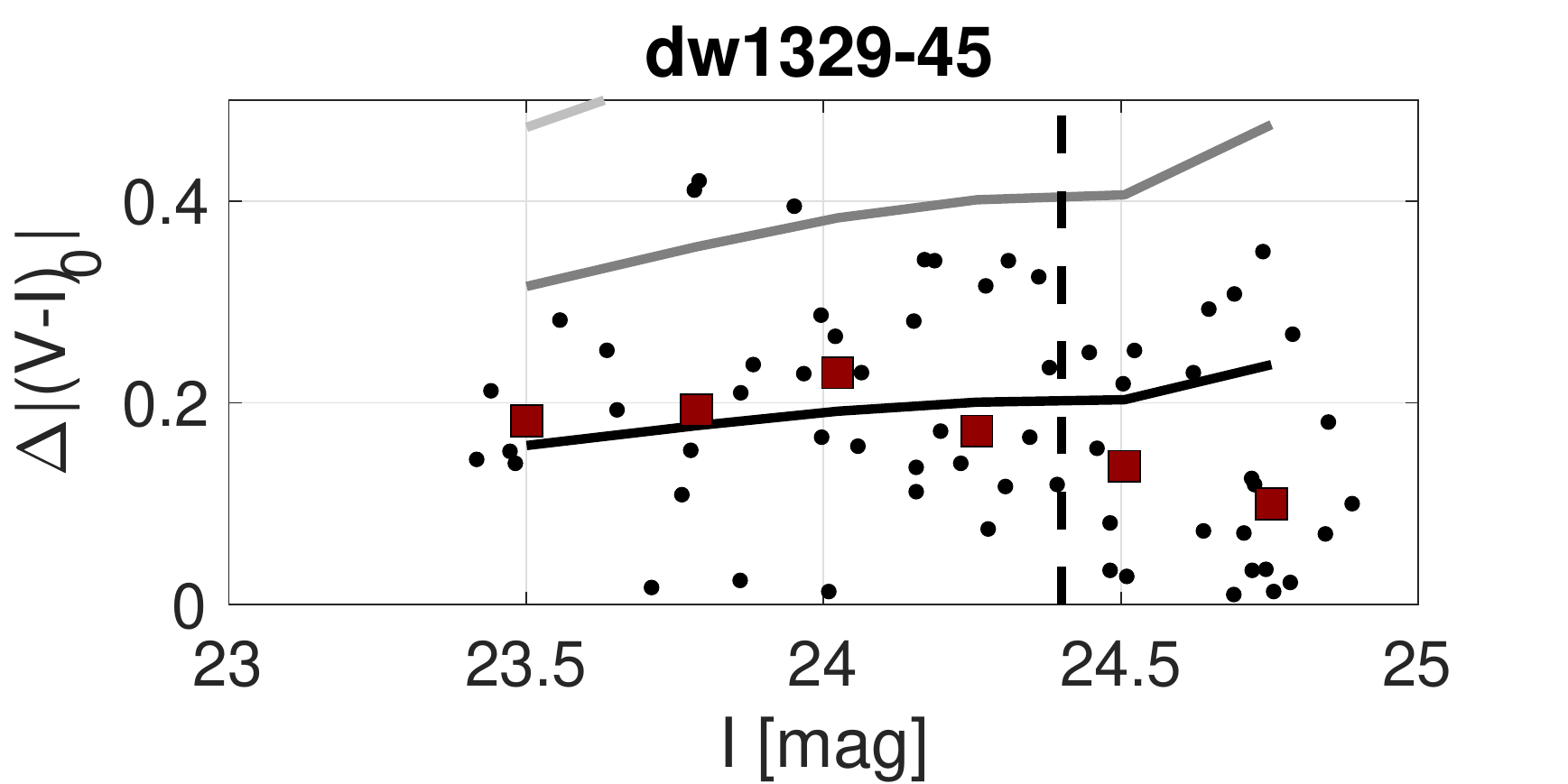}
\includegraphics[width=6cm]{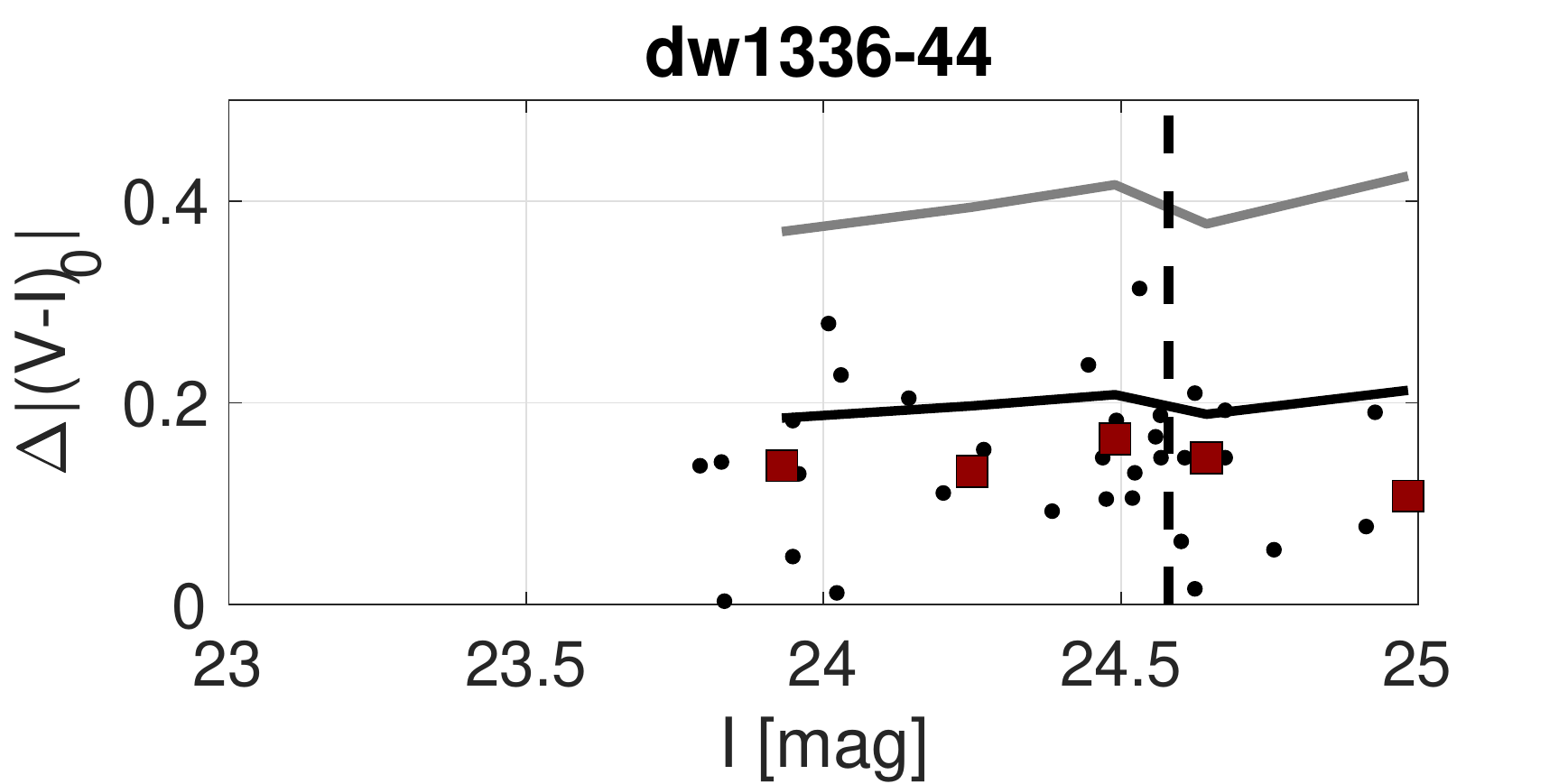}

\includegraphics[width=6cm]{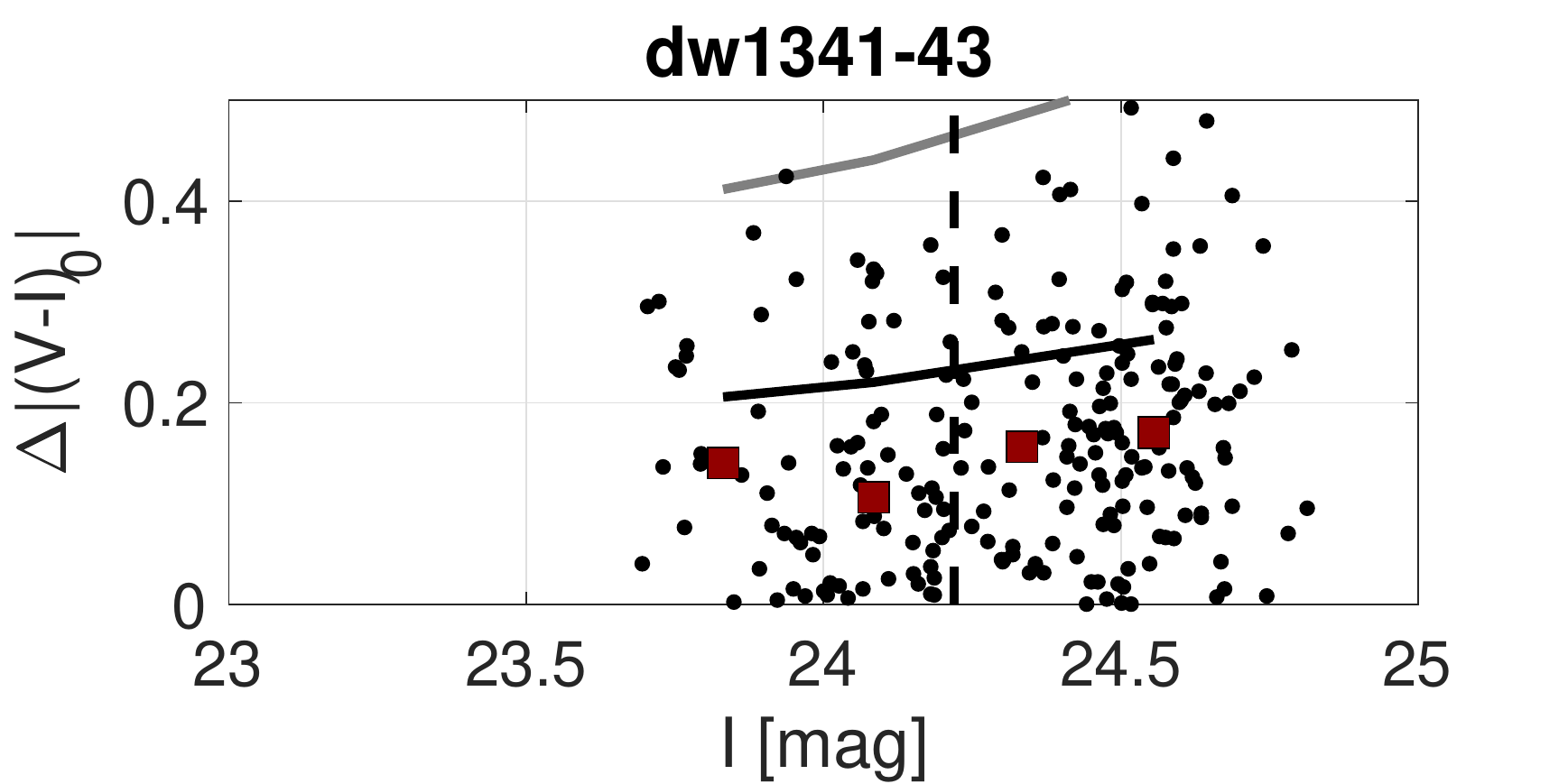}
\includegraphics[width=6cm]{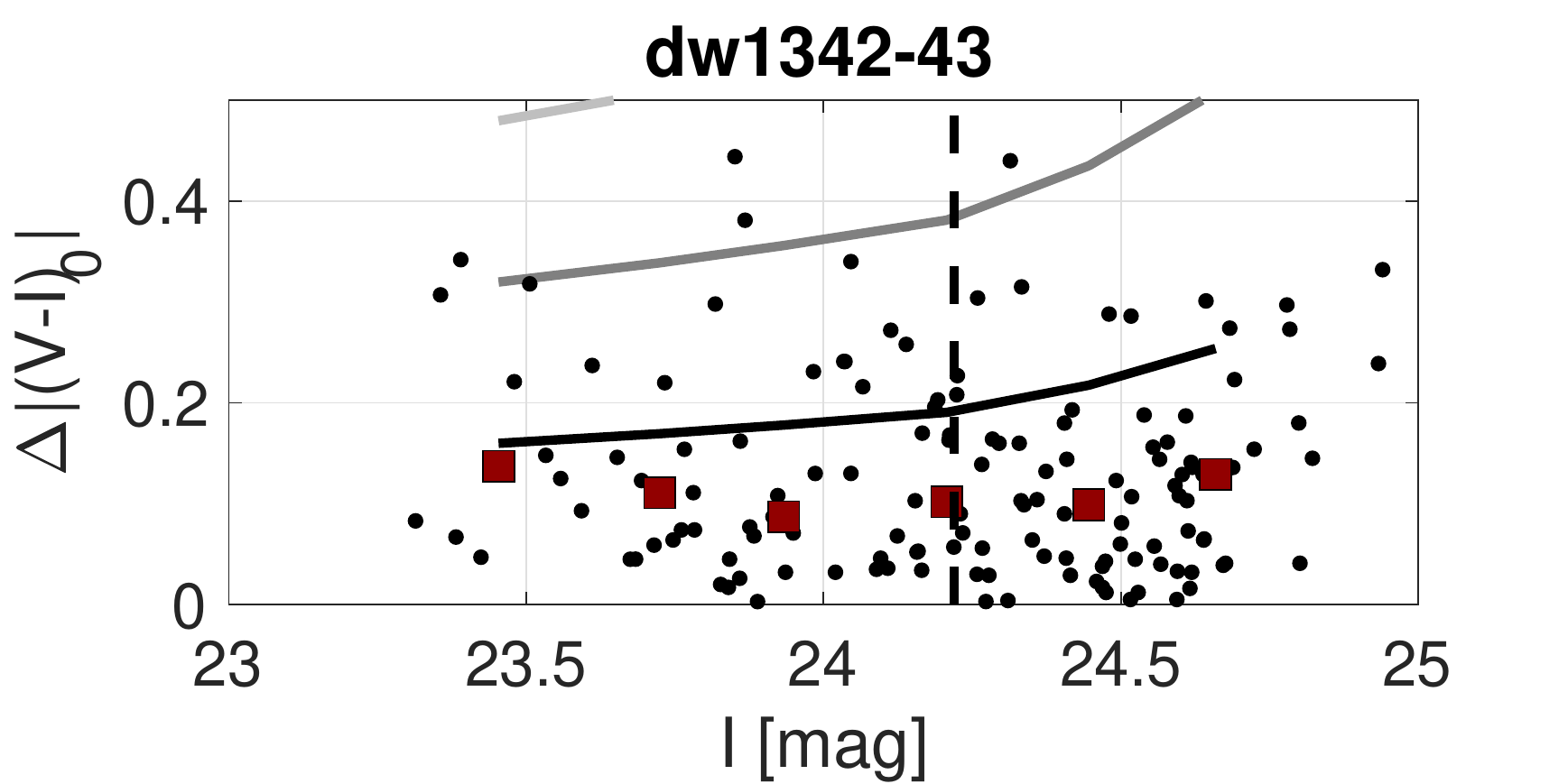}
\includegraphics[width=6cm]{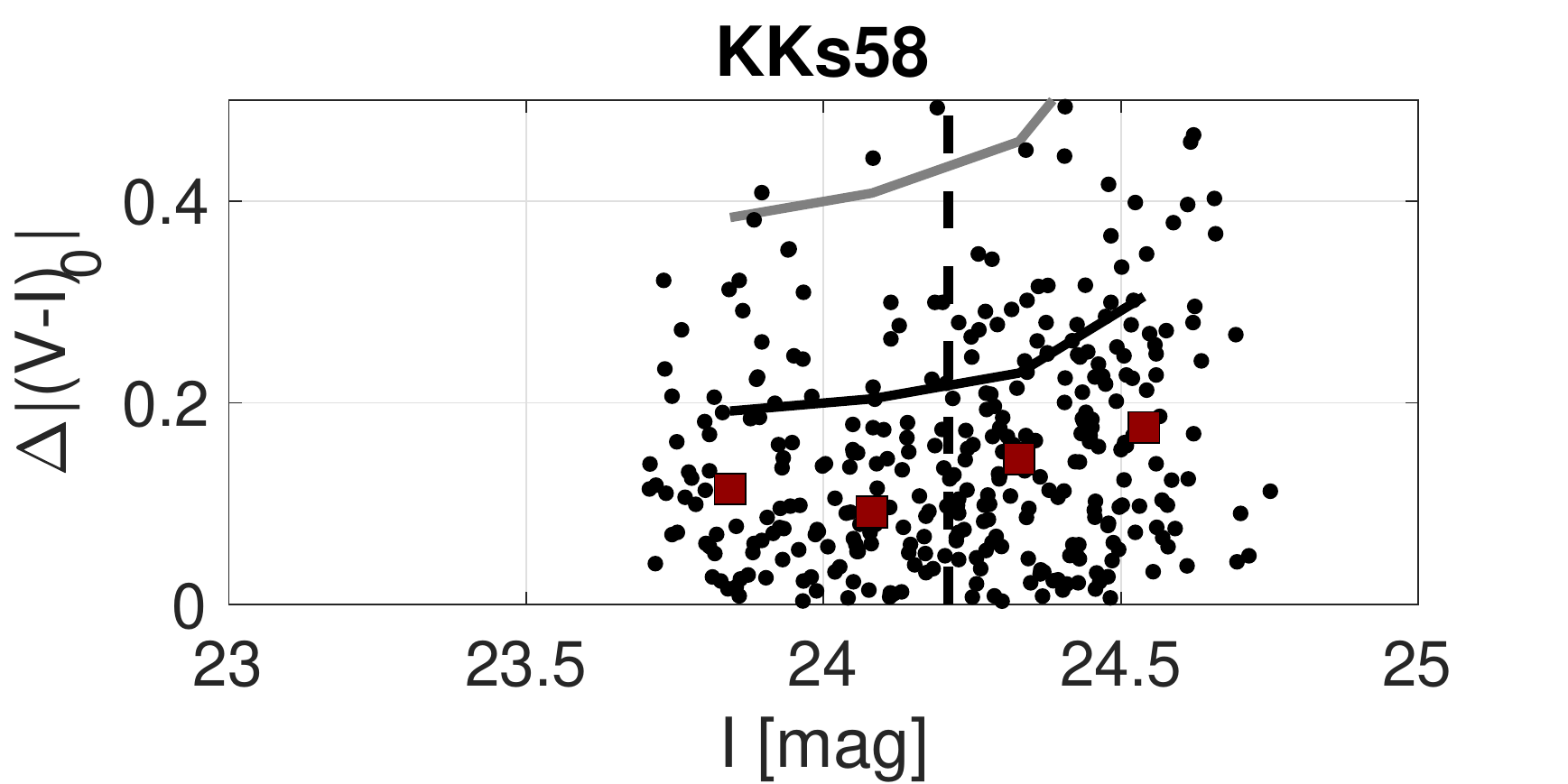}
\caption{Measured $(V-I)_0$ spread as function of $I$-band magnitude for stars observed in {the confirmed galaxies} (black dots) in comparison with  expected uncertainty derived from our artificial star tests. The red squares correspond to the mean values binned over 0.25\,mag steps.  The solid lines corresponds to the 1, 2, and 3 $\sigma$ levels, the dotted line to the 75\% completeness limit. }
\label{app:spread}
\end{figure*}

\begin{figure*}[ht]
\includegraphics[width=9cm]{map_KKs54.pdf}
\includegraphics[width=9cm]{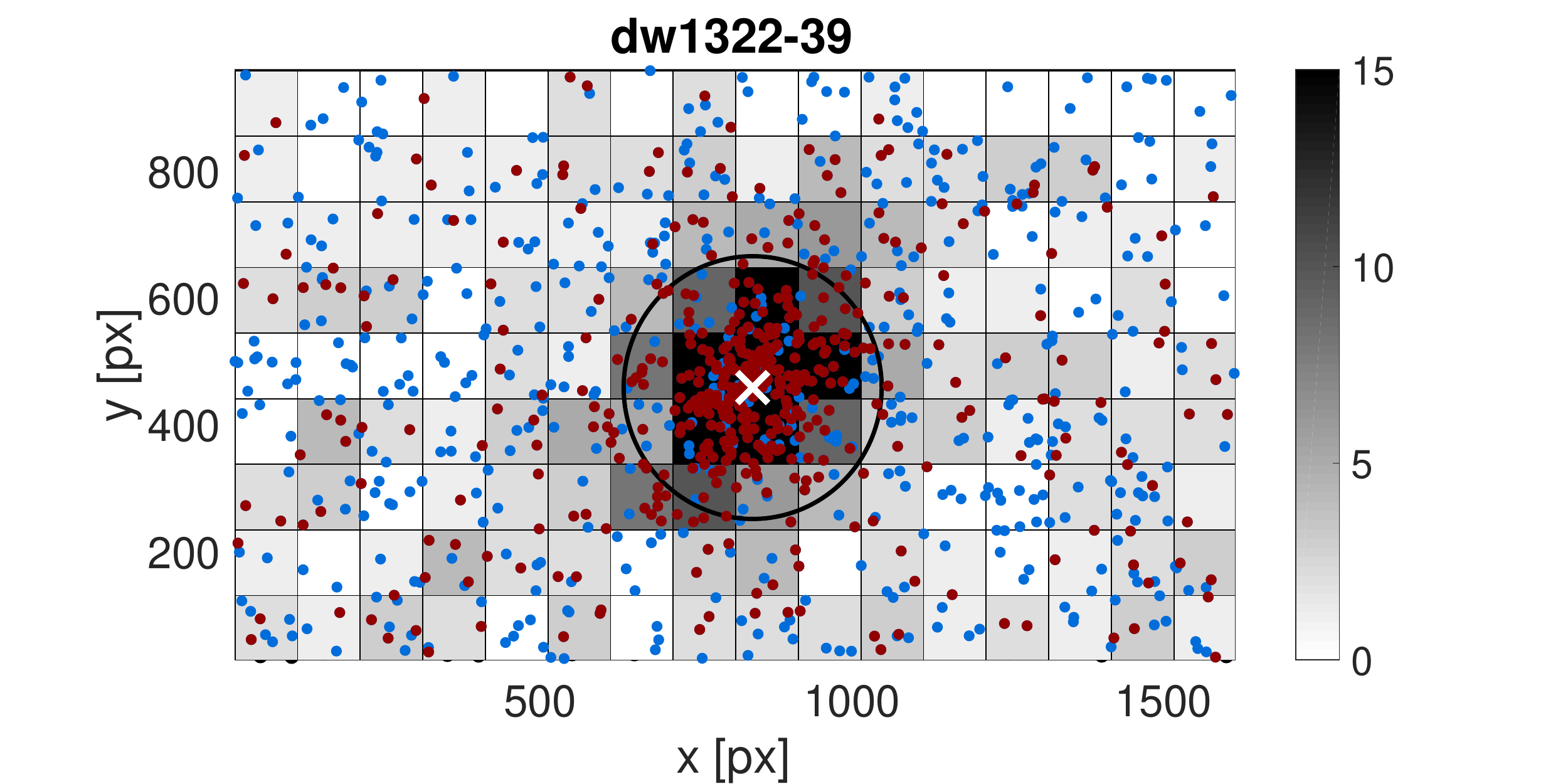}

\includegraphics[width=9cm]{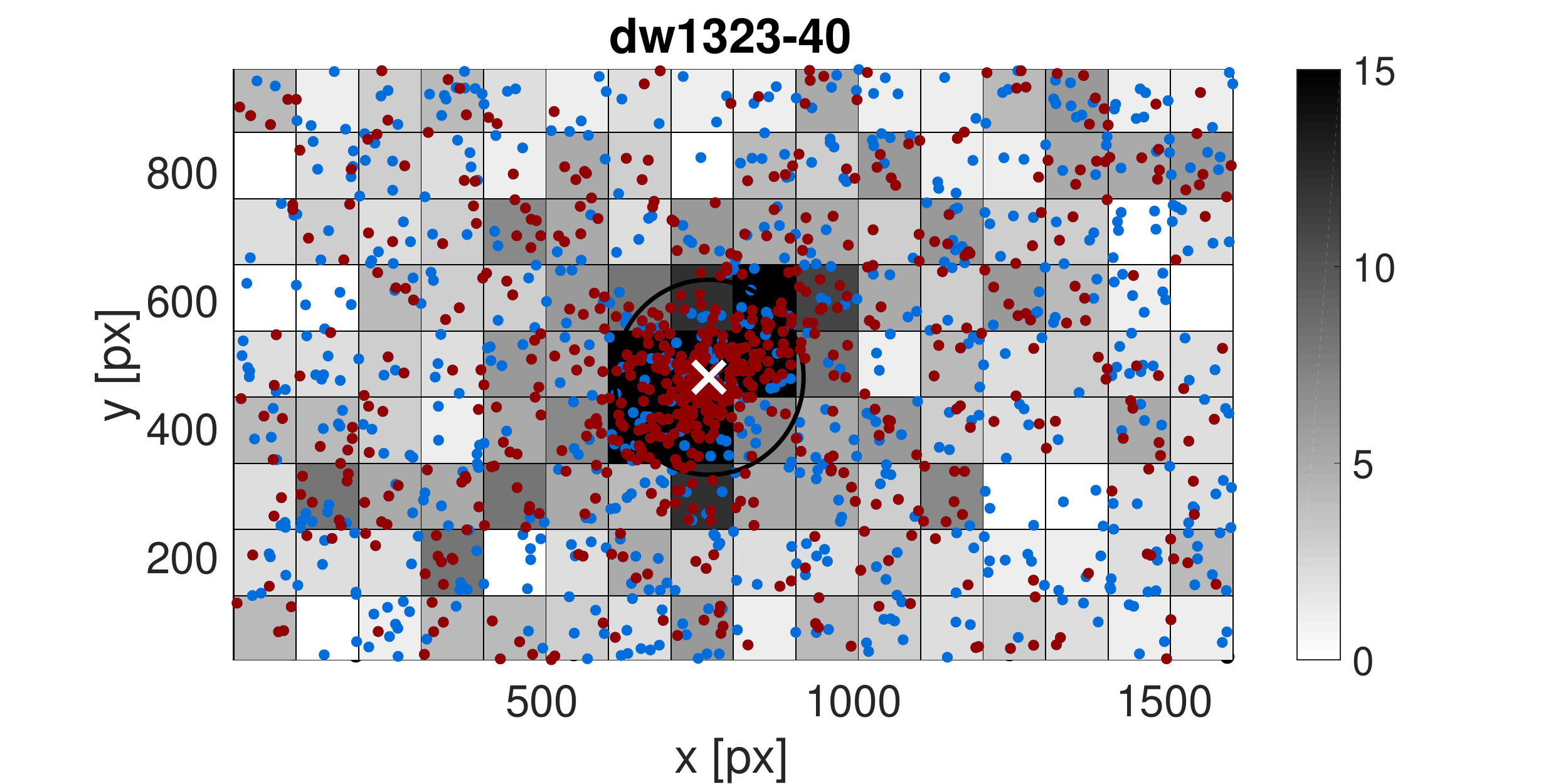}
\includegraphics[width=9cm]{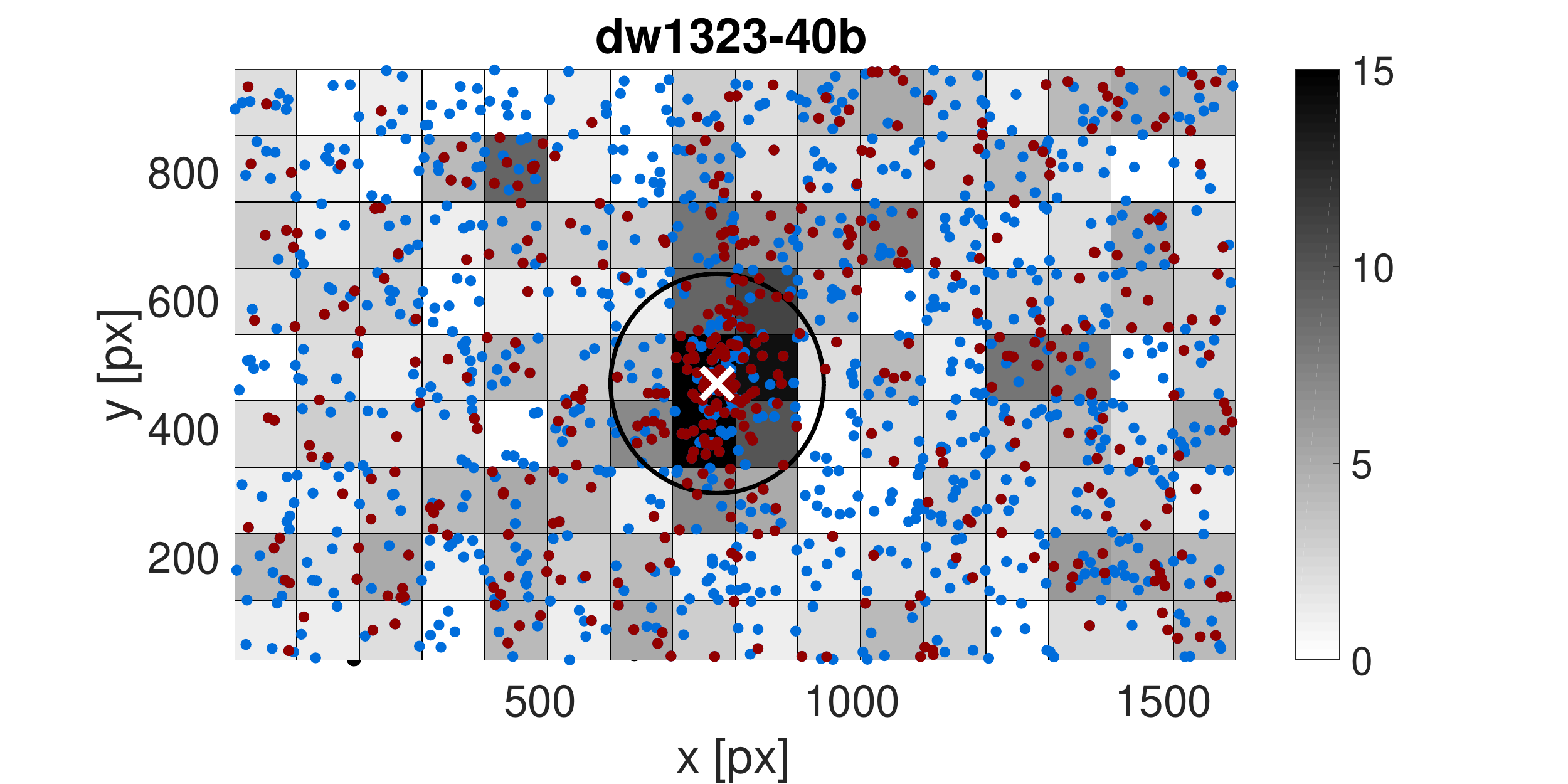}

\includegraphics[width=9cm]{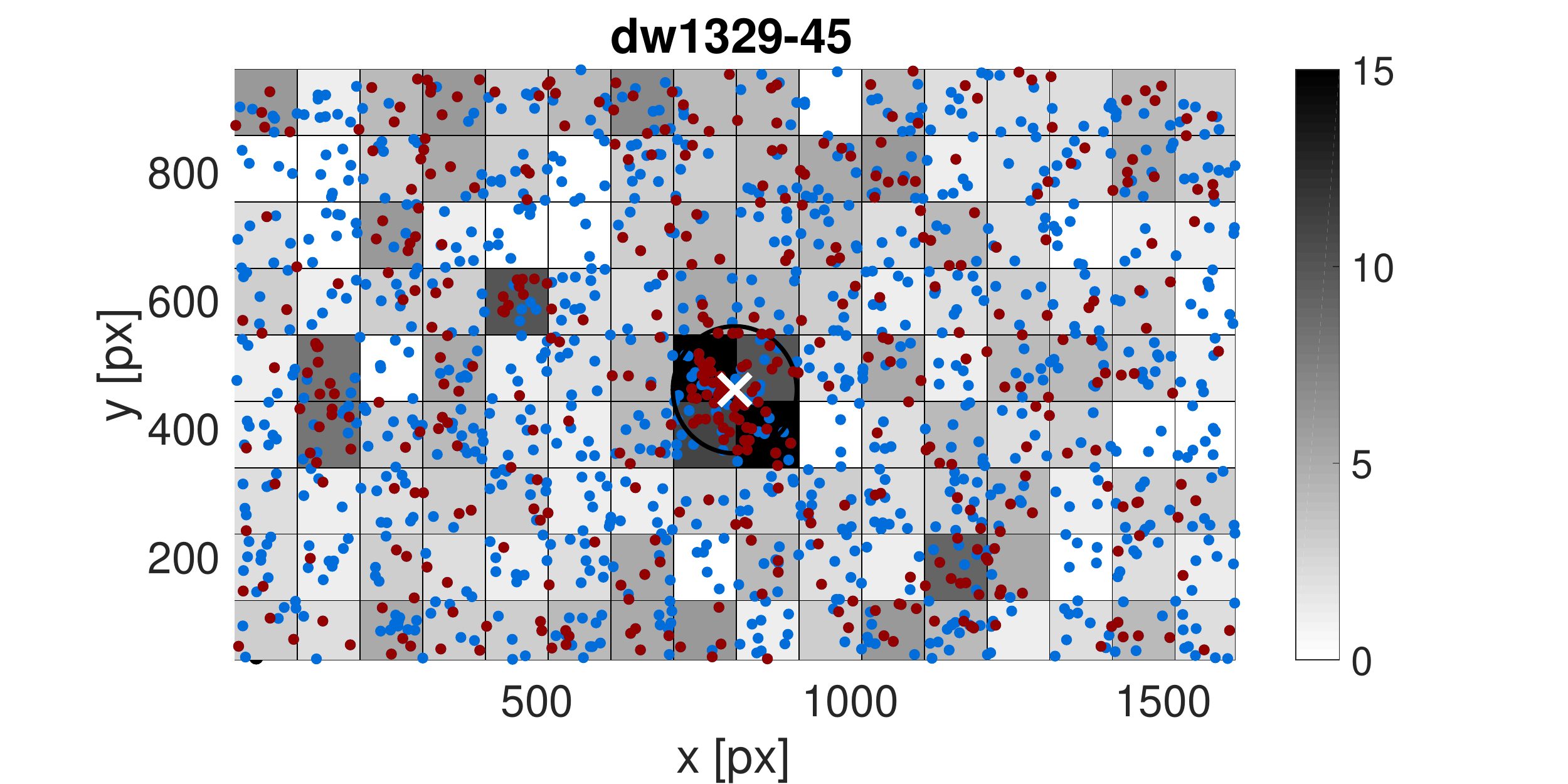}
\includegraphics[width=9cm]{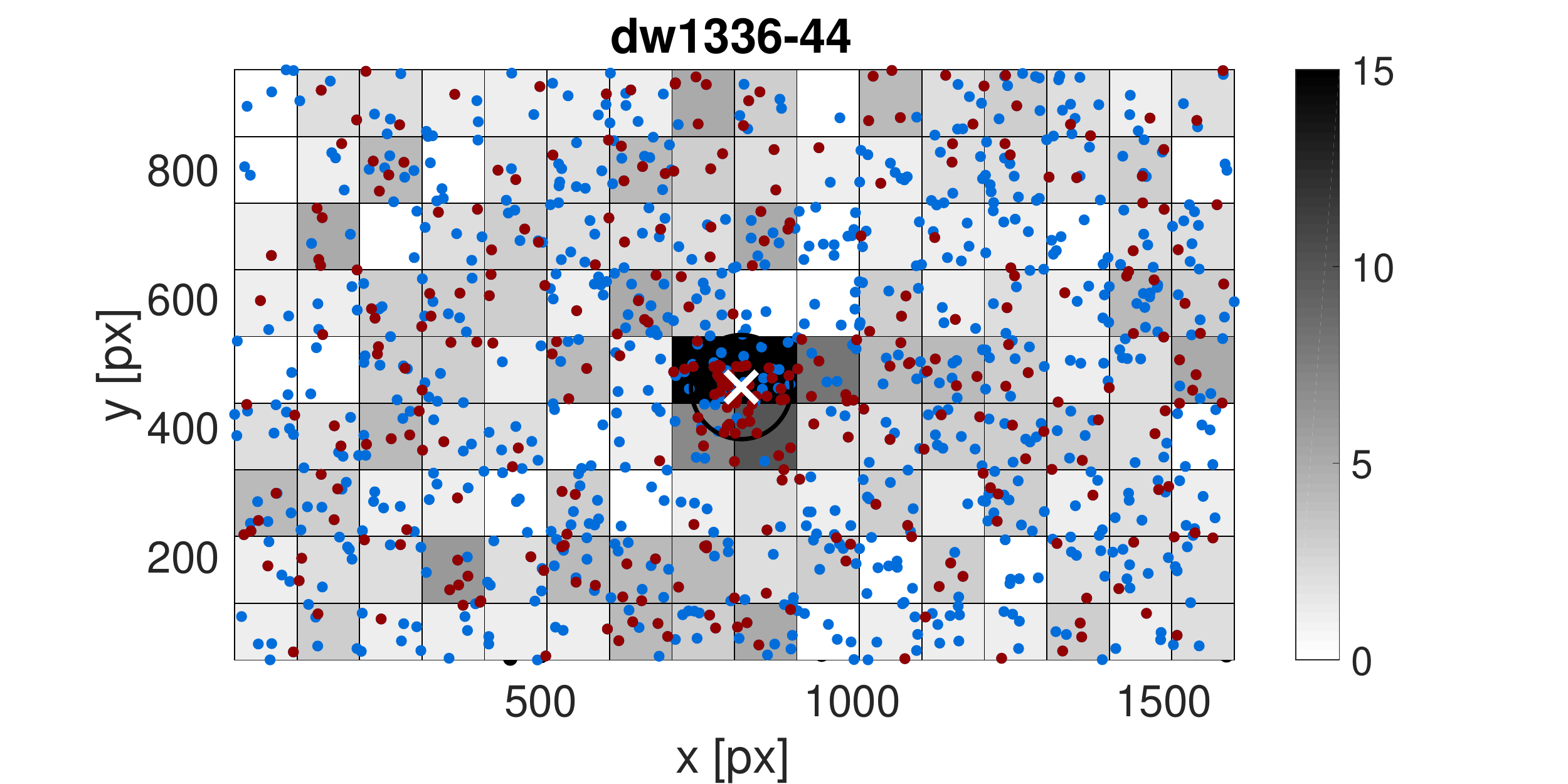}

\includegraphics[width=9cm]{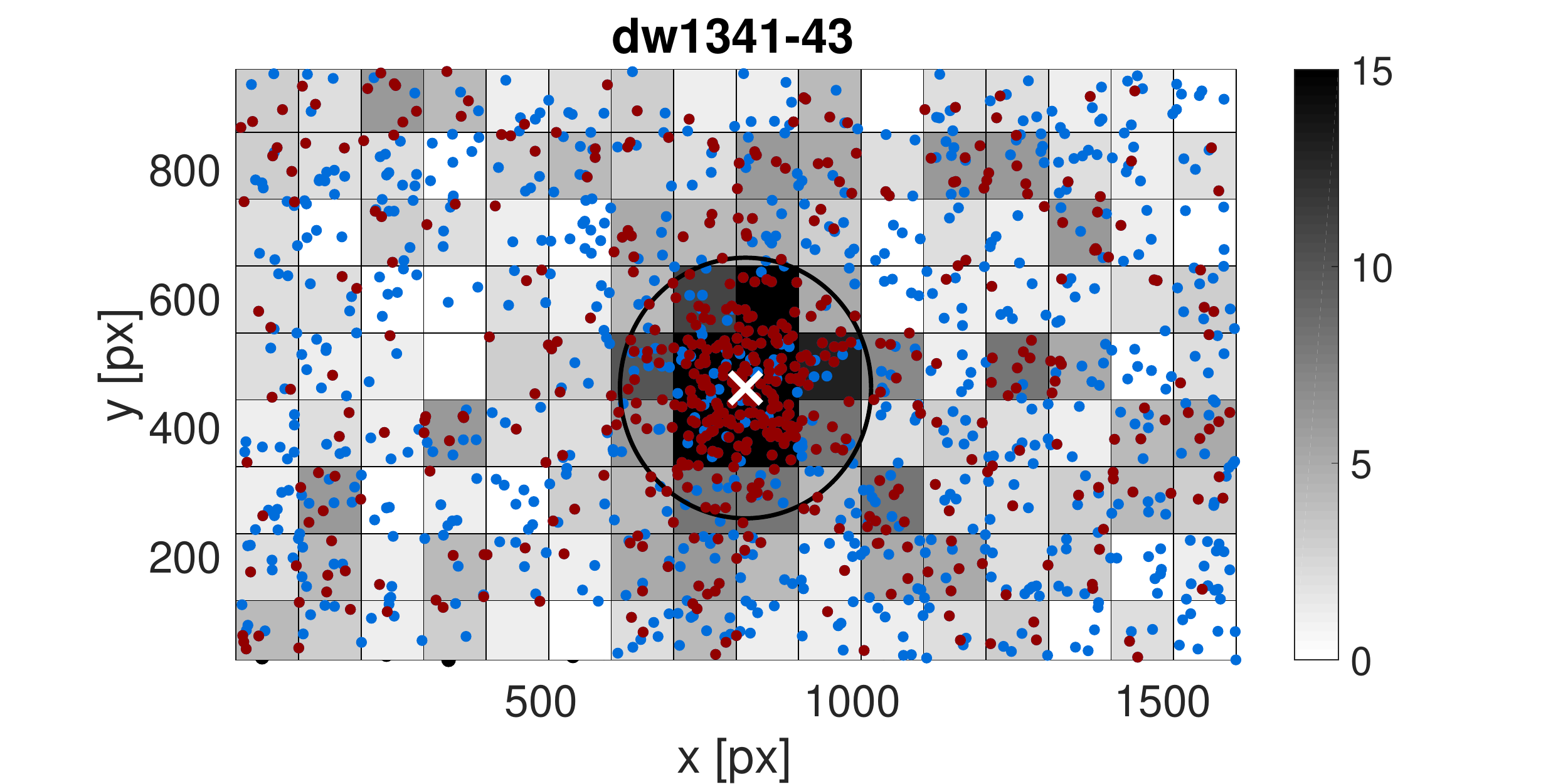}
\includegraphics[width=9cm]{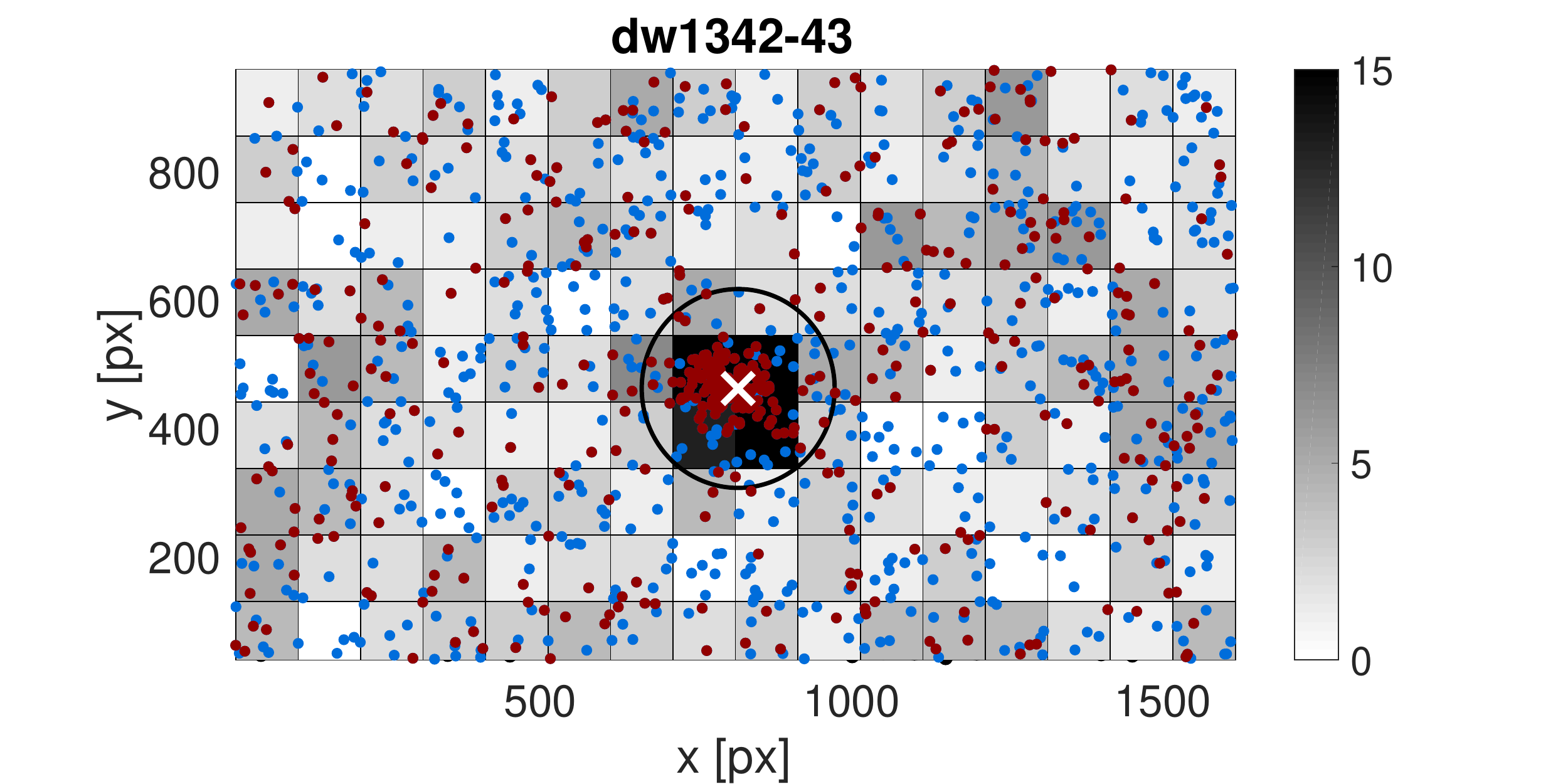}

\includegraphics[width=9cm]{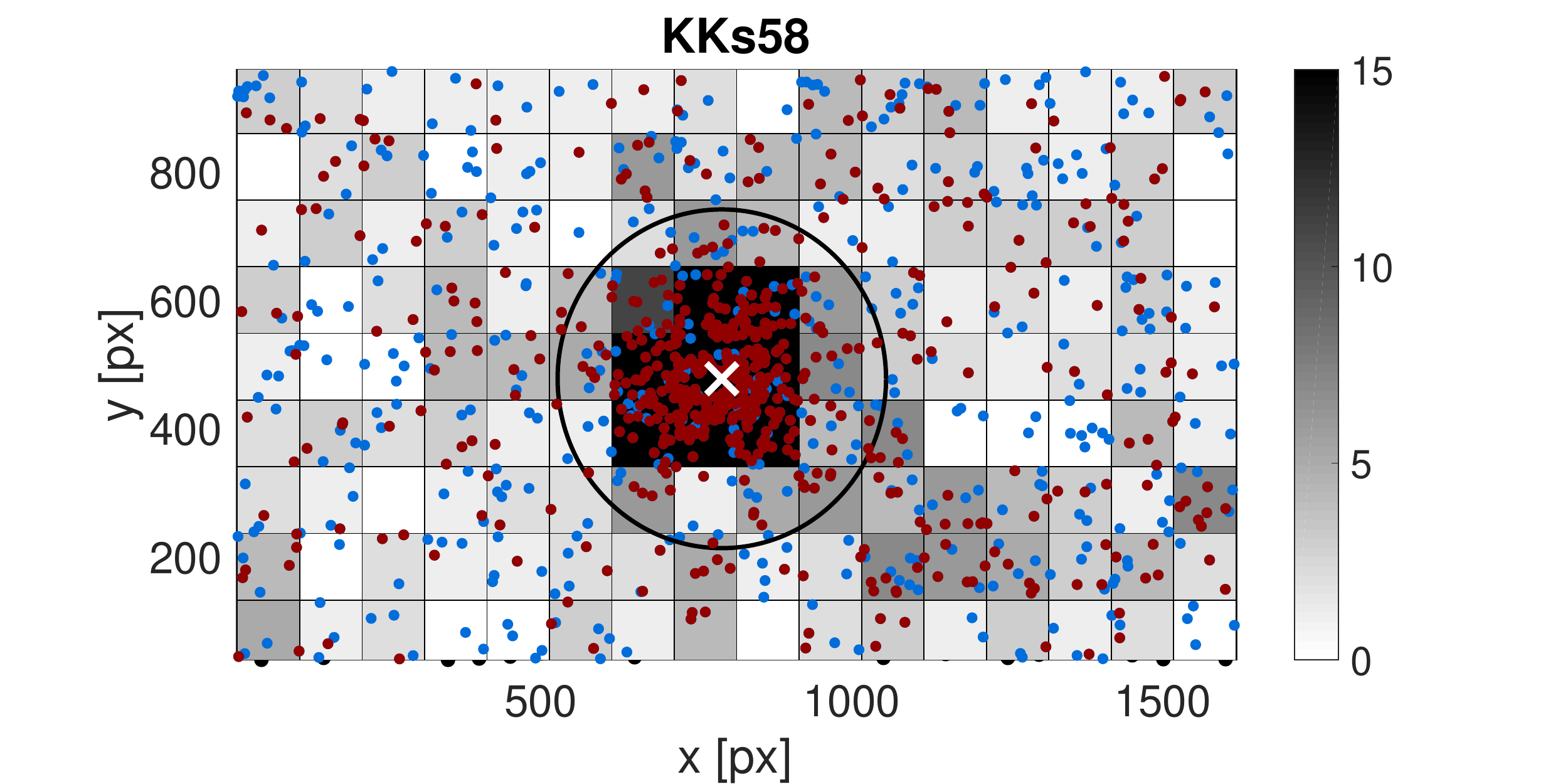}
\caption{Stellar maps of confirmed dwarf galaxies. Red {dots} indicates stars {contained in the RGB mask} according to the best-fitting isochrone. {Blue {dots} indicates the remaining stars.} The circle is the aperture {within} which the CMD is constructed. 
}
\label{app:maps}
\end{figure*}

\begin{figure*}[ht]
\centering
\includegraphics[width=9cm]{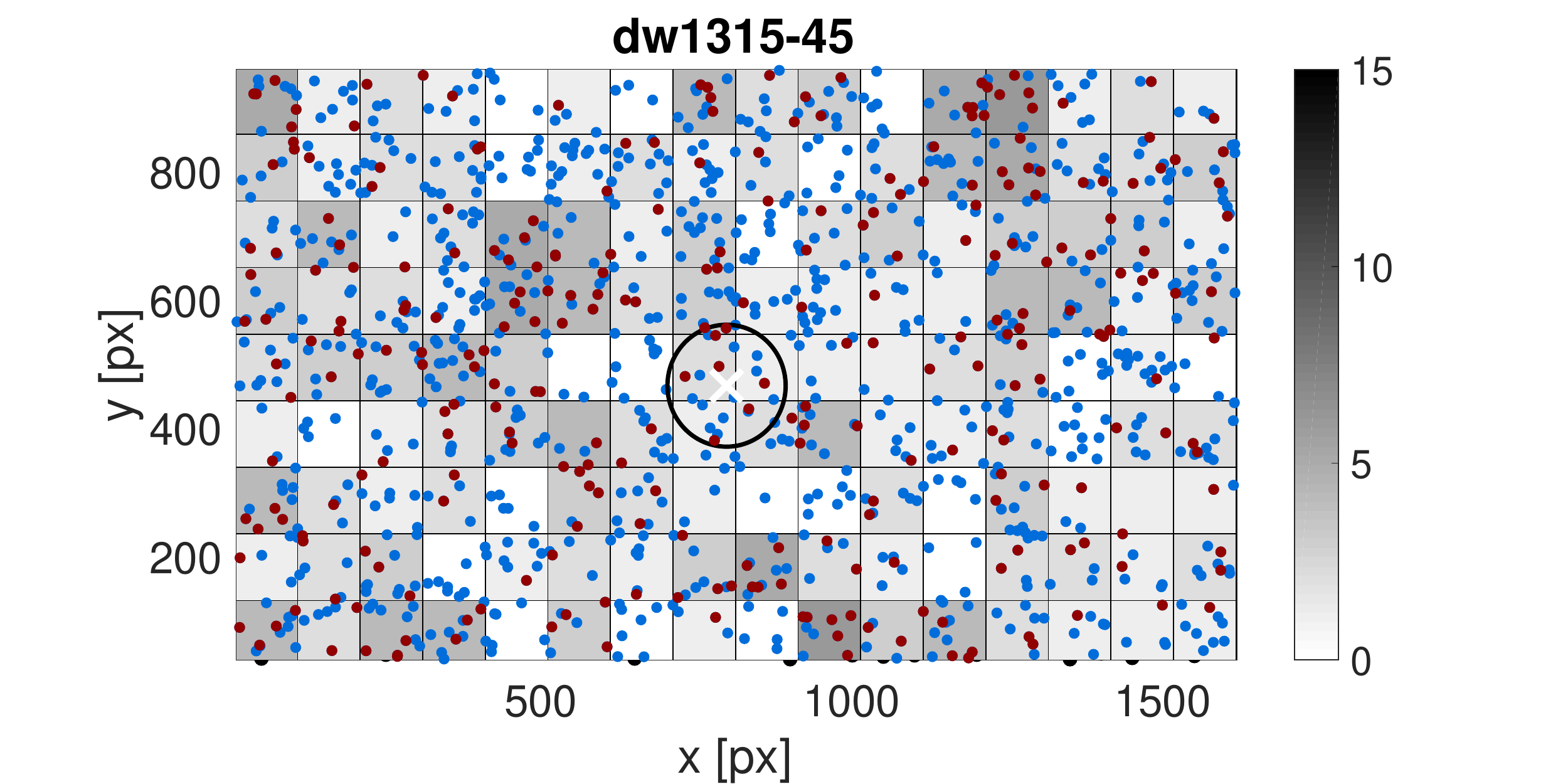}
\includegraphics[width=9cm]{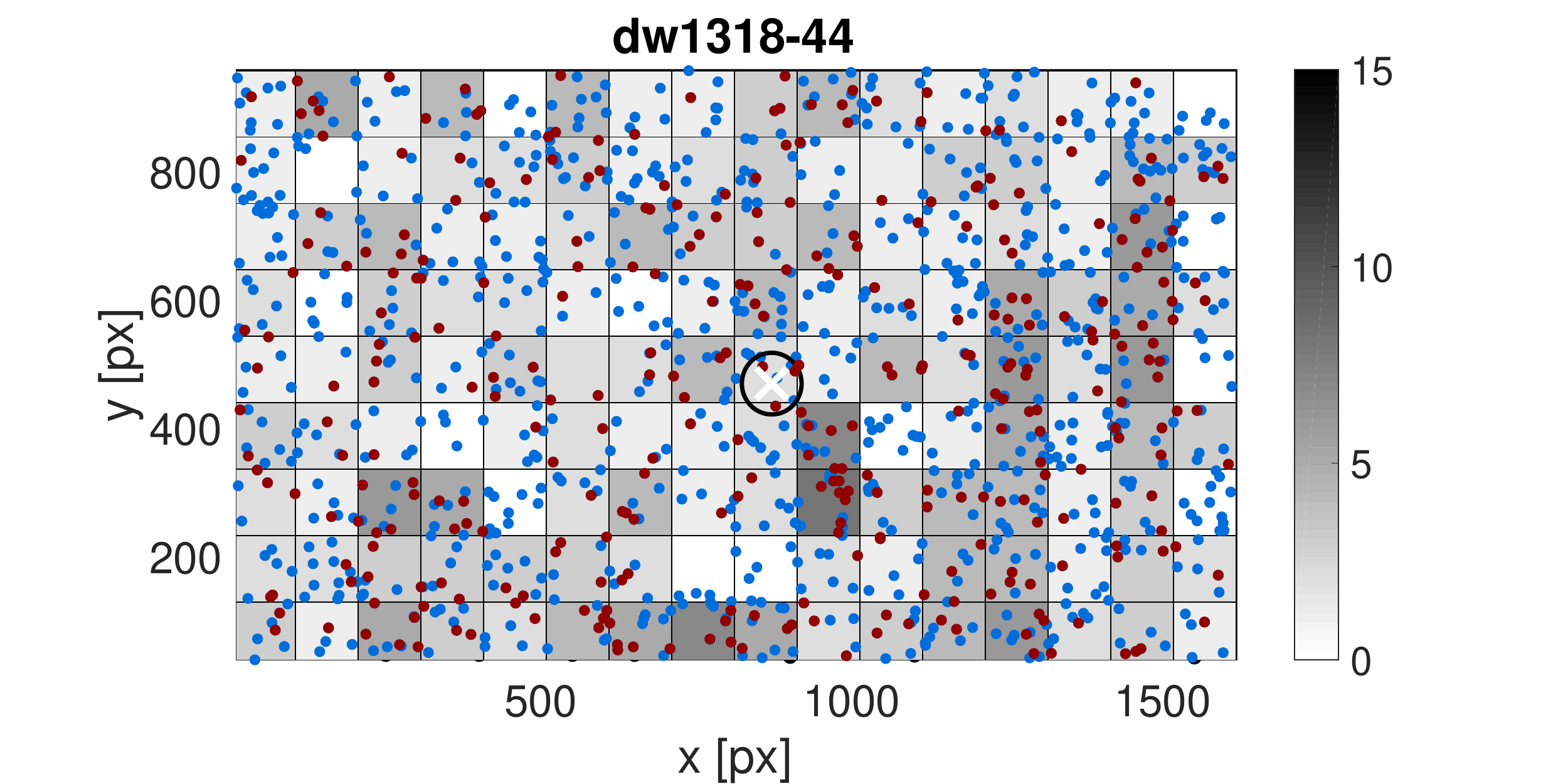}

\includegraphics[width=9cm]{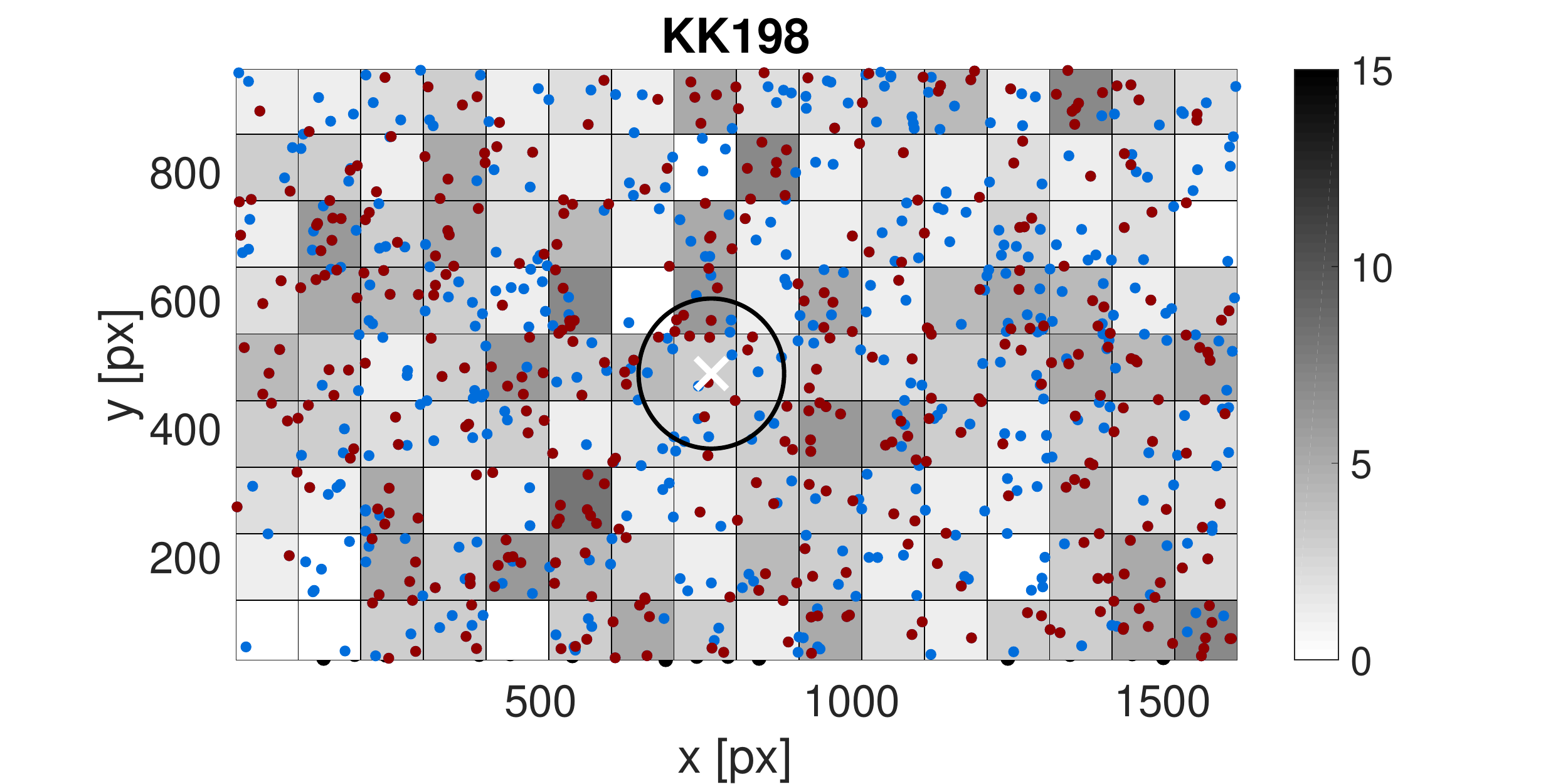}
\includegraphics[width=9cm]{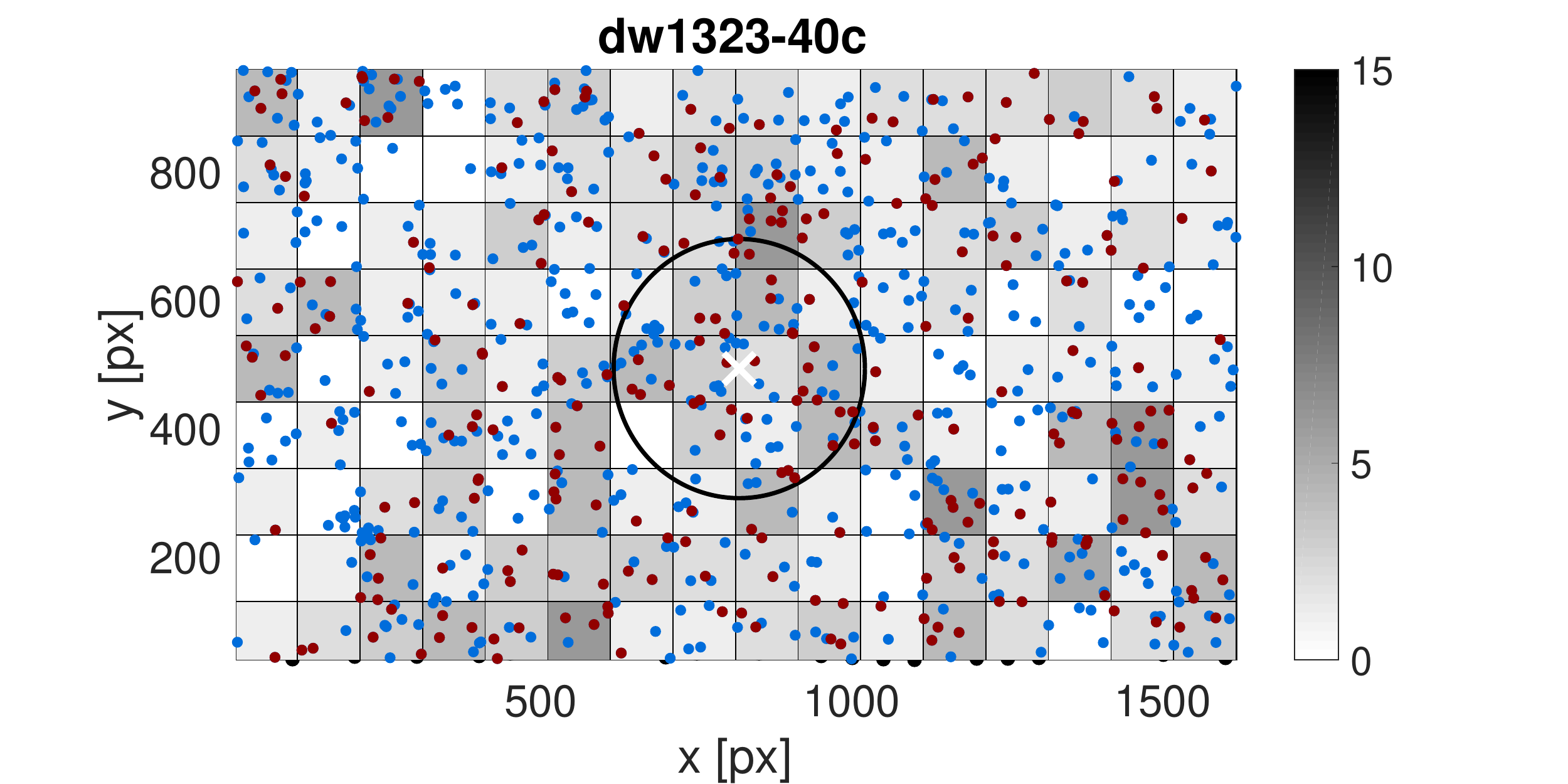}

\includegraphics[width=9cm]{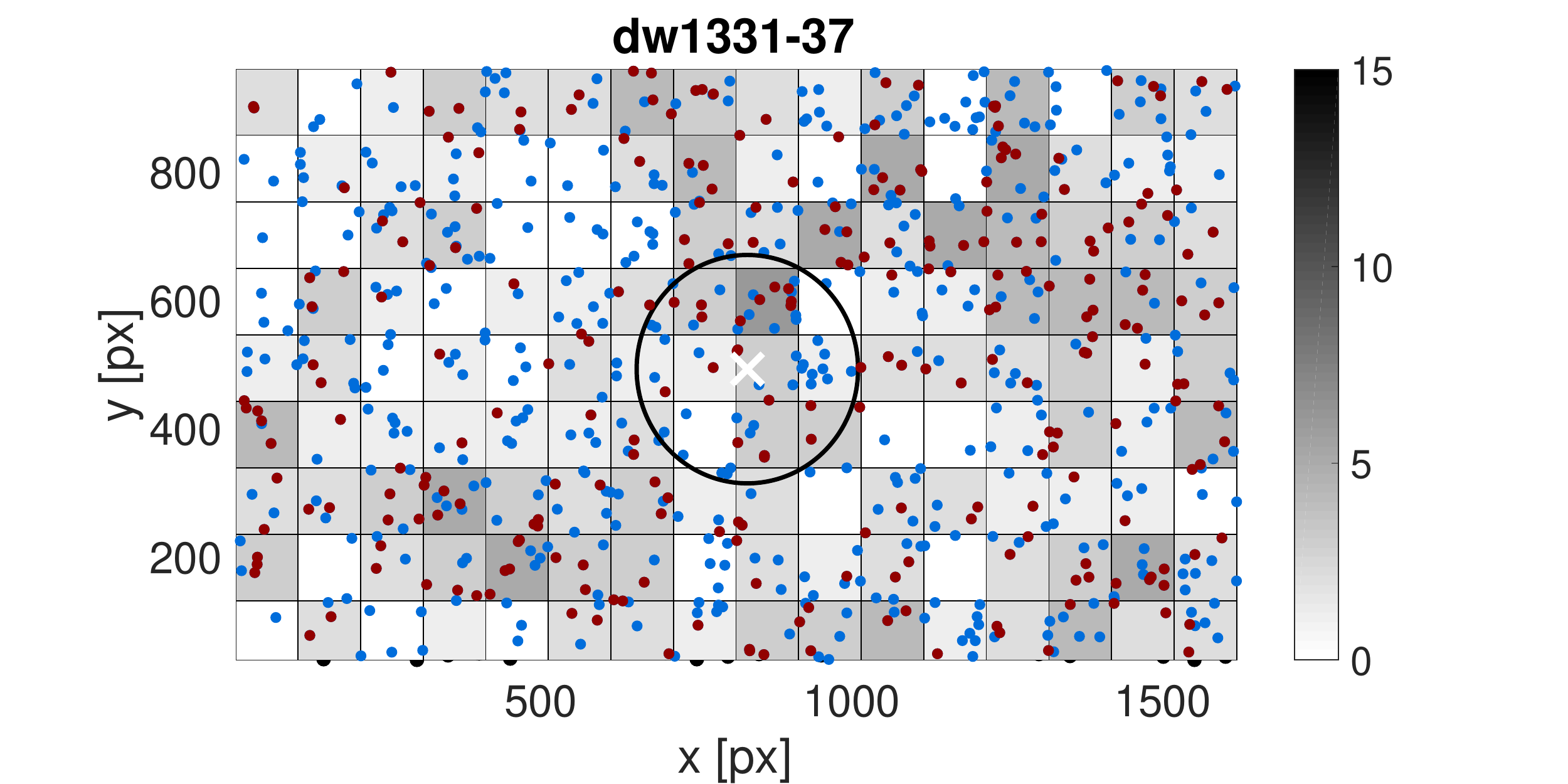}
\includegraphics[width=9cm]{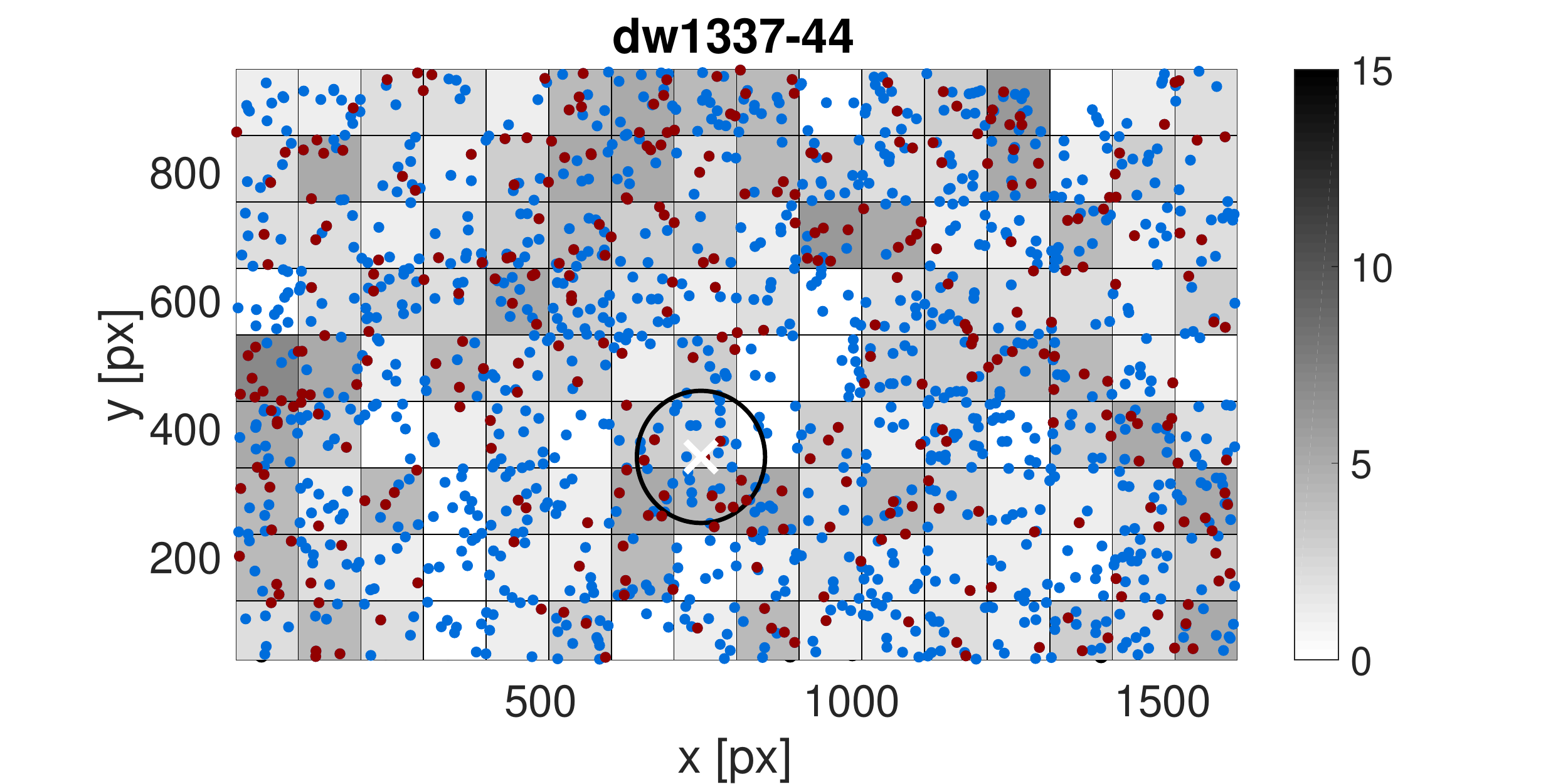}
\caption{Stellar maps of unconfirmed dwarf galaxy candidates. Red indicates stars {contained in the RGB mask} {with $I>23$\,mag and  a color between 0.9 and 1.6\,mag}. {Blue indicates the remaining stars.} The circle is the aperture {within} which the CMD is constructed.}
\label{app:maps2}
\end{figure*}

\end{appendix}

\end{document}